\documentclass[preprint,3p,times]{elsarticle}
\usepackage{tabularx}

 \biboptions{comma,sort&compress}
\usepackage{here}
\usepackage{graphicx}
\usepackage{epsfig}
\usepackage{color}
\usepackage{slashed}
\usepackage{hyperref}
\usepackage{extarrows}

\def\nin{\noindent}
\def\bea{\begin{eqnarray}}
\def\eea{\end{eqnarray}}
\def\nnb{\nonumber}

\newcommand{\intd}{{1\over i}\int{\texttt{d}^dk\over(2\pi)^d}}
\newcommand{\intdd}{{\pi\over i}\int{\texttt{d}^{d+2}k\over(2\pi)^{d+2}}}

\begin{document}

\begin{frontmatter}



\title{Analyses of pion-nucleon elastic scattering amplitudes  up to $O(p^4)$ in extended-on-mass-shell subtraction scheme}


\author[pku,ihep]{Yun-Hua Chen}
\ead{chenyh@ihep.ac.cn}
\author[pku]{De-Liang Yao}
\ead{yaodeliang@pku.edu.cn}
\author[pku]{H.~Q. Zheng}
\ead{zhenghq@pku.edu.cn}
\address[pku]{ Department of Physics and State Key Laboratory of
Nuclear Physics and Technology, Peking University, Beijing 100871,
China P.R.}
\address[ihep]{Institute of High Energy Physics, CAS, P.O. Box 918(4), Beijing
100049, China\footnote{Current address.}}

\begin{abstract}
\noindent We extend the analysis of elastic pion-nucleon scattering
up to $O(p^4)$ level using the extended-on-mass-shell subtraction scheme
within the framework of covariant baryon chiral perturbation theory.
Numerical fits to partial wave phase shift data up to $\sqrt{s}=1.13$~GeV
are performed to pin down the free low energy constants. A good
description of the existing phase shift data is achieved. We find a good convergence for the chiral
series at $O(p^4)$, considerably improved with respect to the $O(p^3)$-level analyses found in
previous literature. Also, the leading order contribution from explicit
$\Delta(1232)$ resonance  and partially-included $\Delta(1232)$ loop contribution are included to describe the phase shift data up
to $\sqrt{s}=1.20$~GeV. As phenomenological applications, we
investigate chiral corrections to the Goldberger-Treiman relation
and find that it converges rapidly, and the $O(p^3)$ correction is
found to be very small:  $\simeq 0.2\%$. We also get a reasonable
prediction of the pion-nucleon sigma term $\sigma_{\pi N}$ up to
$O(p^4)$ by performing fits including both the
pion-nucleon partial wave phase shift data and the lattice QCD data. We report that $\sigma_{\pi N}=52\pm7$~MeV from the fit without $\Delta(1232)$, and $\sigma_{\pi N}=45\pm6$~MeV from the fit with explicit $\Delta(1232)$.
\end{abstract}

\begin{keyword} pion-nucleon scattering\sep chiral perturbation
theory\sep partial wave analysis\sep Goldberger-Treiman relation\sep pion-nucleon sigma term


\end{keyword}

\end{frontmatter}


\section{Introduction}
\label{intro}

{Pion-nucleon scattering is an important process for the understanding of chiral QCD
dynamics and the interpretation of some prominent phenomenology of strong interactions~\cite{Hohler}.
Many efforts have been made to study it.
However, unlike the successfulness of chiral perturbation theory ($\chi$PT)~\cite{ChPT1,ChPT2} in the
pure meson sector, a chiral expansion in the pion-nucleon scattering amplitude suffers
from the power counting breaking (PCB) problem in the traditional
subtraction $\overline{MS}-1$ scheme~\cite{gasser}. Many proposals
have been made to remedy this problem, e.g., heavy baryon (HB) chiral
perturbation theory~\cite{HB}, infrared regularization (IR)
scheme~\cite{becherleutwyler}, extended on mass shell (EOMS)
scheme~\cite{Gegelia99,EOMS}, etc.. }

As a successful nonrelativistic effective field theory, HB chiral
perturbation theory rebuilds a power counting rule through
simultaneous expansions in terms of $1/m_N$ and external momentums.
The pion-nucleon scattering has been investigated up to $O(p^3)$~\cite{Mojzisp3,Meissnerp3} and
$O(p^4)$~\cite{Meissnerp4} with HB approach.
Though the description of $\pi$-N scattering phase shift data is well
described near the threshold region, the nonrelativistic expansion encounters the problem of convergence in many cases~\cite{becherleutwyler,Meissnerp4,Scherer1,Bernard1}, e.g. the scalar form factor of the nucleon does not converge in the region close to the two-pion threshold $t=4M_\pi^2$~\cite{becherleutwyler,Bernard1}.

On the other side, in the framework of relativistic chiral theory,
one may conclude that all the power-violating terms are
polynomials and can thus be absorbed in the low energy constants from
the effective Lagrangian~\cite{Tang2,gegelia94}. Hence  the IR
prescription and EOMS scheme are proposed to retain both correct
power counting and covariance. Nevertheless, they are different in
practice when  removing chiral polynomials, the former subtracts all
the so-called infrared regular part of the loop integrals, which is
always an infinite chiral polynomial of different order, while the
latter only cancels the finite PCB terms. In
Refs.~\cite{Ellis,becherleutwyler2,oller1}  the pion-nucleon
scattering amplitude is analyzed within IR prescription. In
Ref.~\cite{becherleutwyler2}, the $O(p^4)$ calculation was carried out and the analytic
property of the amplitude was discussed. In
Refs.~\cite{Ellis,oller1} the $O(p^3)$ calculation result was used
to fit the phase shift data. However, it has been shown that the
pion-nucleon amplitude in IR prescription is
scale-dependent~\cite{becherleutwyler2} and suffers from
an unphysical cut at u=0~\cite{oller1}. Additionally, a huge violation of
Goldberger-Treiman(GT) relation shows up~\cite{oller1}, which
queries the applicability of covariant baryon chiral perturbation theory.
Hence these  problems lead to the application of the EOMS scheme.

The EOMS scheme provides a good solution to the PCB problem in the
sense that it faithfully respects the analytic structure of the
original amplitudes, e.g., see Ref.~\cite{Geng}, and
being scale independent for merely  making an additional subtraction
of a polynomial of PCB terms with respect to the traditional
substraction. A first attempt of the EOMS scheme on pion-nucleon
scattering was made up to $O(p^3)$ level by Alarc${\rm\acute{o}}$n et
al.~\cite{oller2}. Remarkably, achievements in the description of
violation of the GT relation~\cite{oller3} and the pion-nucleon sigma
term~\cite{oller4} are obtained. However, as pointed in
Ref.~\cite{oller5}, the size of the $O(p^3)$ contribution can be
very large and comparable to those given by the lower-order terms
even at very low energies above threshold. Thus, the applicability
of EOMS scheme to describe the partial wave phase shift at $O(p^3)$
seems to be questionable. The authors of Refs.~\cite{oller2,oller5}
solve the problem by explicitly including the contribution of
$\Delta(1232)$ resonance.

In this paper we extend the analysis up to  $O(p^4)$, and settle down the
convergence problem occurred in the $O(p^3)$-level analysis, even when the
$\Delta(1232)$ resonance is absent. Compared  with the $O(p^4)$-IR
results in Ref.~\cite{becherleutwyler2}, the pion-nucleon scattering
amplitude presented here is the first analytic and complete $O(p^4)$
result.
Especially, we pay great attention to the subtraction of PCB terms,
such that the ``threshold divergence" problem first pointed out
by~\cite{MeissnerIRp3} within the IR prescription never occurs. We perform
fits to partial wave phase shift data and determine all the LECs
involved. Besides, the leading order contribution from an explicit
$\Delta(1232)$ resonance and partially-included $\Delta(1232)$ loop contribution are included to describe phase shift data up
to energies just below the resonance region.

A phenomenological discussion is also made based on the pion-nucleon
scattering amplitude we obtain. We have mainly studied the
Goldberger-Treiman (GT) relation and the pion-nucleon sigma term
$\sigma_{\pi N}$. The GT relation violation is a basic quantity to test
the applicability of EOMS-B$\chi$PT to the pion-nucleon system. Hence we
calculate it up to $O(p^3)$. The prediction of the violation is in
good agreement with other determinations~\cite{DeltaGT1,DeltaGT2}, and its chiral series
converges rapidly. The analysis on $\sigma_{\pi N}$ is important to understand the origin of the mass of the
ordinary matter and can be useful for the study of the supersymmetric dark
matter~\cite{Bottino,JREllis}. Taking both pion-nucleon phase shift
and the lattice QCD data into consideration, we give a reasonable
prediction for the pion-nucleon sigma term: $\sigma_{\pi N}=52\pm7$~MeV from the fit without $\Delta(1232)$, and $\sigma_{\pi N}=45\pm6$~MeV from the fit with an explicit $\Delta(1232)$. The first one is smaller than the $O(p^3)$ result given by Ref.~\cite{oller2} but larger than the recent $O(p^4)$ result in Ref.~\cite{Geng12}, while the latter is in reasonable agreement with previous results found in the
literature~\cite{Gasser90,Weise04,Geng12}.

 \nin

\section{Theoretical discussions on $\pi N\rightarrow\pi N$ in EOMS scheme}
\label{theor}
\nin
\subsection{Kinematics and effective Lagrangian}
\begin{figure}[ht]
\begin{center}
\includegraphics[width=0.3\textwidth]{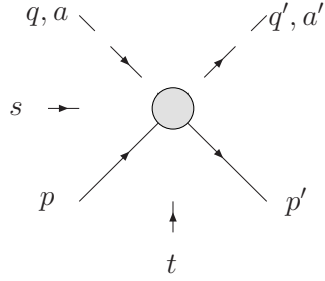}
\caption{Kinematics of elastic $\pi$-N scattering. $p$,$q$
($p^\prime$,$q^\prime$) denote the momenta of the incoming (outgoing)
nucleons and pions, respectively, and $a$ ($a^\prime$) stands for
the isospin index of the incoming (outgoing) pion.}\label{figkinematics}
\end{center}
\end{figure}

In the isospin limit, the standard decomposition of the elastic $\pi$-$N$ amplitude reads~\cite{gasser,becherleutwyler2},
\bea \label{ABform}T_{\pi N}^{a^\prime a}=\chi_{N^\prime}^\dag\left\{\delta_{a^\prime
a}T^++\frac{1}{2}[\tau_{a^\prime},\tau_a]T^-\right\}\chi_N, \,\,\,\,\,\,
T^{\pm}=\bar{u}(p^\prime,s^\prime)\left[A^\pm+\frac{1}{2}\left(\slashed{q}^\prime+\slashed{q}\right)B^\pm\right]u(p,s)\
, \eea
 where $p$,$q$($p^\prime$,$q^\prime$) denote the momenta of the
incoming (outgoing) nucleons and pions, respectively, and $a$
($a^\prime$) stands for the isospin index of the incoming (outgoing)
pion, see Fig.\ref{figkinematics}. $\tau_{a^\prime},\tau_{a}$ are
Pauli matrices and $\chi_{N^\prime}$, $\chi_N$ are the isospinors of the nucleons. For
on-shell elastic scatterings,
$p^2=p^{\prime2}=m_N^2,q^2=q^{\prime2}=M_\pi^2$ and the Mandelstam
variables  $s$,$t$ and $u$ fulfill $s+t+u=2m_N^2+2M_\pi^2$. Eq.~(\ref{ABform}) can be written in an alternative form through the replacement of $A$ by
$D=A+\nu B$ with $\nu=\frac{s-u}{4m_N}$:
\bea
\label{DBform} T^{\pm}=\bar{u}(p^\prime,s^\prime)\left[D^\pm+{i\over
2m_N}\sigma^{\mu\nu}q_\mu^\prime q_\nu B^\pm\right]u(p,s)\ .
\eea
Since the leading order contribution of A and B may cancel each other,
one should better use D and B to perform the low energy expansion of the
scattering amplitude when extracting the PCB terms. Our calculation of
the $\pi$-N scattering amplitude up to $O(p^4)$ level demands the
corresponding calculation of $D^\pm$ and $B^\pm$ up to $O(p^4)$ and
$O(p^2)$, respectively.

In chiral perturbation theory, each graph is assigned an overall
chiral order $D$, which means the graph is of size
$\left(p/\Lambda\right)^D$, where one has the soft scale $p\ll\Lambda$ and $\Lambda$ stands for a ``high energy scale'', i.e., the breakdown scale of the theory. For processes containing one baryon
in the initial and final states, the chiral order for a given graph
with $L$ loops, $V_n$ $n$-th order vertices, $N_M$ meson propagators
and $N_B$ baryon propagators, is given by \bea D=4L+\sum_nn
V_n-2N_M-N_B\ . \eea

The effective Lagrangian relevant to the
one-nucleon sector consists of $\pi$-$N$ and purely mesonic
Lagrangian, \bea {\cal L}_{eff}={\cal L}_{\pi N}^{(1)}+{\cal L}_{\pi
N}^{(2)}+{\cal L}_{\pi N}^{(3)}+{\cal L}_{\pi N}^{(4)}+\cdots+{\cal
L}_{\pi\pi}^{(2)}+{\cal L}_{\pi\pi}^{(4)}+\cdots,\eea where the
superscripts denote the chiral order.
The lowest $\pi$-$N$ Lagrangian takes the standard form
\bea\label{powercounting} {\cal L}_{\pi
N}^{(1)}=\bar{N}\left\{i\slashed{D}-m+\frac{1}{2}g\slashed{u}\gamma^5\right\}N\
. \eea The nucleons are described by an isospin doublet as
$N=(n,p)^T$, and the covariant derivative $D_\mu$
acting on it is defined as $D_\mu=\partial_\mu+\Gamma_\mu$, with
\bea \Gamma_\mu=\frac{1}{2}\left[u^\dag(\partial_\mu-i
r_\mu)u+u(\partial_\mu-i l_\mu) u^\dag\right]\ , \eea where $l_\mu$
and $r_\mu$ are constructed from the external vector and axial
vector currents as $l_\mu=v_\mu-a_\mu$ and $r_\mu=v_\mu+a_\mu$. The
Goldstone bosons are collected in a $2\times2$ matrix-valued field
$u$ in the so-called exponential parametrization \bea\label{exp}
u_\mu=i\left[u^\dag(\partial_\mu-i r_\mu)u-u(\partial_\mu-i l_\mu)
u^\dag\right]\ ,\quad
u=\exp\left(\frac{i\vec{\tau}\cdot\vec{\pi}}{2F}\right)\ , \eea with
$\vec{\tau}$ being the Pauli matrices. The parameters appearing in this lowest-order
$\pi$-$N$ Lagrangian, $m$, $F$, and $g$ are the bare values of the
nucleon mass, the pion decay constant and the axial charge,
respectively.

For the complete form of ${\cal L}_{\pi N}^{(2)}$, ${\cal L}_{\pi
N}^{(3)}$ and ${\cal L}_{\pi N}^{(4)}$, we refer to Ref.~\cite{eff}.
Here we only write down the terms which are relevant to our calculation:
\bea {\cal L}_{\pi
N}^{(2)}&=&c_1\langle\chi_+\rangle\bar{N}N-\frac{c_2}{4m^2}\langle
u^\mu u^\nu\rangle(\bar{N}D_\mu D_\nu N+h.c.)+\frac{c_3}{2}\langle
u^\mu
u_\mu\rangle\bar{N}N-\frac{c_4}{4}\bar{N}\gamma^\mu\gamma^\nu[u_\mu,u_\nu]N\ ,\label{effp2}\\
{\cal L}_{\pi N}^{(3)}&=&\bar{N}\left\{-\frac{d_1+d_2}{4m}\big([u_\mu,[D_\nu,u^\mu]+[D^\mu,u_\nu]]D^\nu+h.c.)\right.\nnb\\
&&+\frac{d_3}{12m^3}([u_\mu,[D_\nu,u_\lambda]](D^\mu D^\nu
D^\lambda+sym.)+h.c.\big)+i\frac{d_5}{2m}([\chi_-,u_\mu]D^\mu+h.c.)\nnb\\
&&+i\frac{d_{14}-d_{15}}{8m}(\sigma^{\mu\nu}\langle[D_\lambda,u_\mu]u_\nu-u_\mu[D_\nu,u_\lambda]\rangle
D^\lambda+h.c.)\nnb\\
&&\left.+\frac{d_{16}}{2}\gamma^\mu\gamma^5\langle\chi_+\rangle
u_\mu+\frac{id_{18}}{2}\gamma^\mu\gamma^5[D_\mu,\chi_-]\right \}N\ ,\label{effp3}\\
{\cal L}_{\pi N}^{(4)}&=&\bar{N}\left\{
e_{14}%
\langle h_{\mu \nu}h^{\mu \nu }\rangle
-\frac{e_{15}}{4m^{2}}%
\left(\langle h_{\lambda \mu }h_{\;\;\nu }^{\lambda }\rangle D^{\mu \nu
} + {\rm h.c.}\right)
+\frac{e_{16}}{48m^{4}}%
\left(\langle h_{\lambda \mu }h_{\nu \rho }\rangle D^{\lambda \mu \nu
\rho } + {\rm h.c.}\right)\right.\nonumber\\
&&+\frac{ie_{17}}{2}%
[h_{\lambda \mu },h_{\;\;\nu }^{\lambda }] \sigma ^{\mu \nu }
-\frac{ie_{18}}{8m^{2}}%
\left([
h_{\lambda \mu },h_{\nu \rho }] \sigma ^{\mu \nu }D^{\lambda \rho } + {\rm h.c.}\right)
+e_{19}%
\langle \chi_{+}\rangle \langle u\cdot u\rangle\nonumber\\
&&-\frac{e_{20}}{4m^{2}}%
\left(\langle \chi _{+}\rangle\langle u_{\mu }u_{\nu
}\rangle D^{\mu \nu } + {\rm h.c.}\right)
+\frac{ie_{21}}{2}%
\langle\chi _{+}\rangle[u_{\mu },u_{\nu }] \sigma ^{\mu \nu }
+e_{22}%
[ D_{\mu },[D^{\mu },\langle \chi _{+}\rangle ]]\nonumber\\
&&-\frac{ie_{35}}{4m^{2}}%
\left(\langle \widetilde{\chi }_{-}h_{\mu \nu }\rangle D^{\mu \nu } + {\rm h.c.}\right)
+ie_{36}\langle u_{\mu }%
[ D^{\mu },\widetilde{\chi }_{-}]\rangle
-\frac{e_{37}}{2}%
[ u_{\mu},[ D_{\nu },\widetilde{\chi }_{-}]] \sigma ^{\mu \nu }
+e_{38}\langle \chi_{+}\rangle \langle \chi _{+}\rangle\nonumber\\
&&+\left.\frac{e_{115}}{4}%
\langle\chi_{+}^{2}-\chi _{-}^{2}\rangle
-\frac{e_{116}}{4}%
\left( \langle \chi_{-}^{2}\rangle -\langle \chi _{-}\rangle
^{2}+\langle \chi _{+}^{2}\rangle -\langle \chi _{+}\rangle ^{2}
\right) \right\}N\ , \label{effp4}\eea where the $c_i$, $d_j$ and $e_k$ are the
low energy constants. The new symbols appearing here are defined as follows,
\bea
\chi^\pm&=&u^\dag\chi u^\dag\pm u\chi^\dag u~,\nnb\\
h_{\mu\nu}&=&[D_{\mu},u_{\nu}] +[D_{\nu},u_{\mu}]~,\nonumber\\
\widetilde{\chi}_{\pm}&=&\chi_{\pm}-\frac12\langle \chi_{\pm}\rangle~,\nonumber\\
D_{\alpha\beta\ldots\omega}&=&\left\{D_{\alpha} D_{\beta}
\ldots D_{\omega}+sym.\right\}~.
\eea
Here $\chi={\cal M}=diag(M^2,M^2)$ and $M$ is the bare pion mass. In the pure meson sector,
 the relevant terms of ${\cal L}_{\pi\pi}^{(2)}$ and ${\cal
L}_{\pi\pi}^{(4)}$  are given by \bea {\cal
L}_{\pi\pi}^{(2)}&=&\frac{F^2}{4}\langle u^\mu
u_\mu+\chi_+\rangle ,\nonumber\\
{\cal L}_{\pi\pi}^{(4)}&=&\frac{1}{8}l_4\langle u^\mu
u_\mu\rangle\langle \chi_+\rangle+\frac{1}{16}(l_3+l_4)\langle
\chi_+\rangle^2, \eea and $l_3,l_4$ are low energy constants that
will appear in our calculation, too. It is noticed that
throughout this paper $m, M, g, F$ represent the bare quantities for
nucleon mass, pion mass, axial coupling constant and pion decay
constant, respectively whereas $m_N, M_\pi, g_A, F_\pi$ the
corresponding physical quantities. For the kinematic region close to the $\pi$N threshold, one has
$$\frac{\sigma}{\Lambda^2}\ll1\ ,\frac{t}{\Lambda^2}\ll1\ ,\frac{M_\pi}{\Lambda}\ll1\ ,$$
or equivalently
$$\frac{\nu}{\Lambda}\ll1\ ,\frac{\nu_B}{\Lambda}\ll1\ ,\frac{M_\pi}{\Lambda}\ll1\ ,$$
where $\sigma=s-m_N^2$, $\nu=\frac{s-u}{4m_N}$, $\nu_B=\frac{t-2M_\pi^2}{4m_N}$. Here the high energy scale $\Lambda=\{4\pi F_\pi,~m_N,~m_\Delta, m_\Delta-m_N\}$ with $m_\Delta$ the mass of $\Delta(1232)$. Hence $\sigma$, $t$, $M_\pi$ (or $\nu$, $\nu_B$, $M_\pi$) are adopted as expansion parameters, and
\bea
\sigma\sim O(p)\ ,t\sim O(p^2)\ ,M_\pi\sim O(p)\ ,\nu\sim O(p)\ ,\nu_B\sim O(p^2)\ ,m_N\sim O(p^0)\ ,m_\Delta\sim O(p^0)\ ,m_\Delta-m_N\sim O(p^0)\ .
\eea

\subsection{Tree amplitudes}\label{treea}

\begin{figure}[ht]
\begin{center}
\includegraphics[width=0.6\textwidth]{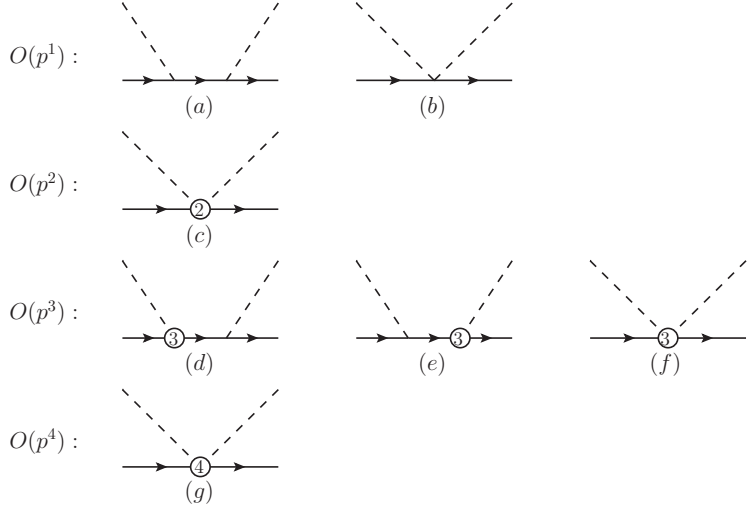}
\caption{Tree graphs for $\pi$-N scattering up to $O(p^4)$. The solid lines correspond to nucleons,
while the dashed lines represent pions. The vertices with circled 2,
3 and 4 stem from ${\cal L}_{\pi N}^{(2)}$, ${\cal L}_{\pi N}^{(3)}$
and ${\cal L}_{\pi N}^{(4)}$, respectively. The nucleon propagates
with $m_4=m-4c_1M^2-2(8e_{38}+e_{115}+e_{116})$. Crossed
diagrams for (a), (d) and (e) are not shown.
}\label{treediagrams}
\end{center}
\end{figure}

We show all the tree graphs which contribute up to $O(p^4)$ in
Fig.~\ref{treediagrams} according to the power counting rule given
by Eq.~(\ref{powercounting}), and list their contributions to
$D^{\pm}$ and $B^{\pm}$ in~\ref{apptree} separately. The nucleon
propagates with a mass parameter
$m_4=m-4c_1M^2-2(8e_{38}+e_{115}+e_{116})M^4$ instead of $m$, so
that the graphs with mass insertions, generated by
$c_1\langle\chi_+\rangle$ in ${\cal L}_{\pi N}^{(2)}$ and
$e_{38}\langle \chi_{+}\rangle \langle \chi
_{+}\rangle+\frac{e_{115}}{4} \langle\chi_{+}^{2}-\chi
_{-}^{2}\rangle -\frac{e_{116}}{4} \left( \langle
\chi_{-}^{2}\rangle -\langle \chi _{-}\rangle ^{2}+\langle \chi
_{+}^{2}\rangle -\langle \chi _{+}\rangle ^{2}\right)$ in ${\cal
L}_{\pi N}^{(4)}$, in the nucleon propagators are automatically
considered. For convenience, one can classify the tree graphs into
two categories: Born-terms and contact terms. In
Fig.~\ref{treediagrams} contributions from Born-term graphs, (a),
(d) and (e), and their crossed diagrams can be summed and rewritten
concisely in terms of the $A$ and $B$ functions as \bea\label{bornterm}
A^\pm&=&A(s)\pm A(u),\qquad A(s)=\frac{g_2^2}{4F^2}\frac{s-m_N^2}{s-m_4^2}(m_4+m_N),\nonumber\\
B^\pm&=&B(s)\mp
B(u),\qquad B(s)=-\frac{g_2^2}{4F^2}\frac{s+2m_Nm_4+m_N^2}{s-m_4^2}, \eea
with\footnote{As for the $O(p^4)$ effective Lagrangian, we adopt the
conventions of Ref.~\cite{eff}, so $m_4$ differs from the one
in~\cite{becherleutwyler2}, where
$e_1^{BL}=-2(8e_{38}+e_{115}+e_{116})$.}
$m_4=m-4c_1M^2-2(8e_{38}+e_{115}+e_{116})M^4$ and
$g_2=g+2M^2(2d_{16}-d_{18})$. Meanwhile, the rest are contact term
graphs without crossed diagrams, and the sum of them is
\bea\label{contactterm}
D^+&=&-\frac{4\hat{c}_1M^2}{F^2}+\frac{\hat{c}_2\left(16m_N^2\nu^2-t^2\right)}{8F^2m^2}+\frac{\hat{c}_3\left(2M_\pi^2-t\right)}{F^2}+
\frac{4}{F_\pi^2}\left\{e_{14}\left(2M_\pi^2-t\right)^2+2e_{15}\left(2M_\pi^2-t\right)\nu^2+4e_{16}\nu^4\right\}\ ,\nonumber\\
D^-&=&\frac{\nu}{2F^2}+\frac{2\nu}{F_\pi^2}\left\{2(d_1+d_2+2d_5)M_\pi^2-(d_1+d_2)t+2d_3\nu^2\right\}\ ,\nonumber\\
B^+&=&\frac{4(d_{14}-d_{15})\nu m_N}{F_\pi^2}\ ,\quad
B^-=\frac{1}{2F^2}+\frac{2\hat{c}_4m_N}{F^2}+\frac{8
m_N}{F_\pi^2}\left\{e_{17}\left(2M_\pi^2-t\right)+2e_{18}\nu^2\right\}\
. \eea
Note that in the graphs of $O(p^3)$ and $O(p^4)$ shown in Fig.~\ref{treediagrams}, the bare constants can be replaced by the physical ones, since the distinction is beyond the accuracy of our calculation. Such replacements have been done in Eqs.~\ref{contactterm} and \ref{apptree}, where the $O(p^3)$ and $O(p^4)$ contributions are expressed only by physical parameters.
According to the discussion in Ref.~\cite{Meissnerp4}, the
terms proportional to $e_k$ ($k=19,20,21,22,35,36,37,38$) only
amount to quark mass corrections of $c_i$ ($i=1,2,3,4$), hence here
we have already adopted in Eq.~(\ref{contactterm}) the following
combinations of LECs, \bea\label{chats}
\hat{c_1}&=&c_1-2M^2\left(e_{22}-4e_{38}\right)\ ,\quad
\hat{c_2}=c_2+8M^2\left(e_{20}+e_{35}\right)\ ,\nonumber\\
\hat{c_3}&=&c_3+4M^2\left(2e_{19}-e_{22}-e_{36}\right)\ ,\quad
\hat{c_4}=c_4+4M^2\left(2e_{21}-e_{37}\right)\ .
\eea

\subsection{Loop amplitudes}\label{loopa}

\begin{figure}[ht]
\begin{center}
\includegraphics[width=0.7\textwidth]{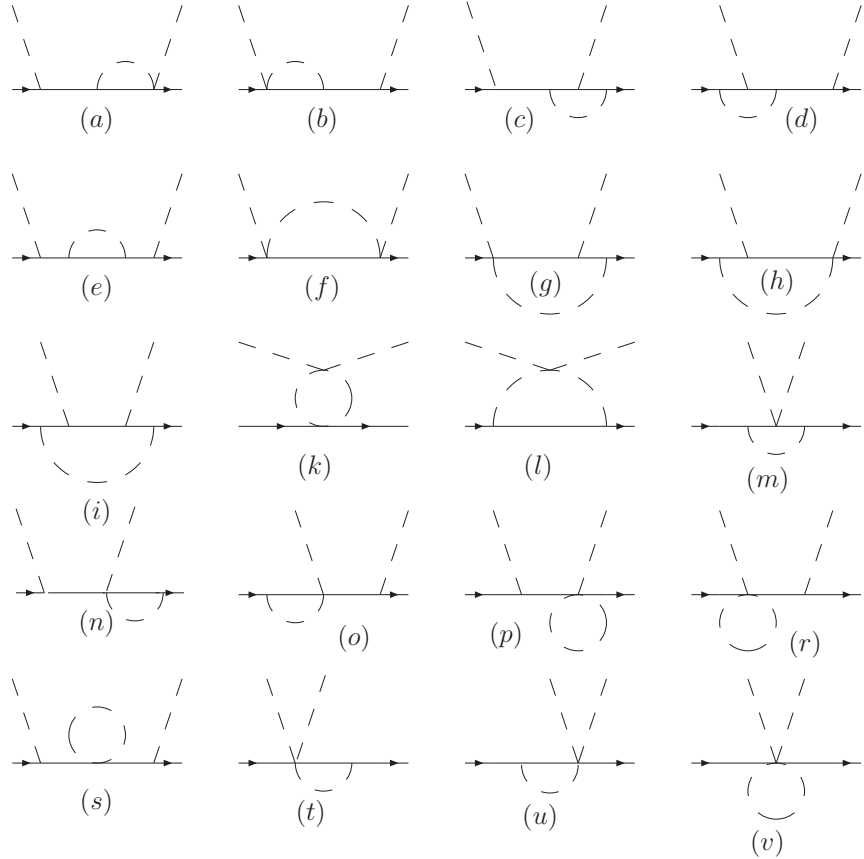}
\caption{\label{loopgraph}Loop graphs contributing to $\pi$-N scattering at $O(p^3)$. The nucleon propagates
with $m_4=m-4c_1M^2-2(8e_{38}+e_{115}+e_{116})$. Crossed graphs for (a)--(i) and (n)--(s) are not shown.}
\end{center}
\end{figure}
\begin{figure}[ht]
\begin{center}
\includegraphics[width=0.8\textwidth]{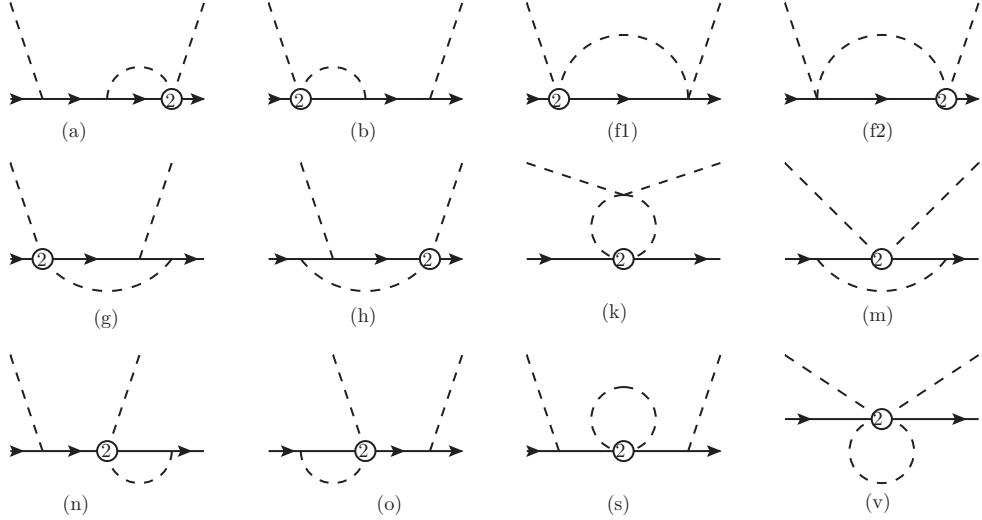}
\caption{\label{loopgraphp4}Loop graphs contributing to $\pi$-N scattering at $O(p^4)$. The vertex with circled `2' stems from ${\cal L}_{\pi N}^{(2)}$. The nucleon propagates
with $m_4=m-4c_1M^2-2(8e_{38}+e_{115}+e_{116})$. Crossed graphs are not shown.}
\end{center}
\end{figure}
To carry out the calculation on $\pi N\rightarrow\pi N$ process up to $O(p^4)$ level,
one ought to include $O(p^3)$ and $O(p^4)$ loop corrections
 with the corresponding Feynman diagrams shown in Fig.~\ref{loopgraph} and Fig~\ref{loopgraphp4}, respectively. The $O(p^4)$ loop graphs in Fig~\ref{loopgraphp4} are simply obtained from the graphs in Fig.~\ref{loopgraph} involving even number of pions by replacing one of the $O(p)$
vertices with corresponding
$O(p^2)$ vertices. The results -- in which no
 subtractions have yet been performed
-- are displayed in~\ref{loopamplp3} and~\ref{loopamplp4}. There the $O(p^3)$ loop
results are listed explicitly for the sake of completeness. The definition of the loop functions
is very similar to that in Ref.~\cite{becherleutwyler2}, which is
presented in~\ref{funct}.
We have checked that our $O(p^3)$ loop results agree with those in
Ref.~\cite{oller5} except a few terms due to the
reason that we have chosen exponential parametrization instead of
sigma parametrization for the pion fields\footnote{All the terms
proportional to $I_B^{(2)}$ in Ref.~\cite{becherleutwyler2} should
be reversed in sign, as it is first pointed out by
Ref.~\cite{oller1}.}\footnote{The physical pion-nucleon scattering amplitudes up to $O(p^3)$ here and in Ref.~\cite{oller5} are the same and are independent of parametrization.}. To our knowledge, the $O(p^4)$ loop
contributions shown here are the first analytic and complete
calculation results, and one can consult in Ref.~\cite{Meissnerp4}
for the $O(p^4)$ results within heavy baryon $\chi$PT and in
Ref.~\cite{becherleutwyler2} for the $O(p^4)$
results within infrared regularization. In
Ref.~\cite{becherleutwyler2} special technical simplification of calculation is
adopted, here we refer the readers to section 6 there for details.

It must be emphasized that the parameter $m$ in both $O(p^3)$ and
$O(p^4)$ loop results should be understood as
$m_4=m-4c_1M^2-2(8e_{38}+e_{115}+e_{116})M^4$ for the following
reasons~\cite{Schindler}:
\begin{itemize}
  \item All loop diagrams with contact interaction insertions in
the nucleon propagators are summed up automatically;
  \item {When performing renormalization, one can directly set $m_4=m_N$ in the one-loop results up to $O(p^4)$
  level, since corrections are at least of two-loop order
  (the lowest chiral order of two-loop contributions is naively $O(p^5)$).}
\end{itemize}

\subsection{EOMS scheme and PCB terms}\label{EOMSPCB}
Since the nucleon mass $m_N$ is nonzero in the chiral limit, the
necessary power counting rule for an effective theory breaks down,
namely PCB problem occurs. To remedy the PCB problem we adopt the
EOMS scheme proposed by T.~Fuchs et al.~\cite{EOMS}, which suggests
performing  renormalization in two steps: the first traditional
$\overline{\rm{MS}}$-1 subtraction to cancel the ultraviolet
divergencies and then EOMS subtraction to remove the PCB terms. The
EOMS subtraction is remarkable in the sense that the renormalized
$\pi N\rightarrow\pi N$ amplitude will possess good analytic and
correct power counting properties since the PCB terms are
polynomials of quark masses and momenta and are absorbed in the LECs
eventually. Especially, as proved by Becher and
Leutwyler~\cite{becherleutwyler}, the PCB terms stem from the
regular part of the loop integrals, which allows us a simple way to
obtain the PCB terms if we have known all the regular parts of the
loop integrals needed -- these are shown in~\ref{regul}.

Taking the $O(p^3)$ loop amplitude for example, one first changes
the amplitude in $A,B$ form to $D,B$ functions\footnote{In
Ref.~\cite{becherleutwyler2} the fact that the leading contribution
from $A$ and $B$ cancels, while not for $D$ and $B$, is pointed out.}
and reduces them to expressions only containing scalar one-loop
integrals. Then those scalar one-loop integrals are substituted by
their regular parts to a given order, and a chiral
expansion in terms of small quantities like $M, t, \sigma=s-m^2$ is
performed\footnote{One can also chose $\sigma=s-s_{th}$, with
$s_{th}=(m+M)^2$, as expanding parameter like
Ref.~\cite{MeissnerIRp3}, here we follow Ref.~\cite{gasser}.}. Finally, for the total $O(p^3)$ loop amplitude, the series whose chiral order are lower than
$O(p^3)$ are regarded as PCB terms which read

\bea \label{pcbp3} D^{(3)+}_{PCB}&=&\frac
{1}{64F^4m\pi^2\sigma^2}\left\{6 g^2 m^2 M^2\sigma^2
+2\sigma^4+g^4\left[2 m^4\left(10 M^4-7M^2t+t^2\right)\right.\right.
\left.\left.+3 m^2\left(-7 M^2+3 t\right)\sigma^2+\sigma^4 \right]\right\}\ ,\nonumber\\
D^{(3)-}_{PCB}&=&\frac{g^4m}{64 F^4\pi^2\sigma^2}\left\{\sigma^2\left(-2M^2+t
+2\sigma\right)-2m^2\left(2M^2-t\right)\left(2M^2-t+2\sigma\right)\right\}\ ,\nonumber\\
B^{(3)+}_{PCB}&=&\frac {g^4m^4}{8F^4\pi^2\sigma^2}\left(2M^2-t
+2\sigma\right)\ ,\,\,\, B^{(3)-}_{PCB}=\frac{g^2m^2}{32
F^4\pi^2\sigma^2}\left\{5\sigma^2+g^2\left[4m^2\left(-5M^2+t\right)+3\sigma^2\right]\right\}\
. \eea The same procedure can be taken to extract the PCB terms of
the total $O(p^4)$ loop amplitude,
\bea
\label{pcbp4}
D^{(4)+}_{PCB}&=&\frac{1}{1152 F^4\pi^2\sigma^3}\left\{864 {c_1} g^2 m^2 M^2\sigma^3+\left[16 {c_4} -(9 {c_2}-216 {c_3}+272 {c_4}) g^2 \right] m^2(t-2 M^2)\sigma^3\right.\nonumber\\
&&\hspace{2cm}-(9 {c_2}+32 {c_3}+32 {c_4})g^2 m^4 (t-2 M^2)^2 (2 M^2-t+\sigma)-2(9{c_2}+32{c_3}+32{c_4})g^2 m^2\sigma^4\nonumber\\
&&\hspace{2cm}\left.+4 \left[2 {c_4}-(9 {c_2}+16 {c_3}+14 {c_4}) g^2\right](t-2 M^2+\sigma)\sigma^4\right\}\ ,\nonumber\\
D^{(4)-}_{PCB}&=&\frac{1}{2304F^4\pi^2\sigma^3}\left\{32 (2 {c_2}+17 {c_3}-19{c_4})g^2 m^2 M^2(2M^2-t) \sigma^2\right.\nonumber\\
&&\hspace{2cm}-4\left[144{c_1}-2{c_3}-{c_4}+4(18{c_1}+{c_2}-{c_3}-7{c_4})g^2\right] t \sigma^4\nonumber\\
&&\hspace{2cm}+8\left[(72{c_1}-2{c_2}-6{c_3}-3{c_4})+(36{c_1}-2{c_2}+9{c_3}-33{c_4}) g^2\right](t-2M^2)\sigma^4\nonumber\\
&&\hspace{2cm}-(9{c_2}+32{c_3}+32{c_4}) g^2 m^2 (t-2M^2)\left[2m^2(2M^2-t)(2M^2-t+\sigma)+\sigma^2(4m^2+\sigma)\right]\nonumber\\
&&\hspace{2cm}\left.+\frac{4}{m^2}\left[(34{c_2}+30{c_3}-3{c_4})-(4{c_2}+15{c_4}) g^2\right]\sigma^6\right\}\ ,\nonumber\\
B^{(4)+}_{PCB}&=&-\frac {m} {576 f^4\pi^2\sigma^3}\left\{ 24{c_4}\sigma^4+g^2\left[32(2{c_2}+17{c_3}
-19{c_4})m^2 M^2\sigma^2+(67{c_2}-56{c_3}+96{c_4}) \sigma^4\right]\right.\nonumber\\
&&\hspace{2cm}\left.+2(9{c_2}+32{c_3}+32{c_4})g^2m^4\left[4 M^4+t^2-t \sigma +2 \sigma ^2+M^2
(-4 t+2 \sigma )\right]\right\}\ ,\nonumber\\
B^{(4)-}_{PCB}&=&\frac{m^3}{576 f^4 \pi ^2 \sigma ^3} \left\{\left[9 {c_2}+32{c_3}+16{c_4}-2 (9{c_2}+16{c_3}
-28{c_4}) g^2\right]\sigma ^3\right.\nonumber\\
&&\hspace{2cm} \left.+2g^2 m^2(9c_2+32c_3+32c_4)(2 M^2-t)(2
M^2-t+\sigma)\right\}\ .
\eea
Those PCB terms will be subtracted, namely absorbed by redefinition of the LECs, when performing the EOMS renormalization of the $\pi N\rightarrow\pi N$ amplitude in section~\ref{secren}.

Before ending this subsection it may be worthwhile to
mention that the PCB terms should be prevented from divergences
induced by prefactors of the type
\bea\label{thredive}\frac{1}{\lambda(s,m^2,M^2)}\,\,\, \rm{and}\,\,\,
\frac{1}{\lambda(s,m^2,M^2)+s\,t},\eea
 respectively, and so should the EOMS-renormalized amplitude be. The obstacle is first noted by Ref.~\cite{MeissnerIRp3}, that the numerical analysis of the IR-renormalized amplitude encounters divergences at threshold $s_{th}=(m+M)^2$ and at $t=-{\lambda(s,m^2,M^2)}/{s}$. Nevertheless, both the EOMS- and IR-renormalized amplitudes should possess good analytic properties at those $s$ and $t$ values, namely singularities caused by~(\ref{thredive}) are canceled by the numerators of the amplitudes. It can be easily seen from  $H_B(s)^{(i)}(i=1,\cdots,6)$ and $H_{13}^{(j)}(j=1,2)$ in~\ref{funct} that the prefactors are actually introduced by the standard Passarino-Veltman decomposition~\cite{PasVel} of tensor integrals, which can be avoided by the new approach developed in Ref.~\cite{davydychev}.

Taking the tensor integral $H_{B}^\mu$ for example, in Passarino-Veltman approach $H_{B}^\mu$ is decomposed into
\bea
H_B^\mu=(p+\Sigma)^\mu H_B^{(1)}+(p-\Sigma)^\mu H_B^{(2)}\ ,\nnb
\eea
where the expressions for $H_B^{(1)}$ and $H_B^{(2)}$ refer to Eq.~(\ref{IB1}) and Eq.~(\ref{IB2}). On the other hand, following the approach in Ref.~\cite{davydychev}, $H_{B}^\mu$ is now decomposed into the new form
\bea
H_{B}^\mu&=& P^\mu H_1+ \Sigma^\mu H_2\ ,\nnb\\
H_1&=&
\intdd\frac{1}{[M^2-k^2]\;[m^2-(P-k)^2]^2\;[m-(\Sigma-k)^2]}\ ,\nnb\\
H_2&=&\intdd\frac{1}{[M^2-k^2]\;[m^2-(P-k)^2]\;[m-(\Sigma-k)^2]^2}\ ,\nnb
\eea
where $H_1$ and $H_2$ are already scalar loop integrals in dimension 6 momentum space, in other words, coefficient like ${1}/{\lambda(s,m^2,M^2)}$ in standard Passarino-Veltman decomposition never occurs. Hence the threshold divergence introduced by ${1}/{\lambda(s,m^2,M^2)}$ disappears, the same conclusion holds for the divergence at $t=-{\lambda(s,m^2,M^2)}/{s}$ when considering $H_{13}^\mu$.

The new approach enables us to reduce the tensor integrals in the amplitude to scalar integrals defined in higher dimension momentum space, without confusion  such as the divergence at threshold. If further regular parts of the scalar integrals are known, the PCB terms can be obtained in a new way. In~\ref{regul}, the method proposed by Refs.~\cite{becherleutwyler,Schindler03} is adopted to calculate the regular parts of the scalar integrals in dimension 4 space, those of the scalar integrals in higher dimension space can also be calculated term by term using the same method. With the aid of the regular parts, the PCB terms for $H_1$ and $H_2$ can be easily obtained, which are $\frac{1}{32\pi^2 m^2}$ and  $\frac{1}{32\pi^2 m^2}$, respectively. Using the
relations $H_B^{(1)}=\frac{1}{2}(H_1+H_2)$ and
$H_B^{(2)}=\frac{1}{2}(H_1-H_2)$, one then gets the PCB terms for
$H_B^{(1)}$ and $H_B^{(2)}$ in this new way, which are $\frac{1}{32\pi^2 m^2}$ and 0
respectively. On the other hand, the PCB terms for $H_B^{(1)}$ and $H_B^{(2)}$ can also be obtained in the usual way adopted in the paper. One first replaces the scalar integrals in Eqs.~\ref{IB1} and~\ref{IB2} by their regular parts shown in~\ref{regul}, then expands $H_B^{(1)}$ and $H_B^{(2)}$ in terms of $\sigma,t, M_\pi$. From the expanded expressions of $H_B^{(1)}$ and $H_B^{(2)}$, one finds that the PCB terms for $H_B^{(1)}$ and $H_B^{(2)}$ are $\frac{1}{32\pi^2 m^2}$ and 0, which are the same as the expressions obtained through the new way above. In this sense, the approach developed in Ref.~\cite{davydychev} provides us a new way to obtain the PCB terms.

\subsection{Renormalization}
\label{secren}
 As an example, we will first show the renormalization of nucleon mass $m_N$ as well as  axial-vector coupling $g_A$
  to interpret the essence
 of EOMS scheme. Noticeably the expressions of $m_N$ and $g_A$ are also needed for replacing the
 corresponding bare quantities in the tree
 amplitude when performing numerical fits. Part of the results are
 already given in Ref.~\cite{YaoMont}.
\subsubsection{Nucleon mass and wave-function renormalization constant}
\begin{figure*}[ht]
\begin{center}
\includegraphics{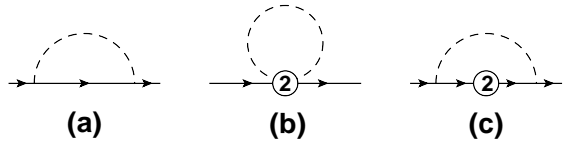}
\caption[pilf]{One-loop Feynman diagrams contributing to the self-energy of the nucleon up to $O(p^4)$. The vertex with circled `2' stems from ${\cal L}_{\pi N}^{(2)}$.}\label{figp4mn}
\end{center}
\end{figure*}
As a simple example to illustrate the EOMS method, we evaluate the nucleon physical mass up to $O(p^4)$. The $O(p^4)$ result of $m_N$ in the EOMS scheme can also be found in Refs.~\cite{EOMS,Pascalutsa05}.
The one-loop Feynman diagrams are depicted in Fig.~\ref{figp4mn}. A straightforward calculation leads to the
primitive expression for the nucleon mass,
\bea
m_N&=&m-4c_1
M^2-2e_mM^4-\frac{3mg^2}{2F^2}\left\{\Delta_N-M^2H(m^2)\right\}+\frac{3M^2}{F^2}\left\{\left(2c_1-c_3\right)
-\frac{c_2}{d}\right\}\Delta_\pi\nonumber\\
&&-\frac{3c_1M^2g^2}{F^2}\left\{-2\left[\Delta_N-M^2H(m^2)\right]+4m^2\left[J_N(0)-M^2H_A(0)\right]\right\}\
,\nonumber \eea
where $e_m=8e_{38}+e_{115}+e_{116}$. The second and
third terms are tree contributions stemming from $O(p^2)$ term involving
$c_1$ and $O(p^4)$ term related to $e_{38},~e_{115}$ and $e_{116}$.
The term with the first, second and third brace bracket represents
the loop contribution from (a), (b) and (c) in Fig.~\ref{figp4mn},
respectively. Definitions of all loop functions appeared here follow
~\ref{funct}. One can perform different renormalization schemes
on the above expression of $m_N$, e.g. IR prescription, etc.. However, we proceed with the EOMS
remormalization by first carrying out traditional
$\rm{\overline{MS}}-1$ subtraction, which gives
\bea\label{p4mnms}
m_N&=&m^r-4c_1^r M^2-2e_m^rM^4-\frac{3mg^2}{2F^2}\left\{\bar{\Delta}_N-M^2\bar{H}(m^2)\right\}+\frac{3M^2}{F^2}\left\{\left(2c_1-c_3\right)-\frac{c_2}{d}\right\}\bar{\Delta}_\pi\nonumber\\
 &&-\frac{3c_1M^2g^2}{F^2}\left\{-2\left[\bar{\Delta}_N-M^2\bar{H}(m^2)\right]+4m^2\left[\bar{J}_N(0)-M^2\bar{H}_A(0)\right]\right\}\ ,
\eea
The bar over the  loop function denotes the finite part of it,
and the LEC with a subscript {\it{r}} means that it is a
$\rm{\overline{MS}}-1$ quantity\footnote{Hereafter we denote the
$\overline{{MS}}-1$ and EOMS renormalized LECs with a superscript
\it{r} (eg. $c_1^r$) and overhead tilde (eg. $\tilde{c_1}$)
respectively.}. The $\rm{\overline{MS}}-1$ subtraction does nothing
but shifts the divergencies in loop functions to the bare mass and
LECs: \bea
m^r&=&m-\frac{3m^3g^2R}{32F^2}\ ,\nonumber\\
c_1^r&=&c_1+\frac{3g^2mR}{128F^2\pi^2}(1-12c_1m)\ ,\nonumber\\
e_m^r&=&e_m+\frac{3R}{128F^2\pi^2}\left[-8c_1(1+3g^2)+c_2+4c_3\right]\
,\nonumber \eea
 where $R=\frac{2}{d-4}+\gamma_E-1-\ln4\pi$. Since $m_N$ is scale-independent
, we now take the renormalization scale
$\mu=m$ in Eq.~(\ref{deltaN}) for simplicity, and therefore $\bar{\Delta}_N=0$. If
the loop functions are replaced by their regular parts, one naively finds that
the term $\frac{3m M^2 g^2}{2F^2}\overline{H}(m^2)$ in
Eq.~(\ref{p4mnms}) should be $O(p^3)$, but actually
contributes an $O(p^2)$ PCB term ${\frac{3m M^2 g^2}{32\pi^2f^2}}$,
and the same thing happens for the last term where a $O(p^2)$ PCB
term $\frac{3c_1m^2M^2g^2}{4F^2\pi^2}$ occurs. Since they are polynomials, they can be absorbed by the
LEC $c_1^r$, \bea
\tilde{c_1}=c_1^r-\frac{3mg^2}{128F^2\pi^2}(1+8c_1m)\ .\nonumber
\eea Finally, we get the expression for $m_N$ in EOMS scheme, which
takes the following form \bea\label{mNdirect}
m_N&=&\tilde{m}-4\tilde{c_1} M^2-2\tilde{e_m}M^4+\frac{3m M^2
g^2}{2F^2}\overline{H}(m^2)+\frac{3M^2}{F^2}\left\{\left(2c_1-c_3\right)-\frac{c_2}{d}\right\}\bar{\Delta}_\pi\nonumber\\
&&-\frac{3c_1M^2g^2}{F^2}\left\{2M^2\bar{H}(m^2)+4m^2\left[\bar{J}_N(0)-M^2\bar{H}_A(0)\right]\right\}-{\frac{3m M^2
g^2}{32\pi^2F^2}}-\frac{3c_1m^2M^2g^2}{4F^2\pi^2}\ ,
 \eea
with $\tilde{m}=m^r$, $\tilde{e_m}=e_m^r$, namely they are unaffected by the PCB terms.

We note that one can also carry out the mass renormalization by
replacing $m$ by $m_4=m-4c_1 M^2-2e_mM^4$ in the nucleon propagator.
In this case, the graph (c) in Fig.~\ref{figp4mn} is absent and
automatically included in graph (a), and the result is
\bea\label{mN4} m_N=m-4c_1
M^2-2e_mM^4-\frac{3m_4g^2}{2F^2}\left\{\Delta_N-M^2H(m_4^2)\right\}+\frac{3M^2}{F^2}
\left\{\left(2c_1-c_3\right)-\frac{c_2}{d}\right\}\Delta_\pi\ , \eea
while the wave-function renormalization constant of nucleon reads
\bea\label{ZN} Z_N=1-\frac{3g^2}{4F^2}\left\{\Delta_\pi-4m_4^2M^2
\frac{\partial}{\partial
s}H(s)\right\}_{\slashed{p}=m_N}-\frac{6c_2}{F^2
m_4}\frac{M^2}{d}\Delta_\pi\ . \eea Hereafter, the $m_4$ related to loop contributions is always taken as $m$
for short. Instead of Eq.~(\ref{mNdirect}),
Eqs.~(\ref{mN4}) and (\ref{ZN}) are adopted for the renormalization
of the $\pi$-N scattering amplitude. Throughout  this paper, we use
this way to simplify our calculation for the reasons discussed in
section~\ref{loopa}.

\subsubsection{Axial-vector coupling constant}
\begin{figure}[!htbp]
\centering
\includegraphics[width=0.7\textwidth]{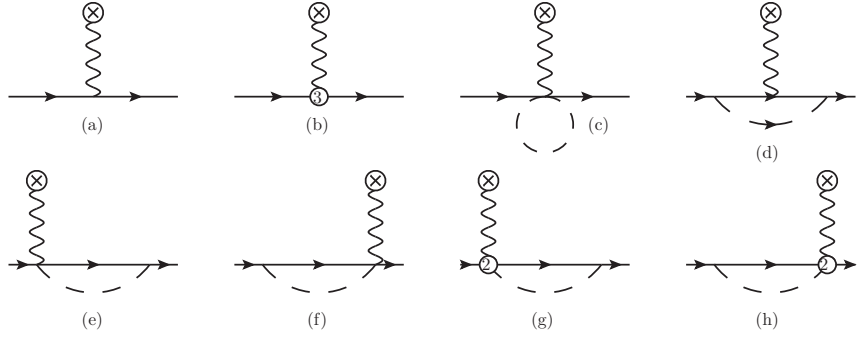}
\caption{\label{figga}Diagrams contributing to the nucleon axial form factor $G_A(t)$ up to $O(p^4)$. The wavy line with a circled cross at the end stands for the axial-vector current. The vertex with circled `2' and `3' stems from ${\cal L}_{\pi N}^{(2)}$ and ${\cal L}_{\pi N}^{(3)}$, respectively.}
\end{figure}
The axial-vector current $A^{\mu,a}(0)$ between one-nucleon states
can be written as \bea \langle N(p^\prime)|A^{\mu,a}(0)|N(p)
\rangle=\bar{u}(p^\prime)\left[G_A(q^2)\gamma^\mu\gamma_5+G_P(q^2)\frac{q^\mu}{2m}\gamma_5\right]\frac{\tau^a}{2}u(p),
\eea where $q_\mu=p^\prime_\mu-p_\mu$ and $a$ is isospin index.
$G_A(q^2)$ is called the axial form factor and $G_P(q^2)$ is the
induced pseudoscalar form factor. The axial-vector coupling constant
$g_A$ is defined as \bea g_A=G_A(q^2=0). \eea Up to the $O(p^4)$
diagrams in Fig.~\ref{figga} are needed. A straightforward
calculation leads to $g_A$: \bea
g_A&=&g+4d_{16}M^2-\frac{g^3m^2}{32F^2\pi^2}+\frac{g(4-g^2)}{2F^2}\Delta_N-\frac{g(2+g^2)}{2F^2}
\Delta_\pi+\frac{g^3(2m^2+M^2)}{4F^2}J_N(0)\nonumber\\
&&-\frac{g(8-g^2)M^2}{4F^2}H(m^2)-\frac{g^2M^4}{4F^2}H_A(0)+\frac{3g^3m^2M^2}{F^2}\frac{\partial}
{\partial s}H(s)\bigg{|}_{\slashed{p}=m_N}-\frac{8c_4m g}{F^2}\left(\Delta_\pi-M^2H(m^2)\right)\nonumber\\
&&-\frac{2g}{m
F^2}\left[c_2\left(4\frac{M^2}{d}\Delta_\pi+\frac{m^2}{d}\Delta_N-M^2H^{(2)}(m^2)\right)-4(c_3+c_4)
m^2H^{(2)}(m^2)\right]\ ,
\eea
which agrees with Refs.~\cite{Schindler06,Ando06}\footnote{Though in Ref.~\cite{Ando06} the mass insertion graphs are calculated directly, the comparison between our result and the one there is easy.}.
Here, the EOMS renormalization procedure is similar to that for $m_N$. The bare $g$ will be redefined as
\bea
\label{gEOMS}\tilde{g}=g^r-\frac{g^3 m^2}{16F^2\pi^2}-\frac{g\,m^3}{576F^2\pi^2}(9c_2+32c_3+32c_4)\ ,\  g^r=g+\frac{gm^2R}{16F^2\pi^2}(2-g^2)-\frac{gm^3R}{96F^2\pi^2}(3c_2+8c_3-40c_4)\ ,
\eea
whereas the redefinition of the LEC $d_{16}$ reads
\bea
\label{d16EOMS}
\tilde{d_{16}}=d_{16}^r+\frac{m g}{288F\pi^2}(c_2+18c_3-18c_4)\ ,\ d_{16}^r=d_{16}+\frac{R}{192F^2\pi^2}\left[3g(1-g^2)-mg(c_2+6c_3-18c_4)\right].
\eea
The final expression for $g_A$ is lengthy but rather straightforward to get with the help of Eqs.~(\ref{gEOMS}) and (\ref{d16EOMS}), so we do not present it here explicitly.
\subsubsection{Full $\pi N\rightarrow\pi N$ amplitude}
In order to present a full $\pi N\rightarrow\pi N$ amplitude, we
also need the formulae for $M_\pi$ and $F_\pi$, and the
corresponding Z factor  for the pion . To $O(p^4)$ level, these read
\bea
M_\pi^2&=&M^2\left(1+2\ell_3\frac{M^2}{f^2}+\frac{1}{2f^2}\Delta_\pi\right)\ ,\nonumber\\
F_\pi&=&F\left[1+\ell_4\frac{M^2}{f^2}-\frac{1}{f^2}\Delta_\pi\right]\ ,\nonumber\\
Z_\pi&=&1+\frac{1}{f^2}\left[\frac{2}{3}\Delta_\pi-2\ell_4M^2\right]\ .
\eea
All of them do not contain PCB terms from loop integrals, and hence can be
treated traditionally.

Since all the necessary preparations are completed, we proceed with
the renormalization of $\pi N\rightarrow\pi N$ amplitude. Unlike the
renormalization of $m_N$ and $g_A$, it is  hard to visualize the
procedure of $\pi N\rightarrow\pi N$ amplitude renormalization for
its extremely lengthy expression. However, the essence is the same,
that is to carry out renormalization procedure in two steps:
$\rm{\overline{{MS}}-1}$ renormalization and EOMS renormalization.
Corresponding to the $\rm{\overline{{MS}}-1}$ renormalization, those LECs appeared in the tree amplitudes are demanded to cancel the
ultraviolet divergences and yield the so-called
$\rm{\overline{{MS}}-1}$ renormalized LECs, \bea
c_i^r(\mu)=c_i-\frac{\gamma_i^c mR}{16
\pi^2F^2},\hspace{1cm}d_j^r(\mu)=d_j-\frac{\gamma_j^dR}{16\pi^2F^2},
\hspace{1cm}e_k^r(\mu)=e_k-\frac{\gamma_k^eR}{16\pi^2F^2m} \
,\nonumber \eea where details of $\gamma_i^c,\gamma_j^d,\gamma_k^e$
can be found in~\ref{coupl}. To absorb the PCB terms~(\ref{pcbp3})
and~(\ref{pcbp4}), $c_i^r(\mu)$ and $d_j^r(\mu)$ are further
redefined as \bea \widetilde{c}_i=c_i^r-\frac{\delta_i^c m}{16
\pi^2F^2},\hspace{1cm}\widetilde{d}_j=d_j^r-\frac{\delta_j^d}{16\pi^2F^2}
\ ,\nonumber \eea whereas $e_k^r(\mu)$ remain the same, ie.
$\tilde{e_k}=e_k^r(\mu)$, since the chiral order
 of PCB terms are lower than $O(p^4)$.
Also $\delta_i^c$ and $\delta_j^d$ are specified in~\ref{coupl}. So
far, we have already completed the renormalization of the $\pi
N\rightarrow\pi N$ amplitude in the EOMS scheme, the main feature of
this method is characterized by additional EOMS subtractions, which
distinguishes EOMS scheme from other prescriptions like IR and HB.
We observe that an amplitude in EOMS scheme differs from
 the full covariant amplitude only by a polynomial of small quantities and hence
 owns the same analytical structure
 but  possesses correct power counting. The validity of the $\pi N\rightarrow\pi N$ description in
 EOMS scheme will also be judged
 by numerical fits to existing experimental data.

\subsection{Partial wave expansions}\label{pwexpantion}

We choose to perform fits to the partial wave phase shift data. The
isospin decomposed amplitudes for $\pi N$ scattering are
 \bea
T^{I=\frac{1}{2}}&=&T^++2T^-,\nonumber\\
T^{I=\frac{3}{2}}&=&T^+-T^-. \eea The final partial wave amplitudes
with  isospin $I$ , orbital momentum $\ell$, and total angular
momentum $J=\ell\pm\frac{1}{2}$ (denoted by $\ell\pm$ concisely)
take the form~\cite{Hohler}, \bea
f^{I}_{\ell\pm}(s)&=&\frac{1}{16\pi\sqrt{s}}\left\{(E_p+m_N)\left[A^I_\ell(s)+\left(\sqrt{s}-m_N\right)B^I_\ell(s)\right]\right.\nonumber\\
&&\left.+(E_p-m_N)\left[-A^I_{\ell\pm1}(s)+\left(\sqrt{s}+m_N\right)B^I_{\ell\pm1}(s)\right]\right\},
\eea
where
\bea \label{eqintegrate}
    A^I_\ell(s)&=&\int^1_{-1}\; A^I(s,t)P_\ell(\cos\theta){\rm d}\cos\theta\ ,\nonumber\\
    B^I_\ell(s)&=&\int^1_{-1}\; B^I(s,t)P_\ell(\cos\theta){\rm d}\cos\theta\ .\nonumber
\eea Here $E_p=\frac{s+m_N^2-M_\pi^2}{2\sqrt{s}}$ and $\theta$ are
the nucleon energy and scattering angle in center-of-mass system,
respectively. $P_\ell(\cos\theta)$ are the conventional Legendre
polynomials. The angular variable $\cos\theta$ relates to the
Mandelstam variables via
$\cos\theta=1+\frac{2s\;t}{\lambda(s,m_N^2,M_\pi^2)}$, with
$\lambda(a,b,c)=a^2+b^2+c^2-2ab-2bc-2ac$ being the
$\rm{K\ddot{a}ll\acute{e}n}$ function. As a straightforward consequence
of unitarity of $S$ matrix, one can further express the partial wave
amplitudes by the phase shift $\delta^I_{\ell\pm}$, \bea\label{phaseshiftuni}
f^{I}_{\ell\pm}(s)=\frac{1}{2i|\vec{p}|}\left[\exp\left(2i\delta^I_{\ell\pm}(s)\right)-1\right]\
, \eea where $\vec{p}$ is the 3-momentum of nucleon in the
center-of-mass frame. Since the phase shift is real for elastic
scattering, we follow Ref.~\cite{Meissnerp3} to related it with our perturbative computation of $f^{I}_{\ell\pm}(s)$ via \bea\label{phaseshiftper}
\delta^I_{\ell\pm}(s)=\arctan\left\{|\vec{p}|\,{\rm
Re}f^{I}_{\ell\pm}(s)\right\}\ . \eea
\section{Phenomenological and numerical studies}
\nin In this section, we first perform fits to partial wave phase
shift data near threshold to pin down the free LECs. In order to
describe the partial wave phase shifts up to a higher energy region, we
include explicitly the leading $\Delta(1232)$ Born-term contribution and partially-included $\Delta(1232)$ loop contribution. The contribution to the LECs from the $\Delta(1232)$ resonance  is also
considered. We proceed with discussing the convergence of the chiral expansion of
the resulting partial wave phase shift. The
improvement of the fourth-order calculation compared with the
third-order is shown. Finally, the deviation ($\Delta_{GT}$) of Goldberger-Treiman relation and the pion--nucleon $\sigma$ term $\sigma_{\pi N}$ are discussed. The $O(p^3)$ analyses are also included for the sake of
comparison with the previous literature.

\subsection{Partial wave phase shift}\label{secpar}
To begin with, we first fit the partial waves at the $O(p^3)$ level. We
denote this fit by ``Fit I-$O(p^3)$''. As input we use the phase
shift data from Ref.~\cite{GW08}, namely the current solution of
George Washington University (GWU) group. Since the GWU group does not
give data errors, we assign them with the method of
Ref.~\cite{oller1}, \bea err(\delta)=\sqrt{e_s^2+e_r^2\delta^2}\ ,
\eea with the systematic error $e_s=0.1^\circ$ and the relative
error $e_r=2\%$. Throughout the numerical analyses, we employ
$g_A=1.267$, $F_\pi=92.4$ MeV, $m_N=939$ MeV, $M_\pi=139$ MeV, and the
renormalization scale $\mu=m_N$. There are 9 free LECs (or
combinations of LECs) in total: ${c}_{1-4}$,
$d_1+d_2,~d_3,~d_5,~d_{14}-d_{15},~d_{18}$. All of them can be
pinned down by fitting two S- and four P- partial waves. The fitting
range is from threshold (1.078 GeV) up to 1.130 GeV in $\sqrt{s}$, and
the interval between two data points is 4 MeV. The 2nd column in Table~\ref{tabp3} collects our fit results at $O(p^3)$
level. In column 3 of Table~\ref{tabp3}, we have also listed the
results from Refs.~\cite{oller3} for comparison. We see that,  in
general, our fit results at $O(p^3)$ level are in good agreement
with those in Refs.~\cite{oller3}, except for the $d_5$ parameter.
Especially, the $d_{18}$, related to $\Delta_{GT}$, is nearly the
same.

\begin{table}[ht]
\begin{center}
\begin{tabular}{|c || r r||r r|| r r|}
\hline
 LEC    &        Fit I-$O(p^3)$      & WI08~\cite{oller3} & Fit II-$O(p^3)$ & WI08~\cite{oller5}\\
\hline\hline
$c_1$   &       $-1.39\pm0.07$     & $-1.50\pm0.06$   & $-0.81\pm0.03$  & $-1.00\pm0.04$\\
$c_2$   &       $4.01\pm0.09$      & $3.74\pm0.09$    & $1.46\pm0.09$   & $1.01\pm0.04$\\
$c_3$   &       $-6.61\pm0.08$     & $-6.63\pm0.08$   & $-3.10\pm0.12$  & $-3.04\pm0.02$\\
$c_4$   &       $3.92\pm0.04$      & $3.68\pm0.05$    & $2.35\pm0.06$   & $2.02\pm0.01$\\
\hline
$d_1+d_2$&      $4.40\pm0.54$      & $3.67\pm0.54$    & $0.79\pm0.09$   &$0.15\pm0.20$\\
$d_3$   &       $-3.02\pm0.51$     & $-2.63\pm0.51$   & $-0.47\pm0.05$  &$-0.23\pm0.27$\\
$d_5$   &       $-0.62\pm0.13$     & $-0.07\pm0.13$   & $-0.17\pm0.04$  &$0.47\pm0.07$\\
$d_{14}-d_{15}$&$-7.15\pm1.06$     & $-6.80\pm1.07$   & $-0.90\pm0.15$  &$-0.5\pm0.5$\\
$d_{18}$&       $-0.56\pm1.42$     & $-0.50\pm1.43$   & $-0.91\pm0.25$  &$-0.2\pm0.8$\\
\hline
$h_A$   &             -            &        -         & $2.82\pm0.04$   & $2.87\pm0.04$\\
\hline
$\chi^2_{d.o.f}$& 0.20             & $0.22$           & 0.35            &0.23\\
\hline
\end{tabular}
\caption{\label{tabp3}LECs given by fit up to ${\cal O}(p^3)$. Fit I is performed up to 1.13GeV, while Fit II up to 1.20GeV including the explicit $\Delta(1232)$ contribution. For comparison, we provide the results from~\cite{oller3,oller5}. The
$c_i$ and $d_j$ have units $\rm{GeV}^{-1}$ and $\rm{GeV}^{-2}$
respectively, and $h_A$ is dimensionless. In Fit II, the results correspond to $c^\prime_i$ and $d^\prime_j$ instead of $c_i$ and $d_j$, respectively.}
\end{center}
\end{table}

The fourth-order analysis of $\pi$-$N$ scattering is denoted by
``Fit I (a)-$O(p^4)$" in Table~2. There are 14 free LECs,  which are
four dimension two LECs:
$\hat{c}_1,~\hat{c}_2,~\hat{c}_3~,\hat{c}_4$ , five dimension three
LECs: $d_1+d_2,~d_3,~d_5,~d_{14}-d_{15},~d_{18}$, and five dimension
four LECs: $e_{14},~e_{15},~e_{16},~e_{17},~e_{18}$. Unlike the
 $O(p^3)$ fit, $d_{18}$ is now fixed at its $O(p^3)$ fitted value,
according to the discussion of $\Delta_{GT}$ below in
Sec.~\ref{secGT}. The $O(p^4)$ fit is performed up to 1.13GeV
too, and the results are shown in  column 2 of Table~\ref{tabp4}.
Also, we have taken the results of HB$\chi$PT from
Ref.~\cite{Meissnerp4} for comparison. Our results show improvements
compared to Ref.~\cite{Meissnerp4}. First, from Table~\ref{tabp4},
one can observe that the $\hat{c}_i$, $d_j$ and $e_k$ are mostly of
natural size in EOMS scheme, but in HB results, especially Fit~2 in
Table 1 of Ref.~\cite{Meissnerp4}, some of the $e_k$ come out fairly
large. Second, our results seems to be  more self-consistent. The
$\hat{c}_i$ change a lot when extending the $O(p^3)$ analysis to the
$O(p^4)$ analysis in Ref.~\cite{Meissnerp4}, while our $\hat{c}_i$
change much more acceptably.

\begin{figure*}[ht]
\begin{center}
\includegraphics[width=1.0\textwidth]{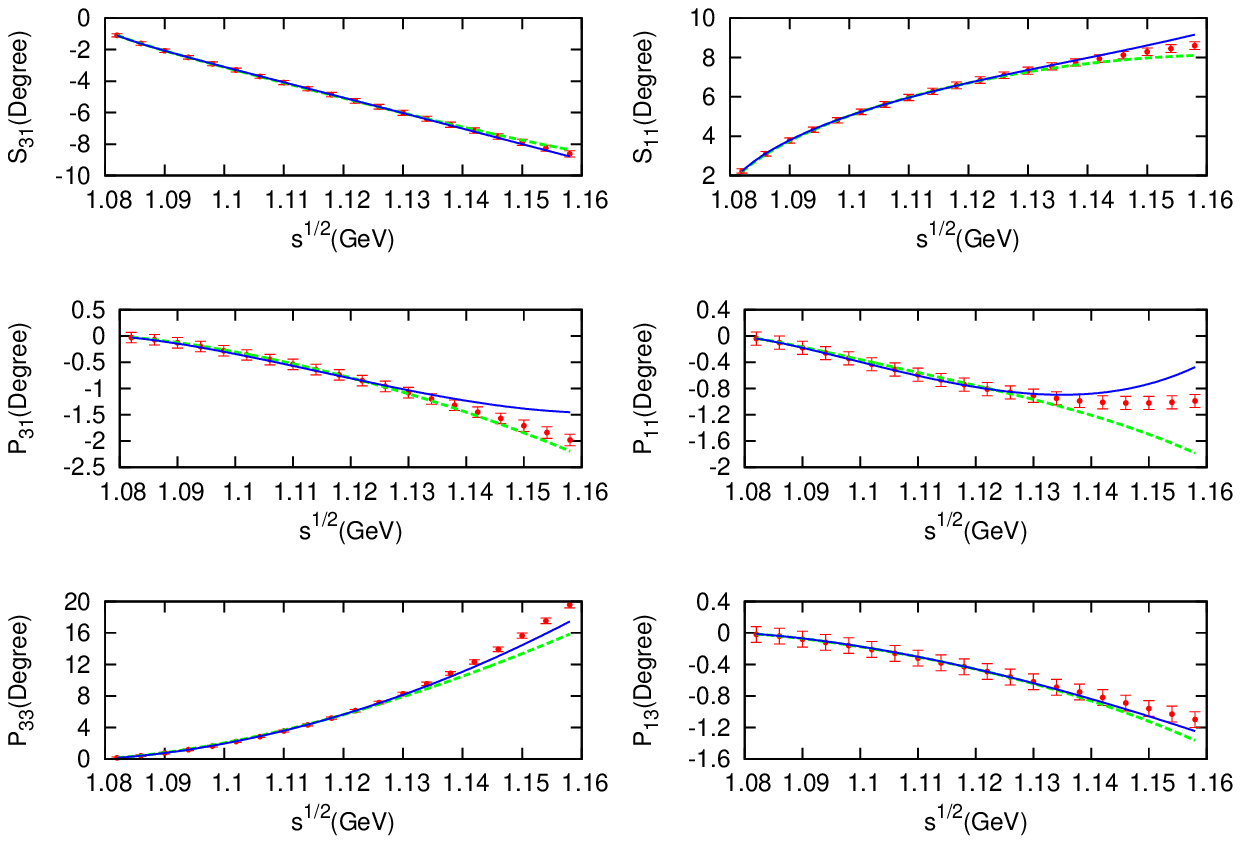}
\caption[pilf]{(Color online) Fit up to 1.13 GeV. The fourth- and
third-order fits are presented by the blue solid and green dash lines respectively. Results of Fit I(a)-$O(p^4)$ in Table~\ref{tabp4} and Fit I-$O(p^3)$ in Table~\ref{tabp3} are adopted for plotting.}\label{figp4}
\end{center}
\end{figure*}

We plot  both $O(p^3)$ and $O(p^4)$ fits in Fig.~\ref{figp4}. Though
fits are performed up $\sqrt{s}=1.13$ GeV, we plot up to $1.16$ GeV.
The conclusions made in Ref.~\cite{Meissnerp4} still hold: the
$P_{33}$ wave is slightly improved compared to the $O(p^3)$
calculation, and the $P_{11}$ partial wave are somewhat off above
$1.14$ GeV.
\begin{table}[ht]
\begin{center}
\begin{tabular}{|c || r r  r || r r|}
\hline
 LEC          &Fit I(a)-$O(p^4)$  &HB$\chi$PT~\cite{Meissnerp4}&Fit II(a)-$O(p^4)$ & Fit I(c)-$O(p^4)$ &Fit II(c)-$O(p^4)$\\
\hline\hline
$\hat{c}_1$   &$-1.08\pm0.06$     &$(-3.31, -0.27)$ & $-1.03\pm0.03$&$-1.09\pm0.08$ &$-0.95\pm0.05$\\
$\hat{c}_2$   &$2.78\pm0.11$      &$(0.13, 3.29)$   & $0.50\pm0.04$ &$2.44\pm0.05$  &$0.10\pm0.06$\\
$\hat{c}_3$   &$-5.26\pm0.14$     &$(-10.37, -1.44)$& $-3.17\pm0.05$&$-5.05\pm0.22$ &$-2.64\pm0.08$\\
$\hat{c}_4$   &$2.43\pm0.19$      &$(2.80, 3.53)$   & $0.79\pm0.03$ &$2.43\pm0.19$  &$0.80\pm0.03$\\
\hline
$d_1+d_2$     &$6.29\pm0.12$      &$(4.45, 5.68)$   & $2.99\pm0.05$ &$6.18\pm0.11$  &$2.93\pm0.05$\\
$d_3$         &$-6.87\pm0.16$     &$(-4.91, -2.96)$ & $-5.04\pm0.05$&$-6.87\pm0.15$ &$-4.90\pm0.04$\\
$d_5$         &$0.51\pm0.11$      &$(-0.95, -0.09)$ & $1.32\pm0.04$ &$0.55\pm0.11$  &$1.24\pm0.03$\\
$d_{14}-d_{15}$&$-12.09\pm0.24$   &$(-11.14, -7.02)$& $-5.61\pm0.09$&$-11.94\pm0.23$&$-5.58\pm0.09$\\
$d_{18}$      &$-0.56^*$          &$(-1.53, -0.85)$ & $1.14\pm0.20$ &$-0.56^*$      &$1.64\pm0.17$\\
\hline
$e_{14}$      &$3.69\pm0.36$      &$(-4.68, 7.83)$  & $-4.53\pm0.09$&$-1.80\pm0.33$ &$-8.22\pm0.08$\\
$e_{15}$      &$-14.99\pm0.55$    &$(-18.41, 9.72)$ & $5.05\pm0.13$ &$-5.41\pm0.57$ &$10.52\pm0.12$\\
$e_{16}$      &$7.35\pm0.35$      &$(6.42, 7.79)$   & $-0.31\pm0.07$&$4.34\pm0.28$  &$-1.50\pm0.05$\\
$e_{17}$      &$-2.29\pm1.34$     &$(-17.79, 14.88)$& $16.98\pm0.15$&$-2.23\pm1.42$ &$15.70\pm0.15$\\
$e_{18}$      & $6.07\pm1.18$     &$(-9.15, 19.66)$ & $-10.99\pm0.12$&$6.00\pm1.26$ &$-9.87\pm0.12$\\
\hline
$h_A$         &  -                 &    -           &$2.90^*$       &-          &$2.90^*$ \\
$e_1$         &              -    &-                &-               &$15.48\pm0.30$&$16.70\pm0.27$\\
$m$           &              -    &-                &-               &$0.88\pm0.02$ &$0.89\pm0.03$\\
\hline
$\chi^2_{d.o.f}$& 0.04            &$(0.008, 0.44)$  & 0.23           &0.51          &0.36\\
\hline
\end{tabular}
\caption{\label{tabp4}LECs given by fit up to ${\cal O}(p^4)$. Fit I(a) and Fit II(a) are performed with phase shift data,  while Fit I(c) and Fit II(c) with both phase shift data and QCD lattice data (see section~\ref{secsigma}). Fit I(a) and Fit I(c) are performed up to 1.13GeV, while Fit II(a) and Fit II(c) up to 1.20GeV including the explicit $\Delta(1232)$ contribution.  In Fit II(a) and Fit II(c), the results correspond to $\hat{c}^\prime_i$, $d^\prime_j$ and $e_k^\prime$ instead of $\hat{c}_i$, $d_j$ and $e_k$, respectively. For comparison, we provide the results from~\cite{Meissnerp4}.  The $c_i$, $d_j$
and $e_k$ have, respectively, units of $\rm{GeV}^{-1}$,
$\rm{GeV}^{-2}$ and $\rm{GeV}^{-3}$, and $h_A$ is dimensionless. The $*$ denotes an input
quantity.}
\end{center}
\end{table}

Note that the effect of unitarity is automatically included through the phase shift formula Eq.~\ref{phaseshiftper}, which is discussed in~\ref{secunitarity}.

\subsection{Contribution of $\Delta(1232)$}
\label{secDel}
In this subsection, the effect of $\Delta(1232)$ is
explicitly included to describe the partial wave shift up to 1.20 GeV.
Pascalutsa et al. discussed how to treat the $\Delta(1232)$ as an
explicit degree of freedom in covariant baryon chiral perturbation
theory in Refs.~\cite{pascalutsa,Pascalutsa00,Pascalutsa06}. The description of $\Delta(1232)$ is
subtle, because the conventional Rarita-Schwinger representation is
a field with 16 components while only 8 of them are physical.
However we adopt the \textit{consistent} formulation here~\cite{Pascalutsa00}.
Additionally, we follow the so-called \textit{$\delta$-counting}
rule~\cite{pascalutsa} which assigns an extra factional suppression of $O(p^{1/2})$ to
the propagator of $\Delta(1232)$. Up to $O(p^4)$ level, there are
three typical $\Delta(1232)$-included Feynman diagrams of different
order: Born-term of $O(p^{3/2})$ and $O(p^{5/2})$, loop graphs of
$O(p^{7/2})$. Refs.~\cite{oller2,oller5} remarked that the
contribution of  Born-term of $O(p^{5/2})$ is negligible. It is rather
complicated  to evaluate loop graphs of $O(p^{7/2})$ in the EOMS scheme,
since the loop diagrams involving both propagator of nucleon and
$\Delta(1232)$ will cause much more subtle PCB effects due to the
heavy masses $m_N$ and $m_\Delta$. So throughout this paper we consider
the leading Born-term contribution of $\Delta(1232)$, whose
expression can be found in~\ref{appborndelta}, together with partially-included $\Delta$ loop contribution illustrated in~\ref{appdeltasat}.
The complete
calculation with $\Delta(1232)$ up to $O(p^{7/2})$ is left as an open
question. It is important to mention that the effect of the $\Delta(1232)$ width is considered through the phase shift formula Eq.~\ref{phaseshiftper}, which is amply discussed in~\ref{secdelwidth}. Likewise, we will have the operators from Eqs.~(\ref{effp2})--(\ref{effp4}) but with couplings different from those in the $\Delta$-less effective field theory. We will mark the analogous coupling of the theory with $\Delta(1232)$ with a prime, e.g., $c_i\to c_i^\prime$.

\begin{figure*}[ht]
\begin{center}
\includegraphics[width=1.0\textwidth]{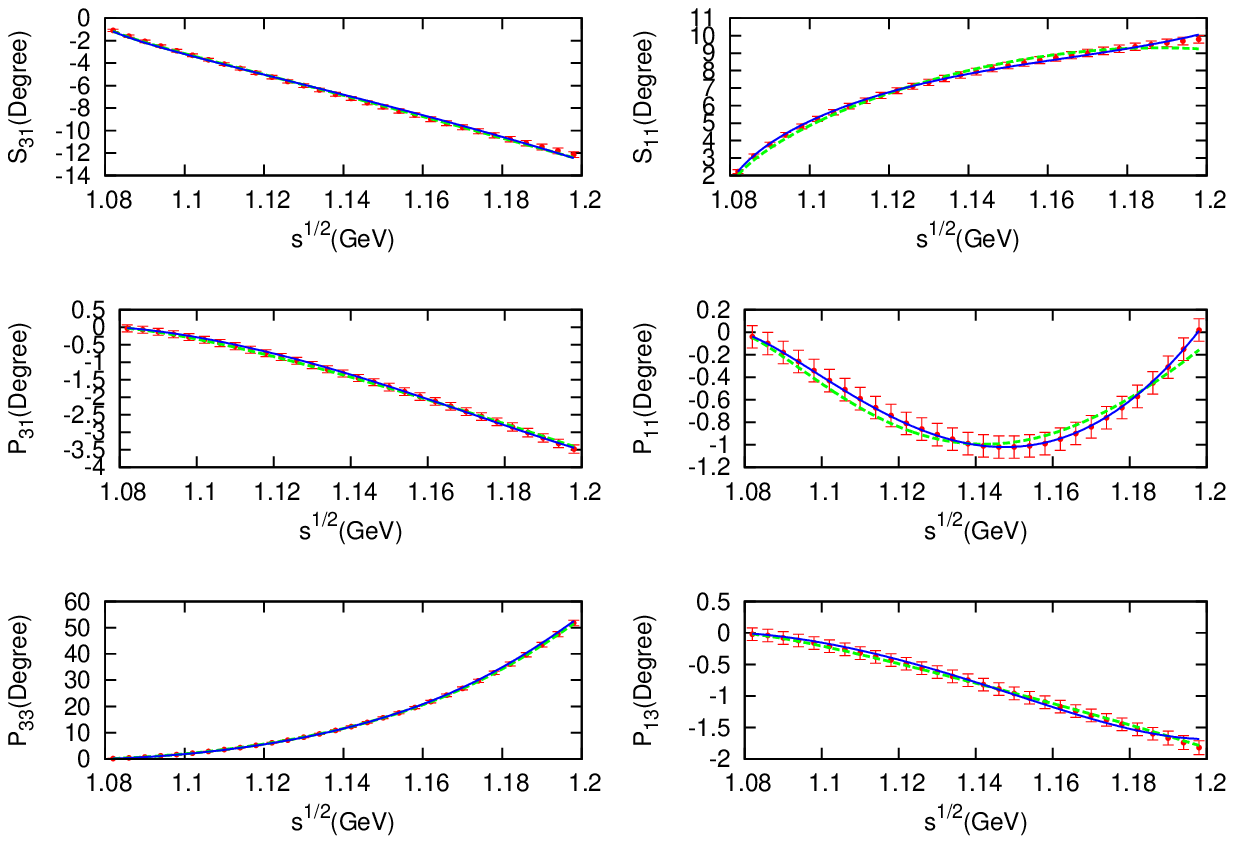}
\caption[pilf]{(Color online) Fit up to 1.20 GeV. The fourth- and
third-order fits are presented by the blue solid and green dash lines respectively. Results of Fit II(a)-$O(p^4)$ in Table~\ref{tabp4} and Fit II-$O(p^3)$ in Table~\ref{tabp3} are adopted for plotting.}\label{figp4del}
\end{center}
\end{figure*}

Corresponding to the two different fits performed in
Sec.~\ref{secpar}, we perform another two fits, ``Fit II-$O(p^3)$"
and ``Fit II(a)-$O(p^4)$'', which explicitly include the $\Delta(1232)$
contribution. The results are shown in the 4th column of
Table~\ref{tabp3} and Table~\ref{tabp4}, respectively. Taking into
consideration the $\Delta(1232)$ contribution to the LECs, the
$d_{18}$ is set free in both fits here. The leading Born-term
contribution of $\Delta(1232)$ is characterized by the $N\Delta$ coupling $h_A$. The value of $h_A=2.90$ is determined
from the Breit-Wigner width $\Gamma_\Delta=118$MeV. In ``Fit
II(a)-$O(p^4)$'', we fix $h_A=2.90$. However, for ``Fit II-$O(p^3)$",
$h_A$ is released as a free parameter, and its fitted value is $2.82$.
In Table~\ref{tabp3}, our result is found mostly compatible with
those of Ref.~\cite{oller5}.
 We plot the $\Delta$-included $O(p^3)$ and $O(p^4)$ fits  together in Fig.\ref{figp4del} for the convenience of comparison.
 We find that, both $O(p^3)$ and $O(p^4)$ calculations with $\Delta(1232)$ contribution
 give a reasonable description to data and the $O(p^4)$ calculation improves the fit quality.

Note that we denote the $O(p^4)$ fits with the notations: Fit I is $\Delta$-less and Fit II includes $\Delta$; Fit (a) and Fit (b) are performed only with pion-nucleon phase shift data whereas Fit (b) has the $c_i$ in $\hat{c}_i$ fixed, and Fit (c) is performed with both phase shift data and lattice QCD data for $m_N$.

\subsection{Convergence properties of partial wave phase shifts}\label{seccon}
We have made  chiral expansions up to $O(p^4)$. It is necessary, at this stage, to
check the  convergence property of the chiral amplitudes. However,
in the fits of Table~\ref{tabp4} the $O(p^2)$ parameters $c_i$ mix with
$O(p^4)$ parameters $e_i$, so it is not suitable for testing the
convergence of the chiral expansion. To overcome the problem, here
we follow the strategy of Ref.~\cite{Meissnerp4} to redo the fits. That
is, we fix $c_{1-4}$ in $\hat{c}_{1-4}$ of Eq.~(\ref{chats}) with their
corresponding fit values at $O(p^3)$ level given in
Table~\ref{tabp3}. In other words, we perform fits with four
dimension-4 combinations: $ e_{22}-4e_{38}$ (in $\hat{c_1}$), $
e_{20}+e_{35}$ (in $\hat{c_2}$), $ 2e_{19}-e_{22}-e_{36}$ (in
$\hat{c_3}$ ), $2e_{21}-e_{37}$ (in $\hat{c_4}$),  instead of
$\hat{c}_{1-4}$.
For clarity, we will denote the modified $O(p^4)$ fits discussed here
by ``Fit I(b)-$O(p^4)$'' and ``Fit II(b)-$O(p^4)$'', respectively.
In this case, the contributions from different orders are separated, so
we can study the convergence of the amplitude. The resulting values for
the LECs are shown in Table~\ref{tabp4s2}.

\begin{table}[ht]
\begin{center}
\begin{tabular}{|c||r r||r|}
\hline
 LEC   & Fit I(b)-$O(p^4)$ &HB$\chi$PT& Fit II(b)-$O(p^4)$\\
\hline\hline
$c_1$   &  $-1.39^*$  &$(-1.47, -1.21)$& $-0.81^*$\\
$c_2$   &  $4.01^*$ &$(3.13, 3.29)$& $1.46^*$  \\
$c_3$   &  $-6.61^*$  &$(-6.14, -5.85)$& $-3.10^*$\\
$c_4$   &  $3.92^*$ &$(3.47, 3.50)$& $2.35^*$\\
\hline
$d_1+d_2$& $7.39\pm0.11$   &$(4.90, 5.32)$&$3.18\pm0.05$\\
$d_3$   &  $-8.04\pm0.13$  &$(-4.37, -3.61)$&$-4.75\pm0.04$\\
$d_5$   &   $0.62\pm0.11$  &$(-1.03, -0.13)$&$1.11\pm0.03$\\
$d_{14}-d_{15}$& $-13.90\pm0.20$  &$(-9.31, -8.70)$&$-5.82\pm0.09$\\
$d_{18}$&  $-0.56^*$  &$(-1.49, -0.84)$&$-0.15\pm0.17$\\
\hline
$e_{14}$&   $3.25\pm0.37$  &$(2.33, 4.19)$&$-9.78\pm0.08$\\
$e_{15}$& $-14.50\pm0.55$  &$(-3.33, 4.54)$&$15.29\pm0.12$\\
$e_{16}$&   $7.65\pm0.35$  &$(2.74, 5.69)$&$-2.76\pm0.07$\\
$e_{17}$&   $8.21\pm1.34$  &$(5.14, 7.20)$&$18.35\pm0.14$\\
$e_{18}$&  $-0.79\pm1.19$  &$(-3.36, -1.27)$&$-11.58\pm0.11$\\
$e_{22}-4e_{38}$&  $-8.19\pm1.79$  &$(7.38, 27.72)$&$10.29\pm0.82$\\
$e_{20}+e_{35}$&  $-12.86\pm0.83$  &$(-17.35, -10.49)$&$-13.12\pm0.28$\\
$2e_{19}-e_{22}-e_{36}$&  $18.18\pm1.72$  &$(-25.12, -1.49)$&$0.83\pm0.55$\\
$2e_{21}-e_{37}$&  $-32.74\pm3.40$  &$(-7.12, -1.66)$&$-25.46\pm0.48$\\
\hline
$h_A$   &  -               &-& $2.90^*$\\
\hline
$\chi^2_{d.o.f}$&  0.03    &$(0.14, 0.58)$&0.11\\
\hline
\end{tabular}
\caption{\label{tabp4s2}LECs given by fit up to ${\cal O}(p^4)$. Fit I(b) and Fit II(b) are performed with $c_i$ in $\hat{c}_i$ are fixed at the corresponding $O(p^3)$ fit values shown in Table~\ref{tabp3}, see the explanation in the text. Fit I(b) is performed up to 1.13GeV, while Fit II(b) up to 1.20GeV including explicit $\Delta(1232)$ contribution. In Fit II(b), the results correspond to $\hat{c}^\prime_i$, $d^\prime_j$ and $e_k^\prime$ instead of $\hat{c}_i$, $d_j$ and $e_k$, respectively. For comparison, we provide the results from~\cite{Meissnerp4}. The $c_i$, $d_j$ and $e_k$ have, respectively, units of $\rm{GeV}^{-1}$, $\rm{GeV}^{-2}$ and $\rm{GeV}^{-3}$, and $h_A$ is dimensionless. The $*$ denotes an input quantity.}
\end{center}
\end{table}

 Comparing ``Fit I(b)-$O(p^4)$'' results with those from
Ref.\cite{Meissnerp4}, which are summed as intervals listed in the
third column of Table~\ref{tabp4s2}, it is found, however, that most
of our fitted $d_j$ and $e_k$ do not locate inside the intervals.
The main reason might be that our primitive values for $c_{2-4}$ as
input are not in the corresponding intervals (see
Table~\ref{tabp4s2}), which cause the incomparability since a small
variation of $c_i$ may lead to big changes of the higher order LECs,
$d_j$ and $e_k$, though both our fit and that of
Ref.~\cite{Meissnerp4} maintain a good convergence property.

\begin{figure*}[ht]
\begin{center}
\includegraphics[width=1.0\textwidth]{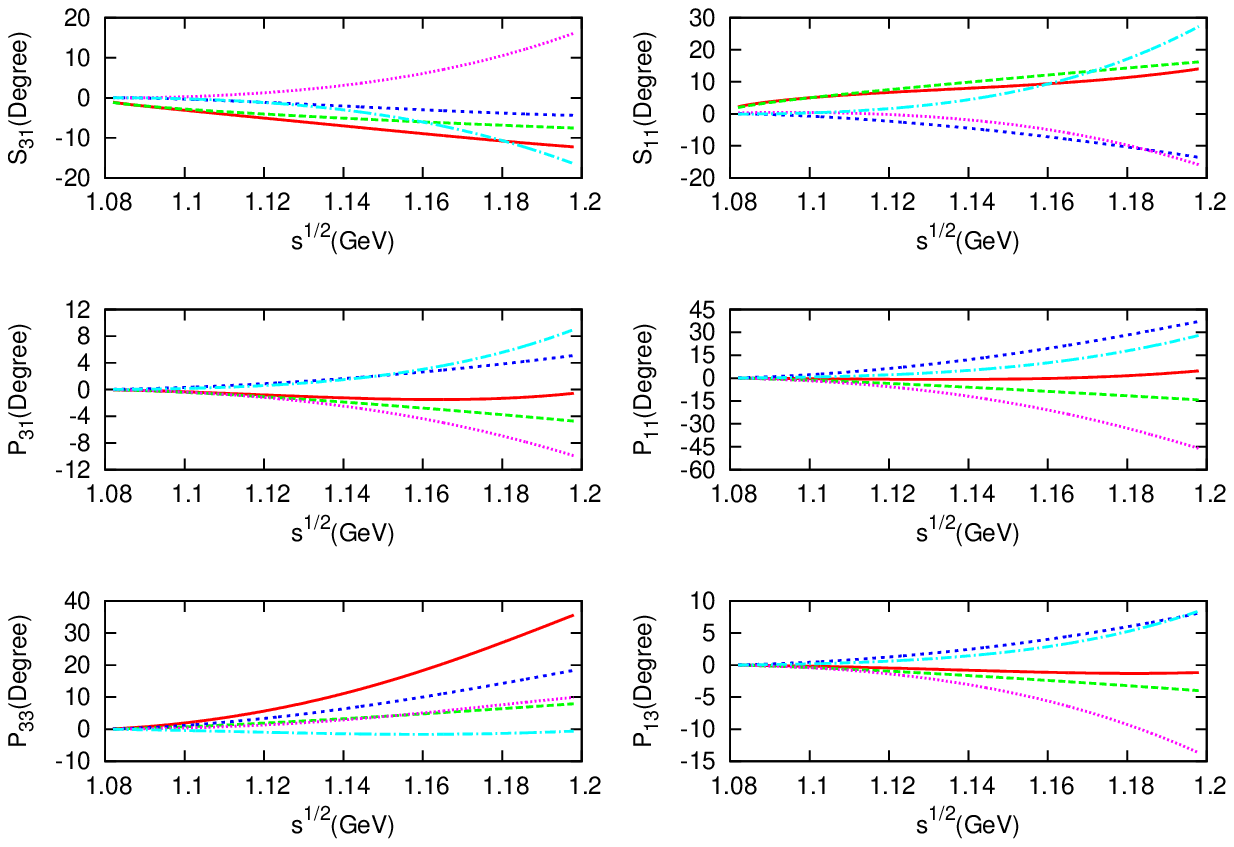}
\caption[pilf]{(Color online) Convergence properties of the chiral
series. The results of Fit I(b) in Table~\ref{tabp4s2} are adopted for plotting. The dashed (green),  short-dashed (blue), dotted (magenta),
dash-dotted(cyan), and solid (red) line represents
${\cal{O}}(p^1)$, ${\cal{O}}(p^2)$, ${\cal{O}}(p^3)$,
${\cal{O}}(p^4)$, and total contribution, respectively.}\label{figp4cov}
\end{center}
\end{figure*}

The convergence can be visualized by plotting
contributions from $O(p)$, $O(p^2)$, $O(p^3)$, $O(p^4)$ separately, and the sum of
them in Fig.~\ref{figp4cov}. Note that we plot up to 1.20GeV, though
fit only up to 1.13 GeV. One can observe that the $O(p^4)$
contributions (cyan dashed-dotted lines in Fig.~\ref{figp4cov}) are
in general small for all the partial waves below 1.13 GeV. The
$O(p^3)$ contributions (magenta dotted lines) are mostly larger than
$O(p)$ contributions (green dashed lines), with the exception of the
$S_{11}$ and $P_{33}$ partial waves. However, there exist
cancellations between the $O(p^2)$ (blue shorted dashed lines) and
$O(p^3)$ contributions. The red solid lines represent the total
contribution up to $O(p^4)$. They describe the existing partial wave
data below 1.13 GeV very well. After all, the convergence property of
the fourth-order calculation is reasonable, while the third-order
calculation is not satisfactory as pointed out by
Ref.~\cite{oller5}.

\begin{figure*}[ht]
\begin{center}
\includegraphics[width=1.0\textwidth]{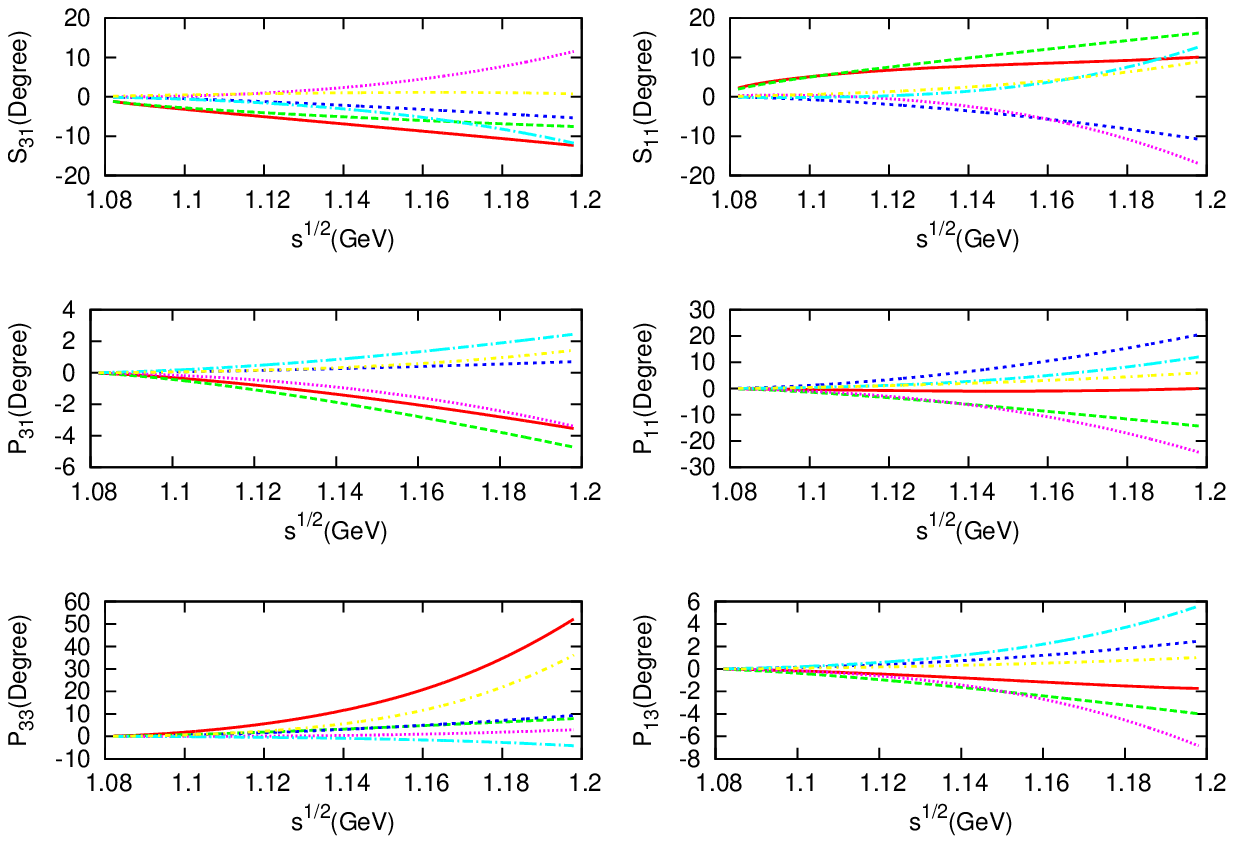}
\caption[pilf]{(Color online) Convergence properties of the chiral
series with explicit $\Delta(1232)$.  The results of Fit II(b) in Table~\ref{tabp4s2} are adopted for plotting. The dashed (green),
short-dashed (blue), dotted (magenta), dash-dotted (cyan),
short-dash-dotted (yellow) and solid (red) line represents
respectively ${\cal{O}}(p^1)$, ${\cal{O}}(p^2)$, ${\cal{O}}(p^3)$,
${\cal{O}}(p^4)$, $\Delta(1232)$ and total contribution.}\label{figp4delcov}
\end{center}
\end{figure*}

In Fig.~\ref{figp4delcov} we include the $\Delta$ contribution and plot the contribution from $\Delta(1232)$ together with the
contributions from $O(p)$, $O(p^2)$, $O(p^3)$, $O(p^4)$, and the sum of
them. From the yellow short dash-dotted
lines in Fig.~\ref{figp4delcov}, we can observe that $\Delta(1232)$
mainly contributes to $P_{33}$-wave while the contribution for other
channels is nonzero but very small. On the other hand, the chiral
series are in general well convergent near threshold almost for all
the partial waves. However when increasing the energy far above
the threshold, the convergence becomes worse. Especially, we can see
from the Fig.~\ref{figp4delcov} that the higher chiral order
contributions grow much more rapidly than the lower chiral order
contributions as the energy increases. This indicates that the chiral
perturbation expansion breaks down in the large energy region and stops being valid.


\subsection{Goldberger-Treiman relation}
\label{secGT} The Goldberger-Treiman (GT) relation~\cite{GTrelation}
is a straightforward result of PCAC and chiral symmetry,  which
connects the $\pi$-$N$ pseudoscalar (Yukawa) coupling constant
$g_{\pi N}$ with the axial vector coupling of nucleon $g_A$. Here its correction up to and including terms of ${\cal
O}{(p^3)}$ is obtained. In our discussion below, one can observe that the ${\cal O}{(p^3)}$ correction, denoted by $\Delta_{loops}^{(3)}$, is negligible compared with the ${\cal O}{(p^2)}$ correction.

The GT relation reads
\bea g_{\pi
N}=\frac{g_Am_N}{F_\pi}\left(1+\Delta_{GT}\right)\ , \eea where
$\Delta_{GT}$ represents the correction which can be divided into
three parts, \bea
\Delta_{GT}=-\frac{2\tilde{d}_{18}}{g_A}M_\pi^2+\Delta_{loop}^{(2)}+\Delta_{loop}^{(3)}\
, \eea with \bea \Delta_{loops}^{(2)}&=&
-\frac{m_N^2g_A^2}{f_\pi^2}\left\{\bar{H}^{(1)}(m_N^2)+\left(M_\pi^2\bar{H}_A(M_\pi^2)-\bar{J}_N(M_\pi^2)\right)-\left(M_\pi^2\bar{H}_A(0)-\bar{J}_N(0)\right)+2\bar{H}_A^{(2)}(0)\right\}\ ,\label{GTloop2}\\
\Delta_{loops}^{(3)}&=&\frac{4\tilde{c}_1M_\pi^2}{m_NF_\pi^2}\left(\bar{\Delta}_\pi-2m_N^2\bar{H}^{(1)}(m_N^2)\right)
+\frac{\tilde{c}_2}{2F_\pi^2m_N^3}
\left[2(2m_N^2-M_\pi^2)M_\pi^2\bar{H}^{(2)}(m_N)-M_\pi^2(m_N^2-\frac{M_\pi^2}{2})\bar{\tilde{H}}^{(1)}(m_N^2)+M_\pi^2\bar{\Delta}^{(2)}_\pi\right.\nonumber\\
&&\left.+\frac{1}{2}M_\pi^4\bar{\Delta}_\pi-\frac{1}{2}M_\pi^6\left(\bar{H}(m_N^2)-\bar{H}^{(1)}(m_N^2)\right)\right]+\frac{2(\tilde{c}_3+\tilde{c}_4)m_NM_\pi^2}{F_\pi^2}\bar{H}^{(3)}(m_N^2)
-\frac{(\tilde{c}_2-\tilde{c}_3-\tilde{c}_4) m_N M_\pi^2}{72 F_\pi^2
\pi ^2}\ .\label{GTloop3} \eea The first term related to
$\tilde{d}_{18}$ is ${\cal O}{(p^2)}$ and generates the main
contribution to $\Delta_{GT}$, e.g.  $1.71\%$ for   Fit I-${\cal
O}{(p^3)}$. In contrast, though $\Delta_{loops}^{(2)}$ is also
${\cal O}{(p^2)}$, it contributes a much smaller value,
$\Delta_{loops}^{(2)}\sim0.36\%$. In addition, we can see from
Eq.~(\ref{GTloop2}) that its contribution is independent of the
LECs. The last term in Eq.~(\ref{GTloop3}) is employed to cancel the
PCB terms generated by the terms before it.
 According to the naive power counting rule $\Delta_{loops}^{(3)}$ should be  ${ O}{(p^3)}$,
 but actually it possesses a chiral order higher than ${\cal O}{(p^4)}$ and including ${\cal O}{(p^4)}$. This can be easily observed if we reduce all the tensor integrals in Eq.~(\ref{GTloop3}) to scalar integrals and a common prefactor $M_\pi^4$ of order four will show up. It can be estimated by evaluating the loop integrals numerically, which leads to $\Delta_{loops}^{(3)}=\left[-7.07\tilde{c}_1+1.79\tilde{c}_2-2.30(\tilde{c}_3+\tilde{c}_4)\right]\times 10^{-4}$.
 Because the fitted $\hat{c}_{1-4}$ in Fit I-$O(p^4)$ are combinations of dimension 2 and 4 LECs, we prefer to
 bring the corresponding Fit I-${ O}{(p^3)}$ results in Table~\ref{tabp3} into $\Delta_{loop}^{(3)}$ to estimate its value, which gives
$\Delta_{loop}^{(3)}\cong 0.23 \%$.

In conclusion, the correction to GT relation can be rewritten much more explicitly as
 \bea
 \Delta_{GT}=\left\{-3.05 \tilde{d}_{18}+0.36+\left[-7.07\tilde{c}_1+1.79\tilde{c}_2-
 2.30(\tilde{c}_3+\tilde{c}_4)\right]\times 10^{-2}\right\}\times10^{-2}\
 ,
 \eea
The first two terms  contribute about $2.07\%$,
while the last term stands for the next order contribution which is $0.23\%$.
This indicates good convergence of the $\Delta_{GT}$. Practically,
the calculation of $\Delta_{GT}$ to at $O(p^2)$ is sufficient, since $\Delta_{GT}^{(3)}$ is negligible
. Hence in our $O(p^4)$ fits without
explicit $\Delta(1232)$, the parameter $d_{18}$ is fixed at the $O(p^3)$-fit value.


\subsection{pion-nucleon $\sigma$ term: $\sigma_{\pi N}$}
\label{secsigma}
In what follows, an explicit expression for $\sigma_{\pi N}$ up to $O(p^4)$ is introduced. Then fits are performed both including $\pi$-N phase shift data and QCD lattice data to fix the unknown LECs related to $\sigma_{\pi N}$. Finally, the fit values are used to predict $\sigma_{\pi N}$: $\sigma_{\pi N}=52\pm7$ MeV for Fit I (c) without $\Delta(1232)$ and $\sigma_{\pi N}=45\pm6$ MeV for Fit II (c) with the explicit $\Delta(1232)$ contribution.

The sigma term is a quantity of great physical importance to
understand the composition of the nucleon mass. It is
defined as the matrix element of the explicit chiral symmetry
breaking part of the QCD Lagrangian situated between the nucleon states at
zero momentum transfer,
 \bea
 \sigma_{\pi N}=\sum_{q=u,d}m_q\frac{dm_N}{dm_q}=<N|m_u \bar{u}u+m_d \bar{d}d|N>.
 \eea
Using the Gell-Mann-Oakes-Renner relation $M_\pi^2=B_0(m_u+m_d)$, the above equation becomes
\bea
\sigma_{\pi N}=M_\pi^2\frac{\partial m_N}{\partial
 M_\pi^2}\ ,
\eea
where $m_N$ takes the following explicit form derived by Eq.~(\ref{mNdirect}),
\bea\label{mNexplicit} m_N
&=&{m}-4\tilde{c}_1
M_\pi^2+\tilde{e}_1M_\pi^4
-\frac{3{m}g_A^2M_\pi^2}{32\pi^2F_\pi^2}\left\{
\frac{M_\pi}{{m}^2}\sqrt{4{m}^2-M_\pi^2}\arccos\frac{M_\pi}{2{m}}+\frac{M_\pi^2}{2{m}^2}\ln\frac{M_\pi^2}{{m}^2}\right\}+\frac{3\tilde{c}_2M_\pi^4}{128\pi^2F_\pi^2}
\nonumber\\
&+&\frac{M_\pi^4}{16\pi^2F_\pi^2}\left\{8\tilde{c}_1-3\tilde{c}_3-\frac{3}{4}\tilde{c}_2
\right\}\ln\frac{M_\pi^2}{{m}^2}-\frac{3\tilde{c}_1 g_A^2 M_\pi^4
}{8\pi^2F_\pi^2}\left\{1-\frac{M_\pi^2-2{m}^2}{2{m}^2}\ln\frac{M_\pi^2}{{m}^2}+\frac{(M_\pi^2-2m^2)M_\pi}{m^2\sqrt{4m^2-M_\pi^2}}
\arccos\frac{M_\pi}{2m} \right\}\ ,\nonumber\\
 \eea
here $m$ is the nucleon mass in the chiral limit and
$\tilde{e}_1\equiv
-2(4e_{22}-8e_{38}+e_{115}+e_{116}-4\tilde{c}_1\tilde{\ell}_3/F_\pi^2)$.
Actually, $\tilde{c}_1$ is the EOMS renormalized quantity for
$\hat{c}_1=c_1-2M^2(e_{22}-4e_{38})$. All the quantities except $m$
on the right-hand side of Eq.~(\ref{mNexplicit}) are substituted by the
physical ones. $\sigma_{\pi N}$ is obtained straightforwardly,
 \bea \sigma_{\pi N}
 &=&-4\tilde{c}_1
M_\pi^2+\tilde{e}_1M_\pi^4 -\frac{3g_A^2M_\pi^3}{16\pi^2F_\pi^2m}\{
\frac{3m^2-M_\pi^2}{\sqrt{4m^2-M_\pi^2}}\arccos\frac{M_\pi}{2m}+M_\pi\ln\frac{M_\pi}{m}\}
+\frac{M_\pi^4}{16\pi^2F_\pi^2}\{8\tilde{c}_1-3\tilde{c}_3-\frac{3}{4}\tilde{c}_2
\}(4\ln\frac{M_\pi^2}{m^2}+1)\nonumber\\
&&+\frac{3\tilde{c}_2M_\pi^4}{64\pi^2F_\pi^2}-\frac{3\tilde{c}_1
g_a^2 M_\pi^4 }{8\pi^2F_\pi^2}\{\frac{4m^2-3M_\pi^2}{m^2}\ln
\frac{M_\pi}{m}+\frac{12m^2-2M_\pi^2}{4m^2-M_\pi^2}+\frac{26m^2M_\pi^3-60m^4M_\pi-3M_\pi^5}{m^2(4m^2-M_\pi^2)^{\frac{3}{2}}}
\arccos\frac{M_\pi}{2m} \}.\label{sigmap4}
 \eea

\begin{figure}[ht]
\begin{center}
\includegraphics[width=0.6\textwidth]{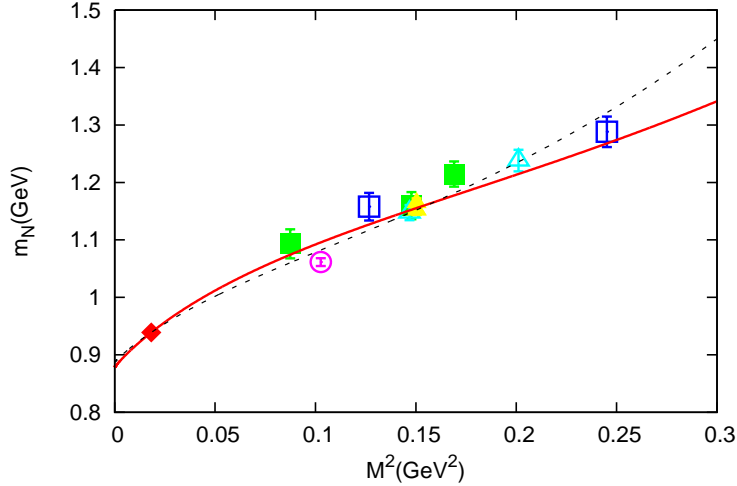}
\caption{The nucleon mass as a function of the pion mass. The solid
line denotes the result from the Fit I(c)-$O(p^4)$ and the dashed
line is Fit II(c)-$O(p^4)$. Sources of different lattice QCD data
are: solid squares (PACS-CS)~\cite{PACS-CS}, open squares (LHPC)
~\cite{LHPC}, open circles (QCDSF-UKQCD) ~\cite{QCDSF-UKQCD}, open
triangles (HSC) ~\cite{HSC}, solid triangles (NPLQCD)
~\cite{NPLQCD}. The solid diamond is the physical
point.}\label{mNLQCD}
\end{center}
\end{figure}

At $O(p^3)$ $\sigma_{\pi N}$ can be determined by the value of
$c_1$. In Ref.~\cite{oller2}, through an analysis on $\pi$-N
scattering partial wave phase shift data using the EOMS-B$\chi$PT, it
predicts $\sigma_{\pi N}=59 \pm 7$ MeV. At $O(p^4)$
$\sigma_{\pi N}$ in Eq.~(\ref{sigmap4}) has the unknown coupling
constants combination $e_1$ which does not appear in the $\pi$-N
scattering amplitude. However, recently the lattice QCD
simulations have gotten many data on the quark mass dependence of
the nucleon mass, which enables us to fix $e_1$ as well as $c_1$. Taking only the lattice
QCD data into consideration, chiral effective field theory have been used to predict $\sigma_{\pi N}$ up to $O(p^4)$~\cite{Weise04,Geng12}. In the current
paper, fits are performed both including the $\pi$-N scattering partial
wave phase shift data and lattice QCD data. In our fits, lattice QCD data are taken from
PACS-CS~\cite{PACS-CS}, LHPC ~\cite{LHPC}, HSC ~\cite{HSC},
 QCDSF-UKQCD ~\cite{QCDSF-UKQCD} and NPLQCD ~\cite{NPLQCD}
collaborations. Following the strategy of Ref.~\cite{Geng12}, in order
to minimize uncertainties of finite volume effects we only use the
data with $M_\pi L>4$, and we also only choose those with
$M_\pi^2<0.25~\rm{GeV}^2$. So there are only 11 lattice data points which meet the
requirements. They are denoted with stars in the tables of the
Appendix A in ~\cite{Geng12}. Note that the physical nucleon mass is
included in the fits as a constraint. Compared with the previous fits to partial
wave phase shift data, two additional fit parameters: $m$ and
$e_1$ are included. The fit results are listed in Table~\ref{tabp4}, where in
Fit II(c) the leading $\Delta$-exchange Born term and the partially-included $\Delta$ loop contribution (see~\ref{appdeltasat}) are
considered whereas Fit I(c) does not. The predicted nucleon mass as a function of pion mass is plotted in Fig.~\ref{mNLQCD}.

From Table~\ref{tabp4}, we find that most
fit parameters in Fit I(c) and Fit II(c) change little compared with Fit I(a) and Fit II(a) in Table~\ref{tabp4}, respectively. The prediction for $\sigma_{\pi N}$ are: $\sigma_{\pi N}=52 \pm 7$ MeV for Fit I(c) and $\sigma_{\pi
N}=45 \pm 6$ MeV for Fit II(c).
Our result $\sigma_{\pi N}=52 \pm 7$ MeV is smaller than the result obtained from the fit
to the $\pi$-N scattering partial wave phase shift data up to
$O(p^3)$ given in Ref.~\cite{oller2}: $\sigma_{\pi N}=59 \pm 7$ MeV. We improve our determination of $\sigma_{\pi N}$ in two ways. On one hand, the fourth-order correction to $\sigma_{\pi N}$ is obtained. On the other hand, to our knowledge, this is the first attempt to treat the $\pi$-N
scattering data and lattice QCD data together using the
EOMS-B$\chi$PT up to $O(p^4)$, and this may constrain the value
of the sigma term better. The result: $\sigma_{\pi
N}=45 \pm 6$ MeV agrees
well with the recent analysis on lattice QCD data using
EOMS-B$\chi$PT up to $O(p^4)$~\cite{Geng12}, which gives $\sigma_{\pi
N}=43(1)(6)$ MeV. However, because the exact $\Delta$-included loop graphs are not considered, $\sigma_{\pi
N}=45 \pm 6$ MeV can still be improved in future.

\section{Conclusions}
\label{secconclusions} \nin In this paper, we performed a calculation of the pion-nucleon elastic
scattering amplitude in the isospin limit within the framework of
covariant baryon $\chi$PT using EOMS scheme up to $O(p^4)$. The amplitude is covariant and possesses correct analyticity and power counting properties. The resultant description of the existing partial wave phase shift
data from Ref.~\cite{GW08} is very good for the energy in center of mass frame up to 1.13~GeV, and up to 1.20~GeV including the leading order $\Delta(1232)$ Born-term and the partially-included $\Delta(1232)$ loop contributions. The dimension-2, -3, and -4 LECs or their combinations are determined. The convergence properties of the chiral series are discussed. The fourth-order calculation without explicit $\Delta(1232)$ displays a good convergence property at $O(p^4)$ in the threshold region--the $O(p^4)$ (NNNLO) contribution is found much smaller than the LO, NLO and NNLO ones for all the partial waves. It is certainly an improvement to the unsatisfactory situation in the
third-order calculation, discussed in previous literature~\cite{oller5}. However, when we explicitly include the $O(p^{3/2})$ Born term contribution of $\Delta(1232)$ in $\delta$-counting~\cite{pascalutsa}, as well as partially the $\Delta(1232)$ loop graphs, the convergence property is not good in the region close to the $\Delta$ resonance, which indicates that the exact $O(p^{7/2})$
 loop contribution may be sizable and should be considered carefully in the future.

As physical applications, first, the correction to GT relation is discussed up to $O(p^3)$. The $O(p^3)$ correction is much smaller (about $0.2\%$) than the
$O(p^2)$ correction (about $2\%$), which implies good convergence property of $\Delta_{GT}$ and confirms the applicability of EOMS-B$\chi$PT to low energy physics. Secondly, a reasonable prediction for the pion-nucleon $\sigma$ term $\sigma_{\pi N}$ is obtained. We find $\sigma_{\pi N}=52\pm7$ MeV from the fit without $\Delta(1232)$, and $\sigma_{\pi N}=45\pm6$ MeV from the fit with explicit $\Delta(1232)$. The two values are obtained by performing fits including the
pion-nucleon partial wave phase shift data and the lattice QCD data for $m_N$.

\section*{Acknowledgements}
\nin
 We would like to thank Li-sheng Geng for very helpful discussions and Juan Jose Sanz Cillero for a careful reading of the manuscript and valuable comments. This work is supported in
part by National Nature Science
Foundations of China under Contract Nos.  10925522 
and
11021092.
 Y.~H.~Chen also acknowledges support by Grant Nos. 11261130311
(CRC110 by DFG and NSFC) and by NNSFC under Contract No. 11035006.
  \appendix

 \section{$\Delta(1232)$ contribution}
 \subsection{Effective Lagrangian and Leading Born-term contribution}
 \label{appborndelta}
\begin{figure}[ht]
\begin{center}
\includegraphics[width=0.3\textwidth]{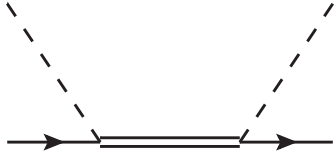}
\caption{Leading $\Delta$-exchange Born-term contribution. Double solid, solid and dashed lines represent $\Delta$, nucleon and pion, respectively. Crossed graph is not shown. }\label{deltaexchange}
\end{center}
\end{figure}

 For our calculation here, the relevant effective Lagrangian with $\Delta(1232)$ as explicit degree of freedom reads
 \bea
 {\cal L}_{eff}^\Delta&=&{\cal L}_{RS}+{\cal L}_{\pi N\Delta}\ ,\nnb\\
 {\cal L}_{RS}&=&\bar{\Delta}_\mu\left\{i\gamma^{\mu\nu\alpha}\partial_\alpha-m_\Delta\gamma^{\mu\nu}\right\}\Delta_\nu\ ,\nnb\\
 {\cal L}_{\pi N\Delta}&=&\frac{ih_A}{2F_\pi m_\Delta}\bar{N}T_a^\dag\gamma^{\mu\nu\lambda}(\partial_\mu\Delta_\nu)\partial_\lambda\pi^a+h.c.\ ,
 \eea
 where $T_a$ are the isospin-$1/2$--isospin-$3/2$ transition matrices satisfying $T_a^\dag T_b=\frac{2}{3}\delta_{ab}-\frac{1}{3}i\varepsilon_{abc}\tau_c$. Conventions for $\gamma^{\mu\nu\alpha}$ and $\gamma^{\mu\nu}$ can be consulted in Ref.~\cite{pascalutsa}. The leading $\Delta$-exchange Born-term contribution to pion-nucleon scattering is $O(p^{3/2})$ in the so called $\delta$-counting rule proposed by Ref.~\cite{pascalutsa}, and the Feynman diagram is shown in Fig.~\ref{deltaexchange}.
The total contribution of leading Born-term reads
\bea\label{deltacon}
A^\pm_\Delta(s,t)=A_{s\Delta}^\pm(s,t)\pm A_{s\Delta}^\pm(u,t)\ ,\nnb\\
B^\pm_\Delta(s,t)=B_{s\Delta}^\pm(s,t)\mp B_{s\Delta}^\pm(u,t)\ ,
\eea
with
\bea
   A^+_{s\Delta}(s,t)&=&\frac{h_A^2}{6F_\pi^2m_\Delta^2}\frac{s}{s-m_\Delta^2}\left\{(m_N+m_\Delta)\left[(M_\pi^2-\frac{t}{2})-\frac{1}{3}(s-m_N^2)\right]-\frac{s-m_N^2+M_\pi^2}{3s}\left[\frac{m_N}{2}(s-m_N^2+M_\pi^2)+m_\Delta M_\pi^2\right]\right\}\ ,\nnb\\
   B^+_{s\Delta}(s,t)&=&\frac{h_A^2}{6F_\pi^2m_\Delta^2}\frac{s}{s-m_\Delta^2}\left\{M_\pi^2-\frac{t}{2}+\frac{1}{3}\left[2m_N(m_N+m_\Delta)-M_\pi^2\right]-\frac{s-m_N^2+M_\pi^2}{3s}\left[\frac{1}{2}(s-m_N^2+M_\pi^2)+m_Nm_\Delta\right]\right\}\ ,\nnb\\
   A^-_{s\Delta}(s,t)&=&-\frac{1}{2}A^+_{s\Delta}(s,t)\ ,\hspace{1cm}
   B^-_{s\Delta}(s,t)=-\frac{1}{2}B^+_{s\Delta}(s,t)\ .\label{LOdeltaBT}
\eea

\subsection{$\Delta$ contribution to LECs and partial inclusion of the $\Delta$ loop}
\label{appdeltasat}
\begin{figure}[ht]
\begin{center}
\includegraphics[width=0.8\textwidth]{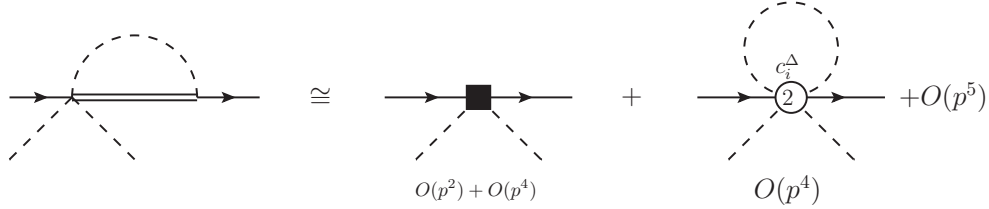}
\caption{Matching between $\Delta$-included and $\Delta$-less loop graphs. The circled `2' vertex is related to $c_{1-4}^\Delta$. Possible polynomial terms (containing PCB terms), denoted by a black square, would be absorbed in the $c_i,~d_j,~e_k$ couplings as it was done in Sec.~\ref{secren}. Barring the latter, the left-hand side $\Delta$-included loop graph differs from the right-hand side $\Delta$-less loop graph by contributions of $O(p^5)$ and higher in the chiral expansion. }\label{deltasat}
\end{center}
\end{figure}

In order to evaluate the tree-level contribution to the LECs $c_{1-4}$ when the $\Delta$-resonance is integrated out, the leading Born-term contribution is expanded in powers of $\sigma=s-m_N^2$, $M_\pi^2$ and $t$,  and then compared with the $O(p^2)$ tree amplitude Eq.~\ref{p2tree},  leading to
\bea\label{deltaci}
c_1^\Delta=0\ ,\quad c_2^\Delta=-c_3^\Delta=2c_4^\Delta=\frac{h_A^2 m_N^2}{9 (m_\Delta-m_N) m_\Delta^2}=1.85\rm{GeV}^{-1}\ ,
\eea
where we used $h_A=2.90$, $m_N=0.939$ GeV, and $m_\Delta=1.232$ GeV for the numerical value at the end of Eq.~\ref{deltaci}. We are not able to exactly calculate pion-nucleon loop diagrams involving $\Delta$ resonance. Nevertheless, this shortcoming can be partially remedied by substituting the $O(p^2)$ vertices in the $O(p^4)$ loop graphs shown in Fig.~\ref{loopgraphp4} by the contributions from $\Delta$ exchanges in Eq.~(\ref{deltaci}).
 The procedure is illustrated in Fig.~\ref{deltasat}. Hence, the `full' one-loop $O(p^4)$ contribution can be given by the diagrams in Fig.~\ref{loopgraphp4} with a replacement\footnote{We thank J.~J.~Sanz Cillero for pointing it out to us.} $c_i=c_i^\prime+c_i^\Delta$. The effects of this replacement include only the $O(p^{7/2})$ $\Delta$-included loop graphs, while higher order graphs, like $\pi$N $\rightarrow$ $\pi\Delta$(loop) $\rightarrow$ $\pi$N of $O(p^{11/2})$, are beyond the accuracy of our calculation and therefore absent. Also, the $O(p^{7/2})$ loop diagrams involving the $\Delta$ propagator contributing to the self energy of nucleon can be estimated in the same way.

 \section{Tree amplitudes}
 \label{apptree}
 For convenience, the two independent scalar kinematical variables, $\nu$ and $\nu_B$,  are defined as
 \bea
 \nu=\frac{s-u}{4m_N}\ ,\quad\nu_B=\frac{t-2M_\pi^2}{4m_N}\nnb\ .
 \eea
 The Feynman graphs in Fig.~\ref{treediagrams} are calculated directly and expressed in the $D$ and $B$ functions, while their crossed graphs are obtained by the following crossing relations:
 \bea\label{DBAcrossrelation}
 D^+(-\nu,\nu_B)&=&D^+(\nu,\nu_B)\ ,\quad D^-(-\nu,\nu_B)=-D^-(\nu,\nu_B)\ ;\nnb\\
 B^+(-\nu,\nu_B)&=&-B^+(\nu,\nu_B)\ ,\quad B^-(-\nu,\nu_B)=B^-(\nu,\nu_B)\ ;\nnb\\
 A^+(-\nu,\nu_B)&=&A^+(\nu,\nu_B)\ ,\quad A^-(-\nu,\nu_B)=-A^-(\nu,\nu_B)\ .
 \eea
 Here the crossing relation for the $A$ function is also displayed for the sake of completeness.  Additionally, if the $D$ and $B$ functions are expressed in terms of arguments, $s$ and $t$, then $s$ should be changed to $u$ on the right-hand side of each relation in Eq.~\ref{DBAcrossrelation}. Note that only the graphs (a), (d) and (e) in Fig.~\ref{treediagrams} have their corresponding crossed diagrams.
 \subsection{$O(p)$}
 \begin{itemize}
  \item Graph (a):
    \bea
    D_{a}^\pm&=&-{g^2\over 4F^2}{2m_N\over s-m_4^2}\left[\nu(\nu-\nu_B)+(m_N+m_4)\nu_B\right]\ ,\nnb\\
    B_{a}^\pm&=&-{g^2\over 4F^2}{2m_N\over s-m_4^2}(\nu-\nu_B+m_N+m_4)\ ,
    \eea
   where $m_4=m-4c_1M^2-2(8e_{38}+e_{115}+e_{116})M^4$.
  \item Graph (b):
    \bea
    D_{b}^+&=&0\ ,\quad D_{b}^-={\nu\over2F^2}\ ,\nnb\\
    B_{b}^+&=&0\ ,\quad B_{b}^-={1\over2F^2}\ .
    \eea
\end{itemize}
 \subsection{$O(p^2)$}
\begin{itemize}
  \item Graph (c):
    \bea\label{p2tree}
    D_{c}^+&=&-{4c_1M^2\over F^2}+{c_2\left(16m_N^2\nu^2-t^2\right)\over 8F^2m^2}+{c_3\left(2M_\pi^2-t\right)\over F^2}\ ,\quad D_{c}^-=0\ ,\nnb\\    B_{c}^+&=&0\ ,\quad B_{c}^-={2c_4m_N\over F^2}\ .
    \eea
\end{itemize}
 \subsection{$O(p^3)$}
\begin{itemize}
  \item Graphs (d)+(e):
    \bea
    D_{de}^\pm&=&-{g_AM_\pi^2\over F_\pi^2}(2d_{16}-d_{18})\left[\nu+{2m_N\nu_B\over \nu-\nu_B}\right]\ ,\nnb\\
    B_{de}^\pm&=&-{g_AM_\pi^2\over F_\pi^2}(2d_{16}-d_{18})\left[1+{2m_N\over \nu-\nu_B}\right]\ .
    \eea
  \item Graph (f):
    \bea
    D_{f}^+&=&0\ ,\quad D_{f}^-={2\nu\over F_\pi^2}\left[2(d_1+d_2+2d_5)M_\pi^2-(d_1+d_2)t+2d_3\nu^2\right]\ ,\nnb\\
    B_{f}^+&=&{4(d_{14}-d_{15})\nu m_N\over F_\pi^2}\ ,\quad B_{f}^-=0\ .
    \eea
\end{itemize}
 \subsection{$O(p^4)$}
\begin{itemize}
\item Graph (g):
  \bea
 D_{g}^+&=&\frac{16}{F_\pi^2}\left\{e_{14}\left(M_\pi^2-\frac{t}{2}\right)^2+e_{15}\left(M_\pi^2-\frac{t}{2}\right)\nu^2+e_{16}\nu^4
 +\frac{1}{2}(2e_{19}-e_{22}-e_{36})M_\pi^2\left(M_\pi^2-\frac{t}{2}\right)\right.\nonumber\\
&&\left.+(e_{20}+e_{35})M_\pi^2\nu^2+\frac{1}{2}(e_{22}-4e_{38})M_\pi^4 \right\}\ ,\quad D_{g}^-=0\ ,\nnb\\
 B_{g}^+&=&0\ ,\quad B_{g}^-=\frac{16 m_N}{F_\pi^2}\left\{e_{17}\left(M_\pi^2-\frac{t}{2}\right)+e_{18}\nu^2+\frac{1}{2}(2e_{21}-e_{37})M_\pi^2\right\}\ .
 \eea
\end{itemize}
 \section{One-loop Amplitudes}
 \subsection{Definitions of loop functions}
 \label{funct}
All the integrals appearing in the scattering amplitude up to ${\cal O}(p^4)$ level can be generalized as
\bea
H_{mn}^{\mu_1\cdots\mu_r}=\intd\frac{k^{\mu_1}\cdots k^{\mu_r}}{A_1\cdots A_m B_1\cdots B_n}\ ,
\eea
where $A_i=M^2-(k-q_i)^2-i\epsilon$ and $B_j=m^2-(P_j-k)^2-i\epsilon$ stems from the meson and nucleon propagator respectively.
A standard approach to evaluate such tensor integrals has been developed by Passarino and Veltman in Ref.\cite{PasVel}. In this approach, the \textit{Passarino-Veltman decomposition} is first carried out by representing the tensor integral as a sum of independent tensor structures multiplied by scalar quantities. Then the scalar quantities are further expressed by means of initial scalar functions of the form:
\bea
H_{mn}=\intd\frac{1}{A_1\cdots A_m B_1\cdots B_n}\ .
\eea

In what follows, we will specify the definitions of  all the loop functions  and  the Passarino-Veltman decomposition formulae, with
the help of the external momenta defined as
\bea
\Sigma_\mu&=&(P+q)^\mu=(P^\prime+q^\prime)^\mu,\nonumber\\
Q^\mu&=&(P^\prime+P)^\mu,\nonumber\\
\Delta_\mu&=&(q^\prime-q)^\mu=(P-P^\prime)^\mu.\nonumber
\eea

\begin{itemize}
\item
1 meson:
$\Delta_\pi=H_{10}$
\bea
&&{\Delta_\pi,\Delta_\pi^\mu,\Delta_\pi^{\mu\nu}}=\intd\frac{\{1,k^\mu,k^{\mu\nu}\}}{M^2-k^2}\ , \nonumber\\
&&\Delta_\pi^\mu=0\ ,\nonumber\\
&&\Delta_\pi^{\mu\nu}=g^{\mu\nu}\Delta^{(2)}_\pi\ ,\nonumber
\eea
where
\bea
\Delta_\pi&=&\frac{M^2}{16\pi^2}\left\{R+\ln\frac{M^2}{\mu^2}\right\}\label{deltapi}\ ,R=\frac{2}{d-4}+\gamma_E-1-\ln 4\pi\ ,\\
\Delta^{(2)}_\pi&=&\frac{M^2}{d}\Delta_\pi=\frac{M^2}{4}\left(\Delta_\pi-\frac{M^2}{32\pi^2}\right)\ .\nonumber
\eea

\item
1 nucleon:
$\Delta_N=H_{01}$
\bea
&&\{\Delta_N,\Delta_N^\mu,\Delta_N^{\mu\nu}\}=\intd\frac{\{1,k^\mu,k^\mu k^\nu\}}{m^2-(\Sigma-k)^2}\ , \nonumber\\
&&\Delta_N^\mu=\Sigma^\mu\Delta_N\ ,\nonumber\\
&&\Delta_N^{\mu\nu}=g^{\mu\nu}\Delta^{(2)}_N+\Sigma^\mu\Sigma^\nu\Delta_N\ ,\nonumber
\eea
where
\bea
\Delta_N&=&\frac{m^2}{16\pi^2}\left\{R+\ln\frac{m^2}{\mu^2}\right\}\label{deltaN}\ ,R=\frac{2}{d-4}+\gamma_E-1-\ln 4\pi\ ,\\
\Delta_N^{(2)}&=&\frac{m^2}{d}\Delta_N=\frac{m^2}{4}\left(\Delta_N-\frac{m^2}{32\pi^2}\right)\ .\nonumber
\eea

\item
2 mesons: $J=H_{20}$
\bea
&&\{J,J^\mu,J^{\mu\nu}\}=\intd\frac{\{1,k^\mu,k^\mu k^\nu\}}{[M^2-k^2]\;[M^2-(k-\Delta)^2]}\ ,\nonumber\\
&&J^\mu(t)=\frac{1}{2}\Delta^\mu J(t)\ .\nonumber\\
&&J^{\mu\nu}(t)=(\Delta^\mu\Delta^\nu-g^{\nu\nu}\Delta^2)J^{(1)}(t)+\Delta^\mu\Delta^\nu
J^{(2)}(t)\ ,\nonumber
\eea
where
\bea
&&J^{(1)}(t)=\frac{1}{(1-d)t}\left\{-\frac{1}{2}\Delta_\pi+\left(M^2-\frac{1}{4}t\right)J(t)\right\}\nonumber\\
&&\hspace{1cm}=-\frac{1}{3t}\left\{-\frac{1}{2}\Delta_\pi+\left(M^2-\frac{1}{4}t\right)J(t)+\frac{1}{16\pi^2}\left( M^2-\frac{1}{6}t\right)\right\}\ ,\nonumber\\
&&J^{(2)}(t)=\frac{1}{2t}\left\{-\Delta_\pi+\frac{1}{2}t J(t)\right\}
\ .\nonumber
\eea

\item
1 meson, 1 nucleon: $H=H_{11}$
\bea
&&\{H,H^\mu,H^{\mu\nu},\tilde{H}^{\mu},H^{\mu\nu\rho}\}=\intd\frac{\{1,k^\nu,k^\mu k^\nu,k^\mu k^2,k^\mu k^\nu k^\rho\}}{[M^2-k^2]\;[m^2-(\Sigma-k)^2]}\ ,\nonumber\\
&&H^\mu(s)=\Sigma^\mu H^{(1)}(s)\ ,\nonumber\\
&&H^{\mu\nu}(s)=g^{\mu\nu}H^{(2)}(s)+\Sigma^\mu\Sigma^\nu
H^{(3)}(s)\ ,\nonumber\\
&&\tilde{H}^\mu(s)=\Sigma^\mu\tilde{H}^{(1)}(s)\ ,\nonumber\\
&&H^{\mu\nu\rho}(s)=(g^{\mu\nu}\Sigma^\rho+g^{\nu\rho}\Sigma^\mu+g^{\rho\mu}\Sigma^\nu)H^{(4)}(s)+\Sigma^\mu\Sigma^\nu\Sigma^\rho H^{(5)}(s)\ ,\nonumber
\eea
where
\bea
H^{(1)}(s)&=&\frac{1}{2s}\left\{\Delta_\pi-\Delta_N+(s-m^2+M^2)H(s)\right\}\ ,\nonumber\\
H^{(2)}(s)&=&\frac{1}{d-1}\left\{-\frac{1}{2}\Delta_N+M^2H(s)-\frac{1}{2}(s-m^2+M^2)H^{(1)}(s)\right\}\ ,\nonumber\\
&=&\frac{1}{3}\left\{-\frac{1}{2}\Delta_N+M^2H(s)-\frac{1}{2}(s-m^2+M^2)H^{(1)}(s)+\frac{M^2+m^2-\frac{1}{3}s}{32\pi^2}\right\}\ ,\nonumber\\
H^{(3)}(s)&=&\frac{1}{s}\left\{-\Delta_N+M^2H(s)-dH^{(2)}(s)\right\}\ ,\nonumber\\
&=&\frac{1}{3s}\left\{-\Delta_N-M^2H(s)+2(M^2+s-m^2)H^{(1)}(s)-\frac{M^2+m^2-\frac{1}{3}s}{32\pi^2}\right\}\ ,\nonumber\\
\tilde{H}^{(1)}(s)&=&-\Delta_N+M^2H^{(1)}(s)\ ,\nonumber\\
H^{(4)}(s)&=&\frac{1}{2s}\left\{\Delta_\pi^{(2)}-\Delta_N^{(2)}+(s-m^2+M^2)H^{(2)}(s)\right\}\ ,\nonumber\\
H^{(5)}(s)&=&\frac{1}{2s}\left\{-\Delta_N+(s-m^2+M^2)H^{(3)}(s)-4H^{(4)}(s)\right\}\ .\nonumber
\eea

\item
2 nucleons: $J_N=H_{02}$
\bea
&&\{J_N,J_N^\mu,J_N^{\mu\nu}\}=\intd\frac{\{1,k^\mu,k^\mu k^\nu\}}{[m^2-(k-P)^2]\;[m^2-(k-P^\prime)^2]}\ ,\nonumber\\
&&J_N^\mu(t)=\frac{1}{2}Q^\mu J_N(t)\ ,\nonumber\\
&&J_N^{\mu\nu}(t)=(\Delta^\mu\Delta^\nu-g^{\nu\nu}\Delta^2)J_N^{(1)}(t)+\Delta^\mu\Delta^\nu
J_N^{(2)}(t)+\frac{1}{2}(P^\mu P^{\prime\nu}+P^\nu P^{\prime\mu})J_N(t)\ ,\nonumber
\eea
where
\bea
&&J_N^{(1)}(t)=\frac{1}{(1-d)t}\left\{-\frac{1}{2}\Delta_N+\left(m^2-\frac{1}{4}t\right)J_N(t)\right\}\nonumber\\
&&\hspace{1cm}=-\frac{1}{3t}\left\{-\frac{1}{2}\Delta_N+\left(m^2-\frac{1}{4}t\right)J_N(t)+\frac{1}{16\pi^2}\left( m^2-\frac{1}{6}t\right)\right\}\ ,\nonumber\\
&&J_N^{(2)}(t)=\frac{1}{2t}\left\{-\Delta_N+\frac{1}{2}t J_N(t)\right\}
\ .\nonumber
\eea

\item
3 nucleons:
\bea
H_{03}=\intd\frac{1}{[m^2-(P-k)^2]\;[m^2-(k-\Sigma)^2]\;[m-(P^\prime-k)^2]}\ .\nonumber
\eea

\item
2 mesons, 1 nucleon:
\bea
&&\{H_{21},H_{21}^\mu,H_{21}^{\mu\nu}\}=\intd\frac{\{1,k^\nu,k^\mu
k^\nu\}}{[M^2-k^2]\;[M^2-(k-\Delta)^2]\;[m-(P-k)^2]}\ ,\nonumber\\
&&H_{21}^\mu(t)=Q^\mu H_{21}^{(1)}(t)+\frac{1}{2}\Delta^\mu
H_{21}(t)\ ,\nonumber\\
&&H_{21}^{\mu\nu}(t)=g^{\mu\nu}H_{21}^{(2)}(t)+Q^\mu Q^\nu
H_{21}^{(3)}(t)+\Delta^\mu\Delta^\nu H_{21}^{(4)}(t)+(\Delta^\mu
Q^\nu+Q^\mu\Delta^\nu)\frac{1}{2}H_{21}^{(1)}(t)\ ,\nonumber
\eea
where
\bea
H_{21}^{(1)}(t)&=&\frac{1}{4m^2-t}\left\{J(t)-H(m^2)+\left(M^2-\frac{1}{2}t\right)H_{21}(t)\right\}\ ,\nonumber\\
H_{21}^{(2)}(t)&=&\frac{1}{(d-2)}\left\{{1\over2}M^2H_{21}(t)-\left(M^2-2m^2\right)H_{21}^{(1)}(t)-{1\over2}J(t)\right\}\ ,\nonumber\\
&=&\frac{1}{2}\left\{{1\over2}M^2H_{21}(t)-\left(M^2-2m^2\right)H_{21}^{(1)}(t)-{1\over2}J(t)\right\}-\frac{1}{64 \pi^2}\ ,\nonumber\\
H_{21}^{(3)}(t)&=&\frac{1}{4m^2-t}\left\{-\frac{1}{2}H^{(1)}(m^2)+\left(M^2-\frac{1}{2}t\right)H_{21}^{(1)}(t)-H_{21}^{(2)}(t)\right\}\ ,\nonumber\\
H_{21}^{(4)}(t)&=&\frac{1}{t}\left\{\frac{1}{2}H^{(1)}(m^2)-\frac{1}{2}H(m^2)+\frac{1}{4}t H_{21}(t)-H_{21}^{(2)}(t)\right\}\ .\nonumber
\eea

\item
1 mesons, 2 nucleon: $H_A$
\bea
&&\{H_{A},H_{A}^\mu,H_{A}^{\mu\nu}\}=\intd\frac{\{1,k^\nu,k^\mu
k^\nu\}}{[M^2-k^2]\;[m^2-(P-k)^2]\;[m-(P^\prime-k)^2]}\ .\nonumber\\
&&H_{A}^\mu(t)=Q^\mu H_{A}^{(1)}(t)\ ,\nonumber\\
&&H_{A}^{\mu\nu}(t)=g^{\mu\nu}H_{A}^{(2)}(t)+Q^\mu Q^\nu
H_{A}^{(3)}(t)+\Delta^\mu \Delta^\nu H_{A}^{(4)}(t)\ ,\nonumber
\eea
where
\bea
H_{A}^{(1)}(t)&=&\frac{1}{4m^2-t}\left\{H(m^2)-J_N(t)+M^2H_A\right\}\ ,\nonumber\\
H_{A}^{(2)}(t)&=&\frac{1}{d-2}\left\{M^2H_A(t)-M^2H_A^{(1)}(t)-{1\over2}J_N(t)\right\}\ ,\nonumber\\
&=&\frac{1}{2}\left\{M^2H_A(t)-M^2H_A^{(1)}(t)-{1\over2}J_N(t)-\frac{1}{32\pi^2}\right\}\ ,\nonumber\\
H_{A}^{(3)}(t)&=&\frac{1}{4m^2-t}\left\{{1\over2}H^{(1)}(m^2)-{1\over2}J_N(t)+M^2H_A^{(1)}(t)-H_A^{(2)}(t)\right\}\ ,\nonumber\\
H_{A}^{(4)}(t)&=&-\frac{1}{t}\left\{{1\over2}H^{(1)}(m^2)+H_A^{(2)}(t)\right\}\ .\nonumber
\eea

\item
1 mesons, 2 nucleon: $H_B$
\bea
&&\{H_{B},H_{B}^\mu,H_B^{\mu\nu}\}=\intd\frac{\{1,k^\mu,k^\mu k^\nu\}}{[M^2-k^2]\;[m^2-(P-k)^2]\;[m-(\Sigma-k)^2]}\ ,\nonumber\\
&&H_{B}^\mu(s)=(P+\Sigma)^\mu H_B^{(1)}(s)+(P-\Sigma)^\mu
H_B^{(2)}(s)\ ,\nonumber\\
&&H_B^{\mu\nu}(s)=g^{\mu\nu}H_B^{(3)}(s)+(P+\Sigma)^\mu(P+\Sigma)^\nu H_B^{(4)}(s)+(P-\Sigma)^\mu(P-\Sigma)^\nu H_B^{(5)}(s)+2(P^\mu P^\nu-\Sigma^\mu\Sigma^\nu) H_B^{(6)}(s)\ ,\nonumber
\eea
where
\bea
H_B^{(1)}(s)&=&-\frac{1}{2\lambda(s,m^2,M^2)}\left\{\left[M^2-(m^2-s)\right]H(s)+\left[M^2+(m^2-s)\right]H(m^2)
-2M^2J_N(M^2)\right.\nonumber\\
&&\left.+\left[(s-m^2+2M^2)M^2-(m^2-s)^2\right]H_B(s)\right\}\ ,\label{IB1}\\
H_B^{(2)}(s)
&=&-\frac{1}{2\lambda(s,m^2,M^2)}\left\{\left[m^2-M^2+3s\right]H(s)+\left[M^2-3m^2-s\right]H(m^2)
+2(m^2-s)J_N(M^2)\right.\nonumber\\
&&\left.+\left(m^2-s\right)\left(3m^2-3M^2+s\right)H_B(s)\right\}\ ,\label{IB2}\\
H_B^{(3)}(s)&=&-\frac{1}{2-d}\frac{1}{2\lambda(s,m^2,M^2)}\left\{M^2 \left(3 m^2-3 M^2+s\right) J_N(M^2)+\left[\left(s-m^2\right)\left(2 m^2-M^2\right)+M^4 \right] H(m^2)\right. \nonumber\\
&&\left.+\left[m^4-s^2+M^4+2 M^2 \left(s-m^2\right)\right] H(s)+2 \left[m^6+2 M^6-2 m^4 s-M^4 s+m^2 \left(s^2-3 M^4\right)\right] H_B(s)\right\}\ ,\nonumber\\
H_B^{(4)}(s)&=&\frac{1}{4\lambda(s,m^2,M^2)}\left\{(s-m^2)\left(H(m^2)-H(s)\right)+(s-m^2-M^2)H^{(1)}(m^2)-(s-m^2+M^2)H^{(1)}(s)\right.\nonumber\\
&&\left.+4M^2J_N(M^2)+\left((s-m^2)^2-4M^4\right)H_B(s)-4(1-d)M^2H_B^{(3)}(s)\right\}\ ,\nonumber\\
H_B^{(5)}(s)&=&\frac{1}{4\lambda(s,m^2,M^2)}\left\{2\left(s+3m^2-3M^2\right)J_N(M^2)+(s-m^2+2M^2)\left(H(m^2)+H(s)\right)+(s+3m^2-M^2)H^{(1)}(m^2)\right.\nonumber\\
&&\left.+(3s+m^2-M^2)H^{(1)}(s)+\left((s-m^2)^2-4M^2(s+3m^2-2M^2)\right)I_B(s)-4(1-d)(2s+2m^2-M^2)H_B^{(3)}(s)\right\}\nonumber\\
H_B^{(6)}(s)&=&\frac{1}{4\lambda(s,m^2,M^2)}\left\{(s-m^2)\left(2J_N(M^2)+H(m^2)+H(s)\right)+(s+3m^2-M^2)H^{(1)}(m^2)\right.\nonumber\\
&&\left.-(3s+m^2-M^2)H^{(1)}(s)+(s-m^2)(s-m^2-2M^2)H_B(s)-4(1-d)(s-m^2)H_B^{(3)}(s)\right\}\ .\nonumber
\eea

\item
1 mesons, 3 nucleon:
\bea
&&\{H_{13},H_{13}^\mu\}=\intd\frac{\{1,k^\mu\}}{[M^2-k^2]\;[m^2-(P-k)^2]\;[m-(\Sigma-k)^2]\;[m^2-(P^\prime-k)^2]}\ ,\nonumber\\
&&H_{13}^\mu(s,t)=Q^\mu H_{13}^{(1)}(s,t)+(\Delta+2q)^\mu
H_{13}^{(2)}(s,t)\ ,\nonumber
\eea
where
\bea
H_{13}^{(1)}(s,t)&=&-\frac{1}{\lambda(s,m^2,M^2)+s t}\left\{(s-m^2+M^2)H_B(s)-\frac{1}{2}(s-u)H_A(t)\right.\nonumber\\
&&\left.-\left[\frac{1}{2}(s+u)-(m^2-M^2)\right]H_{03}(t)+(s-m^2+M^2)\left[M^2-\frac{1}{2}(s-u)\right]H_{13}(s,t)\right\}\ ,\nonumber\\
H_{13}^{(2)}(s,t)&=&\frac{1}{s-u}\left\{H_B(s)-H_{03}(t)+M^2H_{13}(s,t)-(4m^2-t)H_{13}^{(1)}(s,t)\right\}\ .\nonumber
\eea
\end{itemize}
After removing parts proportional to $R=\frac{2}{d-4}+\gamma_E-1-\ln4\pi$, the remaining scalar integrals are finite and denoted by, e.g. $\bar{H}(s)$, $\bar{J}_N(t)$, $\bar{H}_A(t)$, etc..

\subsection{$O(p^3)$ results}
\label{loopamplp3}
The contributions from the $O(p^3)$ loop graphs shown in Fig.~\ref{loopgraph} are displayed below, respectively. The total $O(p^3)$ loop contributions are given by
\bea
A_{total}^\pm&=&\sum_G\big[A_G^\pm(s,t)\pm A_G^\pm(u,t)\big]+\sum_HA_H^\pm(s,t)\ ,\qquad G\in\left\{a,\cdots,i,n,\cdots,s\right\}\ ,\nnb\\
B_{total}^\pm&=&\sum_G\big[B_G^\pm(s,t)\mp B_G^\pm(u,t)\big]+\sum_HB_H^\pm(s,t)\ ,\qquad H\in\left\{k,l,m,t,u,v\right\}\ ,
\eea
where the $A_G^\pm(u,t)$ and $B_G^\pm(u,t)$ ($G\in\left\{a,\cdots,i,\cdots,s\right\}$) are obtained from the graphs $(a),\cdots,(i),(n)\cdots,(s)$ through crossing.

The abbreviation $F(s)$ in the amplitudes is defined as
\bea
F(s)\equiv2({\Delta_N}-M^2 H(s))+(s-m^2)H^{(1)}(s)\ .\nonumber
\eea
\begin{itemize}
\item
Graphs (a)+(b):
\bea
A_{ab}^\pm&=&\frac{mg^2}{2F^4}F(s)\ ,\nonumber\\
B_{ab}^\pm&=&-\frac{g^2}{2F^4}\left\{\frac{2m^2F(s)}{s-m^2}-\left(\Delta_N-M^2
H(s)\right)+F(s)\right\}\ .\nonumber
\eea

\item
Graphs (c)+(d):
\bea
A_{cd}^\pm&=&\frac{mg^4}{8F^4}\left\{-2\Delta_\pi+\left(s-m^2\right)H^{(1)}(s)-8m^2\left[-\left({J_N(M^2)}-M^2
H_B(s)\right)+\left(s-m^2\right)H_B^{(2)}(s)\right] \right\}\ ,\nonumber\\
B_{cd}^\pm&=&\frac{g^4}{8F^4}\left\{\left({\Delta_N}-M^2H(s)\right)+\left(
s-m^2\right)\left(H^{(1)}(s)-4m^2H_B^{(1)}(s)\right)\right.\nonumber\\
&&\quad\quad
\left.+\frac{4m^2}{s-m^2}\left[\Delta_\pi+\left(s+3m^2\right)\left(-\left({J_N(M^2)}-M^2H_B(s)\right)+\left(s-m^2\right)H_B^{(2)}(s)\right)\right]\right\}\nonumber\ .
\eea

\item
Graph (e):\bea
A^\pm_{e}&=&\frac{3g^4m}{16F^4}\left\{\frac{4m^2F(s)}{m^2-s}-3F(s)+2\left({\Delta_N}-M^2H(s)\right)\right\}\ ,\nonumber\\
B^\pm_e&=&\frac{3g^4}{16F^4}\left\{F(s)-\left({\Delta_N}-M^2H(s)\right)-\frac{4m^2}{m^2-s}\left[2F(s)-\left({\Delta_N}-M^2H(s)\right)\right]+\frac{8m^4}{(s-m^2)^2}F(s)\right\}\ .\nonumber
\eea

\item
Graph (f):
\bea
A_{f}^+&=&\frac{1}{2F^4}m\left(s-m^2\right)H^{(1)}(s)\ ,\nonumber\\
B_{f}^+&=&\frac{1}{8F^4}\left\{4\left(s-m^2\right)H^{(1)}(s)+4\left({\Delta_N}-M^2H(s)\right)-\Delta_\pi\right\}\ ,\nonumber\\
A_{f}^-&=&\frac{1}{2}A_{3f}^+\ ,\quad
B_{f}^-=\frac{1}{2}B_{3f}^+\ .\nonumber
\eea

\item
Graphs (g)+(h):
\bea
A_{gh}^+&=&\frac{mg^2}{2F^4}\left(s-m^2\right)\left\{-2H(s)+H^{(1)}(s)+8m^2H_B^{(1)}(s)\right\}\ ,\nonumber\\
B_{gh}^+&=&\frac{g^4}{4F^4}\left\{\left({\Delta_N}-M^2H(m^2)\right)-2\Delta_\pi+8m^2\left[{J_N(M^2)}-M^2H_B(s)\right]\right.\ ,\nonumber\\
&&\quad\quad\left.+2\left(m^2-s\right)\left[H(s)-H^{(1)}(s)-4m^2\left(H_B^{(1)}(s)-H_B^{(2)}(s)\right)\right]\right\}\ ,\nonumber\\
A_{gh}^-&=&0\ ,
\quad B_{gh}^-=0\ .\nonumber
\eea

\item
Graph (i):
\bea
A_i^+&=&\frac{3mg^4}{16F^4}\left\{2M^2\left[H(m^2)-H(s)\right]+\left(s-m^2\right)\left[2H(s)+H^{(1)}(s)\right]\right.\nonumber\\
&&\qquad\quad+8m^2\left[\left({J_N(t)}-M^2H_A(t)\right)+4m^2H_A^{(1)}(t)-(s-u)H_A^{(3)}(t)\right.\nonumber\\
&&\qquad\quad\left.-\left({J_N(M^2)}-M^2H_B(s)\right)-M^2\left(H_B^{(1)}(s)-H_B^{(2)}(s)\right)-\left(m^2+3s\right)H_B^{(1)}(s)+\left(s-m^2\right)H_B^{(2)}(s)\right]\nonumber\\
&&\qquad\quad\left.+32m^4(s-m^2)H_{13}^{(1)}(s,t)\right\}\ ,\nonumber\\
B_i^+&=&\frac{3g^4}{16F^4}\left\{\left(3m^2+s\right)H(s)+4m^2H^{(1)}(m^2)-\left(m^2+s\right)H^{(1)}(s)\right.\nonumber\\
&&\qquad+4m^2\left[-\left({J_N(t)}-M^2H_A(t)\right)-2H_A^{(2)}(t)-2\left({J_N(M^2)}-M^2H_B(s)\right)-2\left(3m^2+s\right)H_B^{(1)}(s)\right.\nonumber\\
&&\qquad\left.\left.-2\left(m^2-s\right)H_B^{(2)}(s)\right]+16m^4\left[\left({H_{03}(t)}-M^2H_{13}(s,t)\right)+2\left(s-m^2\right)H_{13}^{(2)}(s,t)\right]\right\}\ ,\nonumber\\
A_i^-&=&-\frac{1}{3}A_i^+\ ,\quad B_i^-=-\frac{1}{3}B_i^+\ .\nonumber
\eea

\item
Graph (k):
\bea
A_k^{\pm}=B_k^+=0\ ,\quad
B_k^-=\frac{t}{F^4}J^{(1)}(t)\ .\nonumber
\eea

\item
Graph (l):
\bea
A_l^+&=&\frac{mg^2}{6F^4}\left\{4\left[{\Delta_N}-M^2H(m^2)\right]-3\left(M^2-2t\right)\left[J(t)-4m^2H_{21}^{(1)}(t)\right]\right\}\ ,\nonumber\\
B_l^+&=&0\ ,\nonumber\\
A_l^-&=&-\frac{4m^3g^2}{F^4}\left(s-u\right)H_{21}^{(3)}(t)\ ,\nonumber\\
B_l^-&=&-\frac{g^2}{F^4}\left\{tJ^{(1)}(t)+4m^2H_{21}^{(2)}(t)\right\}\ .\nonumber
\eea

\item
Graph (m):
\bea A_m^+&=&B_m^+=0\ ,\nonumber\\
A_m^-&=&-\frac{g^2m^3}{F^4}\left(s-u\right)H_A^{(3)}(t)\ ,\nonumber\\
B_m^-&=&-\frac{g^2}{8F^4}\left\{\Delta_\pi-4m^2\left[H^{(1)}(m^2)-\left({J_N(t)}-M^2H_A(t)\right)-2H_A^{(2)}(t)\right]\right\}\ .\nonumber
\eea

\item
Graphs (n)+(o):
\bea
A_{no}^\pm&=&\frac{mg^2}{F^4}\left\{{\Delta_N}-M^2H(m^2)\right\}\ ,\nonumber\\
B_{no}^\pm&=&A_{no}^\pm\left\{\frac{2m}{m^2-s}-\frac{1}{4m}\right\}\ .\nonumber
\eea

\item
Graphs (p)+(r):
\bea
A_{pr}^\pm&=&-\frac{mg^2}{3F^4}\Delta_\pi\ ,\nonumber\\
B_{pr}^\pm&=&A_{pr}^\pm\left\{\frac{2m}{m^2-s}-\frac{1}{2m}\right\}\ .\nonumber
\eea

\item
Graph (s):
\bea
A_s^\pm=B_s^\pm=0\ .\nonumber
\eea

\item
Graphs (t)+(u):
\bea
A_{tu}^+&=&-\frac{2mg^2}{3F^4}\left\{{\Delta_N}-M^2H(m^2)\right\}\ ,\nonumber\\
B_{tu}^-&=&-\frac{3}{4m}A_{tu}^+\ ,\nonumber\\
A_{tu}^-&=&0\ ,\quad B_{tu}^+=0\ .\nonumber
\eea

\item
Graph (v):
\bea
A_v^\pm&=&B_v^+=0\ ,\nonumber\\
B_v^-&=&-\frac{5}{24F^4}\Delta_\pi\ .\nonumber
\eea
\end{itemize}

\subsection{$O(p^4)$ results}
\label{loopamplp4}
The contributions from the $O(p^4)$ loop graphs shown in Fig.~\ref{loopgraphp4} are displayed below. The total $O(p^4)$ loop contributions are given by
\bea
A_{total}^\pm&=&\sum_G\big[A_G^\pm(s,t)\pm A_G^\pm(u,t)\big]+\sum_HA_H^\pm(s,t)\ ,\qquad G\in\left\{a,b,f1,f2,g,h,n,o,s\right\}\ ,\nnb\\
B_{total}^\pm&=&\sum_G\big[B_G^\pm(s,t)\mp B_G^\pm(u,t)\big]+\sum_HB_H^\pm(s,t)\ ,\qquad H\in\left\{k,m,v\right\}\ ,
\eea
where the $A_G^\pm(u,t)$ and $B_G^\pm(u,t)$ ($G\in\left\{a,b,f1,f2,g,h,n,o,s\right\}$) are obtained from the graphs $(a),(b),(f1),(f2),(g),(h),(n),(o),(s)$ through crossing.
\begin{itemize}
\item Graphs (a)+(b):
\bea
A_{ab}^\pm&=&\frac{2c_1g^2M^2}{F^4}\left\{(s-m^2)H^{(1)}(s)+\Delta_N-M^2H(s)\right\}\nonumber\\
&&-\frac{c_2 g^2}{8 F^4 m^2}\left\{2 (s+3 m^2)\Delta_N^{(2)}-M^2(s-m^2+M^2)\Delta_\pi-2 (M^2-4 m^2)\Delta_\pi^{(2)}\right.\nonumber\\
&&\hspace{1cm}+2 (s-m^2-M^2)\left[\frac{1}{2}(s-m^2+M^2)\left(\Delta_N-M^2 H(s)+(s-m^2) H^{(1)}(s)\right)\right.\nonumber\\
&&\hspace{1cm}\left.\left.-(m^2-M^2)\widetilde{H}^{(1)}(s)
+(s+7 m^2-2M^2)H^{(2)}(s)+(m^2-M^2)(s-m^2)H^{(3)}(s)
\right]
\right\}\nonumber\\
&&-\frac{(c_3+c_4) g^2}{F^4}\left\{-\frac{1}{2} (s-m^2+M^2)\widetilde{H}^{(1)}(s)+(M^2-4 m^2)H^{(2)}(s)+\frac{1}{2}(s-m^2)(s-m^2+M^2)H^{(3)}(s)\right\}\nonumber\\
&&+\frac{c_4g^2}{F^4}\left\{(s-m^2)(s+m^2)H^{(1)}(s)+(s+3 m^2) \left(\Delta_N-M^2 H(s)\right)\right\}\nonumber\\
B_{ab}^\pm&=&-\frac{2c_1 g^2 m M^2 }{F^4(s-m^2)}\left\{2 \left(\Delta_N-M^2 H(s)\right)+(s-m^2)H^{(1)}(s)\right\}\nonumber\\
&&-\frac{c_2 g^2}{2 F^4 m(s-m^2)}\left\{-2(s+m^2)\Delta_N^{(2)}+\frac{1}{2}M^2 (s-m^2+M^2)\Delta_\pi-(s+3 m^2-M^2)\Delta_\pi^{(2)}\right.\nonumber\\
&&\hspace{2cm}+\frac{1}{2}(m^2-s+M^2)\left[\frac{1}{2}(s-m^2+M^2)\left(2(\Delta_N-M^2 H(s))+(s-m^2)H^{(1)}(s)\right)\right.\nonumber\\
&&\hspace{2cm}\left.\left.-2(m^2-M^2)\widetilde{H}^{(1)}(s)+(5s+11 m^2-4 M^2) H^{(2)}(s)+(m^2-M^2)(s-m^2)H^{(3)}(s)\right]\right\}\nonumber\\
&&-\frac{(c_3+c_4)g^2m}{F^4(s-m^2)}\left\{(s-m^2+M^2)\widetilde{H}^{(1)}(s)+2(s+3 m^2-M^2) H^{(2)}(s)-\frac{1}{2}(s-m^2)(s-m^2+M^2)H^{(3)}(s)\right\}\nonumber\\
&&-\frac{c_4g^2 m}{F^4(s-m^2)}\left\{4(s+m^2)(\Delta_N-M^2 H(s))+(s-m^2)(3s+m^2)H^{(1)}(s)\right\}
\eea

\item Graphs (f1)+(f2):
\bea
A_{f}^+&=&\frac{c_4}{F^4}\left\{(m^2-s) \left[\Delta_N+\Delta_\pi+(s-m^2-M^2) H(s)\right]+2 m^2(s-m^2)H^{(1)}(s)+(m^2-s+M^2)H^{(2)}(s)\right.\nonumber\\
&&\hspace{1cm}\left.+\frac{1}{2}(s-m^2+M^2)\left[\Delta_N+(s-m^2-M^2)H^{(1)}(s)\right]+m^2(s-m^2+M^2)H^{(3)}(s)
\right\}\nonumber\\
A_{f}^-&=&\frac{1}{2}A_{f}^+-\frac{2c_1M^2}{F^4}\left\{\Delta_N+(s-m^2-M^2)H(s)+2m^2H^{(1)}(s)\right\}\nonumber\\
&&+\frac{c_2}{2F^4m^2}\left\{(s-m^2)\Delta_N^{(2)}+\frac{1}{2}(s-m^2-M^2) \left[\frac{1}{2}(s-m^2+M^2) \Delta_\pi+\Delta_\pi^{(2)}\right]\right.\nonumber\\
&&\hspace{0.cm}+\frac{1}{4}(s-m^2-M^2)(s-m^2+M^2)\left[\Delta_N+(s-m^2-M^2)H(s)+2 m^2 H^{(1)}(s)\right]\nonumber\\
&&\hspace{0.cm}+\frac{1}{2}(s-m^2-M^2)
\left[(m^2-M^2)\left(\Delta_N+(s-m^2-M^2)H^{(1)}(s)\right)+2 (s-M^2) H^{(2)}(s)\left.+2 m^2 (m^2-M^2)H^{(3)}(s)\right]\right\}\nonumber\\
&&+\frac{c_3}{F^4}\left\{\frac{1}{2}(s-m^2+M^2) \left[\Delta_N+(s-m^2-M^2) H^{(1)}(s)\right]-(s-m^2-M^2)H^{(2)}(s)+m^2(s-m^2+M^2)H^{(3)}(s)\right\}\nonumber\\
B_{f}^+&=&\frac{c_4}{F^4}\left\{2 m \left[\Delta_N+\Delta_\pi+(s-m^2-M^2) H(s)\right]-2 m(s+m^2)H^{(1)}(s)+4 m H^{(2)}(s)+m(s-m^2+M^2)H^{(3)}(s)\right\}\nonumber\\
B_{f}^-&=&\frac{1}{2}B_{f}^+-\frac{4c_1 m M^2}{F^4}H^{(1)}(s)+\frac{c_2}{2F^4 m^2}\left\{-2 m (\Delta_N^{(2)}+\Delta_\pi^{(2)})+\frac{1}{2} m (s-m^2-M^2)(s-m^2+M^2)H^{(1)}(s)\right.\nonumber\\
&&\hspace{0.cm}\left.+\frac{1}{2}(s-m^2-M^2)\left[-6mH^{(2)}(s)+2 m (m^2-M^2)H^{(3)}(s)\right]\right\}+\frac{2c_3 m }{F^4}\left\{2H^{(2)}(s)+\frac{1}{2}(s-m^2+M^2) H^{(3)}(s)\right\}
\eea

\item Graphs (g)+(h):
\bea
\hspace{-1.5cm}A_{gh}^+&=&-\frac{2c_1g^2M^2}{F^4}\left\{(s-m^2)H(s)+2m^2 H^{(1)}(m^2)-(s+m^2) H^{(1)}(s)+4 m^2M^2 (H_B^{(2)}(s)-H_B^{(1)}(s))\right\}\nonumber\\
&&+\frac{c_2g^2}{2F^4}\left\{\frac{s-m^2}{2m^2} (\Delta_N^{(2)}+ \Delta_\pi^{(2)})+\frac{s-m^2}{4m^2} (s-m^2+M^2)\Delta_\pi+(2M^2-t) \Delta_N\right.\nonumber\\
&&\hspace{1.5cm}-\frac{1}{2}M^2(2M^2-t)J_N(M^2)+M^2(2(s-m^2-M^2)+t)J_N^{(1)}(M^2)+M^2(2M^2-t) J_N^{(2)}(M^2)\nonumber\\
&&\hspace{1.5cm}+2(2M^2-t)H^{(2)}(m^2)+\left[M^2 (m^2+M^2-u)+\frac{1}{2}(s-m^2-M^2)(2 m^2-t) \right]H^{(3)}(m^2)\nonumber\\
&&\hspace{1.5cm}+\frac{1}{2} (s-m^2+M^2)\left[-2M^2H^{(1)}(m^2)+(s+m^2-M^2) H^{(3)}(m^2)\right]-(2M^2+s-u) H^{(4)}(m^2)\nonumber\\
&&\hspace{1.5cm}- \frac{1}{2}(m^2-M^2+s)(m^2+M^2-u)H^{(5)}(m^2)+\frac{(s-m^2-M^2)}{4 m^2}\left[(s-m^2)(s-m^2+M^2)H(s)\right.\nonumber\\
&&\hspace{1.5cm}\left.+2(s-3m^2) H^{(2)}(s)-\left((s-m^2)^2+M^2(3s-m^2)\right)H^{(1)}(s)-2(m^2-M^2)(s+m^2)H^{(3)}(s)\right]\nonumber\\
&&\hspace{1.5cm}-(s-m^2-M^2)\left[M^2(s-m^2+M^2) (H_B^{(1)}(s)-H_B^{(2)}(s))-2(2u-2m^2-M^2)H_B^{(3)}(s)\right.\nonumber\\
&&\hspace{1.5cm}\left.\left.-M^2(s-5m^2+M^2+2t)(H_B^{(4)}(s)-H_B^{(6)}(s))+M^2(s-m^2-3 M^2+2 t)(H_B^{(5)}(s)-H_B^{(6)}(s))\right]\right\}\nonumber\\
&&+\frac{(c_3-c_4)g^2}{F^4} \left\{(s-m^2) \left(\frac{1}{2}(s-m^2+M^2) H^{(1)}(s)-H^{(2)}(s)\right)+m^2(m^2+M^2-u)H^{(3)}(m^2)\right.\nonumber\\
&&\hspace{1.5cm}-\frac{1}{2}(s+m^2)(s-m^2+M^2)H^{(3)}(s)-2 m^2\left[2(m^2-s+2 M^2-t) H_B^{(3)}(s)\right.\nonumber\\
&&\hspace{1.5cm}\left.\left.+M^2 (2M^2+s-u) (H_B^{(4)}(s)-H_B^{(6)}(s))-M^2(2m^2-s-u) (H_B^{(5)}(s)-H_B^{(6)}(s))\right]\right\}\nonumber\\
&&+\frac{c_4g^2}{F^4}(s-m^2)\left\{\Delta_\pi-(s-m^2)(H^{(1)}(s)-H(s))-4 m^2 \left(J_N(M^2)-M^2H_B(s)\right)-4 m^2(s-m^2) (H_B^{(1)}(s)-H_B^{(2)}(s))\right\},\nonumber\\
\hspace{-1.5cm}A_{gh}^-&=&-A_{gh}^+\;(c_4=0).\nonumber\\
\hspace{-2.5cm}B_{gh}^+&=&-\frac{4c_1g^2mM^2}{F^4} \left\{J_N(M^2)-M^2 H_B(s)-\frac{1}{2}H^{(1)}(s)+(s-m^2-M^2) (H_B^{(1)}(s)-H_B^{(2)}(s))\right\}\nonumber\\
&&+\frac{c_2g^2}{2F^4 m} \left\{-(t-2M^2)\Delta_N-2\Delta_N^{(2)}+\frac{1}{4} (2m^2+M^2)(t-2M^2)J_N(M^2)-4 m^2 M^2 J_N^{(1)}(M^2)\right.\nonumber\\
&&\hspace{1.5cm}-M^2(t-2M^2)J_N^{(2)}(M^2)-\frac{1}{2}M^2 (s-m^2+M^2) H(m^2)+\frac{1}{2}M^2(m^2+M^2-u)H^{(1)}(m^2)\nonumber\\
&&\hspace{1.5cm}+(4 m^2+4M^2-t)H^{(2)}(m^2)-(2m^2+s-u)H^{(4)}(m^2)+\frac{(s-m^2-M^2)}{2}\left[-H^{(2)}(s)\right.\nonumber\\
&&\hspace{1.5cm}-\frac{1}{2}(s-m^2+M^2)H^{(1)}(s)-(m^2-M^2)H^{(3)}(s)+\frac{1}{2}(s+3m^2+M^2-2t) J_N(M^2)\nonumber\\
&&\hspace{1.5cm}-M^2(s-m^2+M^2)H_B(s)-M^2(u-t)(H^{(1)}_B(s)+H^{(2)}_B(s))- M^2(m^2-M^2)(3H^{(1)}_B(s)-H^{(2)}_B(s))\nonumber\\
&&\hspace{1.5cm}+((s-m^2)^2-M^4)(H^{(1)}_B(s)-H^{(2)}_B(s))+ 2(9 m^2-s+M^2-2 t)H^{(3)}_B(s)\nonumber\\
&&\hspace{1.5cm}+(s-m^2-M^2)(5 m^2-s-M^2-2 t) \left(H^{(4)}_B(s)-H^{(6)}_B(s)\right)\nonumber\\
&&\hspace{1.5cm}\left.\left.-(s-m^2-M^2)(m^2-s+3M^2-2t) \left(H^{(5)}_B(s)-H^{(6)}_B(s)\right)\right]\right\}\nonumber\\
&&+\frac{(c_3-c_4) m g^2 }{F^4}\left\{2 H^{(2)}(m^2)-\frac{1}{2}(s-m^2+M^2) H^{(3)}(s)+\frac{1}{2}(2M^2+s-u) J_N(M^2)\right.\nonumber\\
&&\hspace{1.5cm}-M^2 \left[(2M^2+s-u)H_B^{(1)}(s)+(2m^2-s-u)H_B^{(2)}(s)\right]- 2(5 m^2-s+M^2-t) H_B^{(3)}(s)\nonumber\\
&&\hspace{1.5cm}+(s-m^2-M^2)( 2M^2+s-u) \left(H_B^{(4)}(s)-H_B^{(6)}(s)\right)\left.-(s-m^2-M^2)(2m^2-s-u)\left(H_B^{(5)}(s)-H_B^{(6)}(s)\right)\right\}\nonumber\\
&&+\frac{c_4m g^2 }{F^4}\left\{-2 \Delta_\pi+(s-m^2) \left(H^{(1)}(s)-2H(s)\right)+2 (s+3 m^2) \left(J_N(M^2)-M^2 H_B(s)\right)\right.\nonumber\\
&&\hspace{1.5cm}\left.+2 (s-m^2)(s+3 m^2-M^2) \left(H_B^{(1)}(s)-H_B^{(2)}(s)\right)\right\},\nonumber\\
\hspace{-1.5cm}B_{gh}^-&=&-B_{gh}^+\;(c_4=0).
\eea

\item Graph (k):
\bea
A_{k}^+&=&-\frac{1}{6F^4}\left\{4 c_1 M^2 \left[4 \Delta_\pi+3 (-M^2+2 t) J(t)\right]+2{c_3} \left[-4 M^2 \Delta_\pi+3 (2t-M^2) \left(\Delta_\pi+\left(\frac{t}{2}-M^2\right) J(t)\right)\right]\right.\nonumber\\
&&\hspace{0.5cm}\left.+\frac{c_2}{m^2}\left[-8 m^2 \Delta_\pi^{(2)}+3(2t-M^2) \left(\frac{1}{4} t^2 J(t)+2 \left(m^2-\frac{t}{4}\right) t J^{(1)}(t)-\frac{1}{2} t^2 J^{(2)}(t)\right)\right]\right\},\nonumber\\
A_{k}^-&=&\frac{2 {c_4} }{F^4}\left(m^2+M^2-s-\frac{t}{2}\right) t J^{(1)}(t),\nonumber\\
B_{k}^+&=&0\nonumber\\
B_{k}^-&=&\frac{4{c_4} m t J^{(1)}(t)}{F^4}
\eea

\item Graph (m):
\bea
A_{m}^+&=&\frac{3{c_1}g^2M^2}{F^4} \left\{\Delta_\pi-4 m^2 H^{(1)}(m^2)-4 m^2 \left(J_N(t)-M^2 H_A(t)\right)\right\}\nonumber\\
&&-\frac{3{c_2} g^2}{4 F^4 } \left\{\frac{2M^2-t}{m^2}\Delta_\pi^{(2)}+(s-u)^2 H^{(3)}(m^2)-4(2M^2-t)H^{(4)}(m^2)-2M^2(s-u)^2H_A^{(1)}(t)\right.\nonumber\\
&&\hspace{1.0cm}+2(s-m^2-M^2)(m^2+M^2-u)\left[\frac{1}{4m^2}\left(\Delta_\pi-4m^2 H^{(1)}(m^2)\right)-H^{(5)}(m^2)+M^2H_A(t)\right]\nonumber\\
&&\hspace{1.0cm}\left.+2t(4M^2 -t)J_N^{(1)}(t)+2 t^2J_N^{(2)}(t)+M^2 \left[4 (2M^2-t)H_A^{(2)}(t)+2(s-u)^2H_A^{(3)}(t)-2 t^2 H_A^{(4)}(t)\right]\right\}\nonumber\\
&&-\frac{3c_3g^2}{4F^4}(2M^2-t) \left\{\Delta_\pi-4 m^2 H^{(1)}(m^2)-4 m^2 \left(J_N(t)-M^2 H_A(t)\right)\right\}\nonumber\\
A_{m}^-&=&-\frac{c_4 g^2 m^2 t (s-u)H_A^{(3)}(t)}{F^4}-\frac{s-u}{4m}B_{m}^-.\nonumber\\
B_{m}^+&=&-\frac{6c_2 g^2 }{F^4 m}\left\{\frac{1}{2}(s-m^2-M^2)+\frac{1}{2} (m^2-u+M^2)\right\} \left(H^{(2)}(m^2)-H^{(4)}(m^2)\right)\nonumber\\
B_{m}^-&=&\frac{c_4m g^2}{2F^4}\left\{\Delta_\pi+4 m^2 \left[-\frac{1}{2}J_N(t)-H^{(1)}(m^2)+M^2 H_A(t)-M^2 H_A^{(1)}(t)-2 H_A^{(2)}(t)\right.\right.\nonumber\\
&&\hspace{2.0cm}\left.\left.+(4 m^2-t)H_A^{(3)}(t)-tH_A^{(4)}(t)\right]\right\}
\eea

\item Graphs (n)+(o)
\bea
A_{no}^\pm&=&\frac{2g^2}{F^4}\left\{(c_1M^2+2c_4m^2)\left[\Delta_N-M^2H(m^2)\right]+(c_3+c_4)m^2\left[2H^{(2)}(m^2)-\frac{1}{2}(s-m^2-M^2) H^{(3)}(m^2)\right]\nonumber\right.\\
&&\hspace{0cm}+\frac{c_2 }{16m^2}\left[ \left(s^2-(m^2-M^2)^2\right) \left(\Delta_N-M^2 H^{(1)}(m^2)\right)-8m^2\Delta_N^{(2)}\right.-M^2 (s-m^2-M^2) \left(\Delta_\pi-M^2\left(H(m^2)-H^{(1)}(m^2)\right)\right)\nonumber\\
&&\hspace{1cm}\left.\left.-2(s+3 m^2-M^2)\left(\Delta_\pi^{(2)}-(s-m^2+2M^2) H^{(2)}(m^2)\right)\right]\right\}\nonumber\\
B_{no}^\pm&=&-\frac{2mg^2}{(s-m^2)F^4}\,\left\{(2c_1M^2+c_4(s+3m^2))\left[\Delta_N-M^2 H(m^2)\right]\right.\nonumber\\
&&\hspace{0cm}-\frac{c_2}{8 m^2}\left[2(s+3m^2) \Delta_N^{(2)}-\left(s^2-(m^2-M^2)^2\right)\left(\Delta_N-M^2 H^{(1)}(m^2)\right)\right.+M^2 (s-m^2-M^2) \left(\Delta_\pi-M^2\left(H(m^2)-H^{(1)}(m^2)\right)\right)\nonumber\\
&&\hspace{0cm}\left.+2(2s+2 m^2-M^2) \left(\Delta_\pi^{(2)}-(s-m^2+2M^2)H^{(2)}(m^2)\right)\right]\nonumber\\
&&\hspace{0cm}\left.+(c_3+c_4)\left[(s+3m^2)H^{(2)}(m^2)-m^2(s-m^2-M^2) H^{(3)}(m^2)\right]\right\}
\eea

\item Graph (s):
\bea
A_{s}^\pm&=&\frac{3 g^2(s+3 m^2)}{4 F^4(s-m^2)}\left\{2c_1M^2 \Delta_\pi-c_2\frac{s}{m^2}\Delta_\pi^{(2)}-c_3 M^2\Delta_\pi\right\},\nonumber\\
B_{s}^\pm&=&-\frac{3g^2m(s+m^2)}{F^4(s-m^2)^2} \left\{2c_1 M^2 \Delta_\pi-c_2\frac{s }{m^2}\Delta_\pi^{(2)}-c_3 M^2\Delta_\pi\right\}.
\eea

\item Graph (v)
\bea
A_{v}^+&=&\frac{2}{3f^4}\left\{5c_1M^2\Delta_\pi-2 c_2 \left[\frac{1}{4 m^2}(s-m^2-M^2)(m^2-u+M^2) \Delta_\pi+\Delta_\pi^{(2)}\right]-c_3(4M^2-t) \Delta_\pi\right\},\nonumber\\
A_{v}^-&=&\frac{c_4 (s-u) \Delta_\pi}{3 F^4},\nonumber\\
B_{v}^+&=&0,\hspace{1cm}B_{v}^-=-\frac{4 c_4 m \Delta_\pi}{3 F^4}.
\eea

\end{itemize}

\section{Regular parts of scalar one-loop integrals}
\label{regul}
Following the method proposed by Refs.~\cite{becherleutwyler,Schindler03}, we have derived the regular parts of scalar integrals to the order needed by the $O(p^4)$ calculation. All the results are listed except for the scalar integrals whose regular parts are equal to zero.
\begin{itemize}
  \item 1 nucleon:
  \bea
  R_{01}=\frac{m^2}{16\pi^2}\left\{R+\ln\frac{m^2}{\mu^2}\right\}.
  \eea
  \item 1 meson, 1 nucleon:
        \bea
     R_{11}&=&R_{11}^{(0)}+R_{11}^{(1)}+R_{11}^{(2)}+R_{11}^{(3)}+{\cal O}(p^4)+\cdots,
     \eea
     where
     \bea
     R_{11}^{(0)}&=&-\frac{1}{16\pi^2}\left\{R-1+\ln\frac{m^2}{\mu^2}\right\},\nonumber\\
     R_{11}^{(1)}&=&\frac{s-m^2}{2m^2}\frac{1}{16\pi^2}\left\{R-1+\ln\frac{m^2}{\mu^2}\right\},\nonumber\\
     R_{11}^{(2)}&=&\frac{1}{2}\left\{\frac{M_\pi^2}{m^2}\frac{1}{16\pi^2}\left[R+3+\ln\frac{m^2}{\mu^2}\right]-\left(\frac{s-m^2}{m^2}\right)^2\frac{1}{16\pi^2}\left[R+\ln\frac{m^2}{\mu^2}\right]\right\}.\nonumber\\
     R_{11}^{(3)}&=&-\frac{1}{16\pi^2}\left\{\frac{(s-m^2)M_\pi^2}{2m^4}\left[R+3+\ln\frac{m^2}{\mu^2}\right]+\frac{(s-m^2)^3}{2m^6}\left[R+\frac{1}{2}+\ln\frac{m^2}{\mu^2}\right]\right\}.\nonumber
     \eea
  \item 2 nucleons:
        \bea
     R_{02}&=&R_{02}^{(0)}+R_{02}^{(2)}+{\cal O}(p^4)+\cdots,
     \eea
     where
     \bea
     R_{02}^{(0)}&=&-\frac{1}{16\pi^2}\left\{R+1+\ln\frac{m^2}{\mu^2}\right\},\nonumber\\
     R_{02}^{(2)}&=&\frac{t}{96\pi^2m^2}.\nonumber
     \eea
  \item 2 mesons, 1 nucleon:
        \bea
     R_{21}&=&R_{21}^{(0)}+R_{21}^{(2)}+{\cal O}(p^4)+\cdots,
     \eea
     where
     \bea
     R_{21}^{(0)}&=&-\frac{1}{32\pi^2m^2}\left\{R+3+\ln\frac{m^2}{\mu^2}\right\},\nonumber\\
     R_{21}^{(2)}&=&\frac{1}{32\pi^2m^2}\left\{\frac{2M_\pi^2}{3m^2}-\frac{t}{6m^2}\left[R+\frac{11}{3}+\ln\frac{m^2}{\mu^2}\right]\right\}.\nonumber
     \eea
  \item 1 meson, 2 nucleons (Case A):
        \bea
     R_{A}&=&R_{A}^{(0)}+R_{A}^{(2)}+{\cal O}(p^4)+\cdots,
     \eea
     where
     \bea
     R_{A}^{(0)}&=&\frac{1}{32\pi^2m^2}\left\{R+1+\ln\frac{m^2}{\mu^2}\right\},\nonumber\\
     R_{A}^{(2)}&=&\frac{1}{32\pi^2m^2}\left\{\frac{t}{6m^2}\left[R+\ln\frac{m^2}{\mu^2}\right]-\frac{M_\pi^2}{m^2}\right\}.\nonumber
     \eea
  \item 1 meson, 2 nucleons (Case B):
        \bea
     R_{B}&=&R_{B}^{(0)}+R_{B}^{(1)}+R_{B}^{(2)}+{\cal O}(p^3)+\cdots,
     \eea
     where
     \bea
     R_{B}^{(0)}&=&\frac{1}{32\pi^2m^2}\left\{R+1+\ln\frac{m^2}{\mu^2}\right\},\nonumber\\
     R_{B}^{(1)}&=&-\frac{s-m^2}{2m^2}\frac{1}{32\pi^2m^2}\left\{R+2+\ln\frac{m^2}{\mu^2}\right\},\nonumber\\
     R_{B}^{(2)}&=&\frac{1}{32\pi^2m^2}\left\{\frac{M_\pi^2}{m^2}\left[\frac{1}{6}\left(R+\ln\frac{m^2}{\mu^2}\right)-1\right]-\frac{1}{6}\left(\frac{s-m^2}{m^2}\right)^2\left[R+\frac{3}{2}+\ln\frac{m^2}{\mu^2}\right]\right\}.\nonumber
     \eea
\item 3 nucleons:
        \bea
     R_{03}&=&R_{03}^{(0)}+R_{03}^{(2)}+{\cal O}(p^4)+\cdots,
     \eea
     where
     \bea
     R_{03}^{(0)}&=&\frac{1}{32\pi^2m^2},\nonumber\\
     R_{03}^{(2)}&=&\frac{1}{6}\frac{1}{32\pi^2m^2}\left\{\frac{t}{2m^2}+\frac{M_\pi^2}{m^2}\right\}.\nonumber
     \eea
  \item 1 meson, 3 nucleons:
        \bea
     R_{13}&=&R_{13}^{(0)}+R_{13}^{(1)}+{\cal O}(p^2)+\cdots,
     \eea
     where
     \bea
     R_{13}^{(1)}&=&-\frac{1}{32\pi^2m^4},\nonumber\\
     R_{13}^{(2)}&=&\frac{s-m^2}{m^2}\frac{1}{32\pi^2m^4}.\nonumber
     \eea
\end{itemize}

\section{Renormalization of the effective couplings}
\label{coupl}
\begin{itemize}
\item{ $\overline{\rm{MS}}-1$ renormalized LECs:}
{
\bea
c_i=c_i^r(\mu)+\frac{\gamma_i^c mR}{16 \pi^2F^2},\hspace{1cm}d_j=d_j^r(\mu)+\frac{\gamma_j^dR}{16\pi^2F^2},\hspace{1cm}e_k=e_k^r(\mu)+\frac{\gamma_k^eR}{16\pi^2F^2m}
\ ,\nonumber
\eea
}
{
\bea
\gamma_1^c&=&-\frac{3}{8}g^2+\frac{9}{2} {c_1} g^2 m\ ,\nonumber\\
\gamma_2^c&=&\frac{1}{2}-g^2+\frac{g^4}{2}+\left[-\frac{2c_4}{3}+\frac{1}{6} (3c_2+8c_3+4c_4) g^2\right] m\ ,\nonumber\\
\gamma_3^c&=&\frac{1}{4}-\frac{3 g^2}{2}+\frac{g^4}{4}+\left[\frac{5 {c_4}}{3}+\frac{1}{12} (-3 {c_2}+54 {c_3}-52 {c_4}) g^2\right] m\ ,\nonumber\\
\gamma_4^c&=&\frac{1}{4}-\frac{g^2}{2}+\frac{g^4}{4}+\left[{c_4}+\frac{1}{12} (-13 {c_2}+8 {c_3}-12 {c_4}) g^2\right] m\ ,\nonumber
\eea
\bea
\gamma_1^d+\gamma_2^d&=&\frac{1}{48}-\frac{g^2}{12}+\frac{g^4}{16}+\left[\frac{1}{24} (7 {c_2}+16 {c_3}-10 {c_4})+\frac{1}{48} (34 {c_2}-32 {c_3}+4 {c_4}) g^2\right] m\ ,\nonumber\\
\gamma_3^d&=&\frac{5}{6} {c_2} \left(1-g^2\right) m\ ,\nonumber\\
\gamma_5^d&=&\frac{1}{48}-\frac{g^2}{48}+\left[\frac{1}{48} (-72 {c_1}-3 {c_2}+8 {c_3}+4 {c_4})+\frac{1}{48} (48 {c_1}-4 {c_3}+2 {c_4}) g^2\right] m\ ,\nonumber\\
\gamma_{14}^d-\gamma_{15}^d&=&\frac{1}{4}-\frac{g^2}{2}+\frac{g^4}{4}+\left[{c_4}+\frac{1}{12} (-13 {c_2}+8 {c_3}-12 {c_4}) g^2\right] m\ ,\nonumber\\
\gamma_{18}^d&=&\frac{1}{12}(24{c_1}+{c_2}-4{c_3}-4{c_4})g m\ .\nonumber
\eea
\bea
\gamma_{14}^e&=&\frac{1}{64}-\frac{g^2}{32}+\frac{g^4}{64}+\left[\frac{1}{48} (-2 {c_2}-12 {c_3}-5 {c_4})+\frac{1}{192} (-13 {c_2}+8 {c_3}+20 {c_4}) g^2\right] m\ ,\nonumber\\
\gamma_{15}^e&=&\frac{1}{12} {c_2} g^2 m\ ,\nonumber\\
\gamma_{16}^e&=&0\ \,\nonumber\\
\gamma_{17}^e&=&\frac{1}{192}-\frac{g^2}{48}+\frac{g^4}{64}+\left[\frac{1}{96} (-{c_2}+2 {c_4})+\frac{7}{96} (3 {c_2}-2 {c_4}) g^2\right] m\ \,\nonumber\\
\gamma_{18}^e&=&\frac{1}{8} {c_2} \left(1-g^2\right) m\ ,\nonumber\\
2\gamma_{19}^e-\gamma_{22}^e-\gamma_{36}^e&=&\frac{1}{16}-\frac{g^2}{8}+\frac{g^4}{16}+\left[{c_1}-\frac{5 {c_2}}{48}+\frac{3 {c_3}}{8}+\frac{{c_4}}{3}+\frac{1}{48} (21 {c_2}+26 {c_3}-16 {c_4}) g^2\right] m\ ,\nonumber\\
\gamma_{20}^e+\gamma_{35}^e&=&\frac{1}{24} {c_2} \left(6-g^2\right) m\ ,\nonumber\\
2\gamma_{21}^e-\gamma_{37}^e&=&\frac{1}{48}-\frac{g^2}{48}+\left[\frac{1}{48} (-24 {c_1}+{c_2}+16 {c_3})+\frac{1}{24} (-3 {c_2}-6 {c_3}+14 {c_4}) g^2\right] m\ ,\nonumber\\
\gamma_{22}^e-4\gamma_{38}^e&=&-\frac{1}{32}+\frac{g^2}{16}-\frac{g^4}{32}+\left[\frac{1}{24} (-36 {c_1}+3 {c_2}+9 {c_3}+{c_4})+\frac{1}{96} (-72 {c_1}-21 {c_2}-4 (2 {c_3}+{c_4})) g^2\right] m\ .\nonumber
\eea
}
\item{EOMS renormalized LECs:}
{
\bea
c_i^r=\widetilde{c}_i+\frac{\delta_i^c m}{16 \pi^2F^2},\hspace{1cm}d_j^r=\widetilde{d}_j+\frac{\delta_j^d}{16\pi^2F^2}
\ ,\nonumber
\eea
}
{
\bea
\delta_1^c&=&\frac{3 g^2}{8}+3 {c_1} g^2 m\ ,\nonumber\\
\delta_2^c&=&-1-\frac{g^4}{2}+\left[-\frac{2 {c_4}}{9}+\frac{1}{9} (9 {c_2}+16 {c_3}+14 {c_4}) g^2\right] m\ ,\nonumber\\
\delta_3^c&=&\frac{9 g^4}{4}+\left[\frac{2 {c_4}}{9}+\frac{1}{72} (-9 {c_2}+216 {c_3}-272 {c_4}) g^2\right] m\ ,\nonumber\\
\delta_4^c&=&-\frac{5 g^2}{4}-\frac{g^4}{4}+\left[-\frac{1}{72} (9 {c_2}+32 {c_3}+16{c_4})+\frac{1}{72} (9 {c_2}-88 {c_4}) g^2\right] m\ ,\nonumber
\eea
\bea
\delta_1^d+\delta_2^d&=&-\frac{1}{36}\left[ (4 {c_2}+10 {c_3}+5 {c_4})+(8 {c_2}-22 {c_3}+38 {c_4}) g^2\right] m\ ,\nonumber\\
\delta_3^d&=&\frac{1}{18}\left[ (-34 {c_2}-30 {c_3}+3 {c_4})+ (4 {c_2}+15 {c_4}) g^2\right] m\ ,\nonumber\\
\delta_5^d&=&\frac{1}{72} \left[(144 {c_1}-2 {c_3}-{c_4})+ (72 {c_1}-38 {c_3}+10 {c_4}) g^2\right] m\ ,\nonumber\\
\delta_{14}^d-\delta_{15}^d&=&\left[\frac{{c_4}}{3}+\frac{1}{72} (67 {c_2}-56 {c_3}+96 {c_4}) g^2\right] m\ ,\nonumber\\
\delta_{18}^d&=&\frac{1}{9} (c2 - c3 - c4) g m\ .\nonumber
\eea
}
\end{itemize}

\section{The effect of unitarity}\label{secunitarity}
The phase shift formula Eq.~\ref{phaseshiftper} for a perturbative amplitude automatically includes the effect of unitarity in an obscure way. This is illustrated in this section.

The partial wave phase shift for an amplitude, satisfying the elastic unitarity relation exactly, can be obtained via Eq.~\ref{phaseshiftuni}, which reads
\bea
\delta_{\ell\pm}^I=\arctan\left\{\frac{{\rm Im} f_{\ell\pm}^I}{{\rm Re} f_{\ell\pm}^I}\right\}\ .
\eea
Nevertheless, for a perturbative amplitude the unitarity relation may be violated. In section~\ref{pwexpantion}, we follow Ref.~\cite{Meissnerp3} to define the phase shift presented by Eq.~\ref{phaseshiftper}. Hence, starting from a perturbative amplitude, one may have several  methods to calculate the phase shift. For instance:
\begin{itemize}
  \item Method 1: perturbative ampl. $f^P\xrightarrow{{\rm Eq.}~\ref{phaseshiftper}} \delta^P =\arctan\left\{|\vec{p}|{\rm Re}f^P\right\}$.
  \item Method 2: perturbative ampl. $f^P\xrightarrow{\texttt{unitarization using K-Matrix method}}$ unitarized ampl. $f^K\rightarrow\delta^K =\arctan\left\{\frac{{\rm Im} f^K}{{\rm Re} f^K}\right\}$.
\end{itemize}
Here the K-Matrix approach is employed to unitarize the perturbative amplitude, one can also adopt other approaches of unitarization. $f^P$ and $f^k$ denote the perturbative amplitude and the K-Matrix unitarized amplitude, respectively. $\delta^P$ stands for the phase shift calculated via Eq.~\ref{phaseshiftper} with $f^P$, while $\delta^K$ presents the one calculated via Eq.\ref{phaseshiftuni} with $f^K$. Note that the indices for isospin and angular momentum are suppressed hereafter. In our paper, Method 1 has been adopted to calculate the phase shift. Below we take the perturbative $\pi$-N scattering amplitude up to $O(p^4)$ without the $\Delta(1232)$ contribution for example to demonstrate that $\delta^P=\delta^K$.

The chiral perturbative $\pi$-N scattering partial wave amplitude at $O(p^4)$ is expressed by
\bea
f^P(s)=f^{(1)}(s)+f^{(2)}(s)+f^{(3)}(s)+f^{(4)}(s)\ .
\eea
Using the K-Matrix approach, the unitarized amplitude that obeys the unitarity relation takes the following form,
\bea\label{unifK}
f^K(s)=\frac{f^{(1)}(s)+f^{(2)}(s)+{\rm Re}f^{(3)}(s)+{\rm Re}f^{(4)}(s)}{1-i|\vec{p}|\left(f^{(1)}(s)+f^{(2)}(s)+{\rm Re}f^{(3)}(s)+{\rm Re}f^{(4)}(s)\right)}\ ,
\eea
where $f^{(1)}(s)$ and $f^{(2)}(s)$ stand for the contributions from the $O(p)$ and $O(p^2)$ tree amplitudes, respectively. The fact that ${\rm Re}f^{(1)}(s)=f^{(1)}(s)$ and ${\rm Re}f^{(2)}(s)=f^{(2)}(s)$ has been used to get Eq.~\ref{unifK}.
The phase shift obtained through Method 1 reads
\bea
\delta^P=\arctan\left\{|\vec{p}|{\rm Re}f^{P}(s)\right\}=\arctan\left\{|\vec{p}|\left[f^{(1)}(s)+f^{(2)}(s)+{\rm Re}f^{(3)}(s)+{\rm Re}f^{(4)}(s)\right]\right\}\ ,
\eea
while the one obtained through Method 2 reads
\bea
\delta^K=\arctan\left\{\frac{{\rm Im} f^K(s)}{{\rm Re} f^K(s)}\right\}=\arctan\left\{|\vec{p}|\left[f^{(1)}(s)+f^{(2)}(s)+{\rm Re}f^{(3)}(s)+{\rm Re}f^{(4)}(s)\right]\right\}\ .
\eea
Thus, $\delta^P=\delta^K$. A numerical calculation we performed also supports this observation. Hence in our work the effect of unitarity has already been included  when performing fits to the partial wave shift data in Section~\ref{secpar}. The phase shift formula Eq.~\ref{phaseshiftper} for the perturbative amplitude is reasonable in the sense that it takes the effect of unitarity into consideration automatically. The phase shift calculated via Eq.~\ref{phaseshiftper} using the perturbative amplitude and the one via Eq.~\ref{phaseshiftuni} using the K-Matrix unitarized amplitude are the same, which is true at least for the calculation up to $O(p^4)$ in this paper.

\begin{figure}[!ht]
\centering
\includegraphics[width=0.497\textwidth]{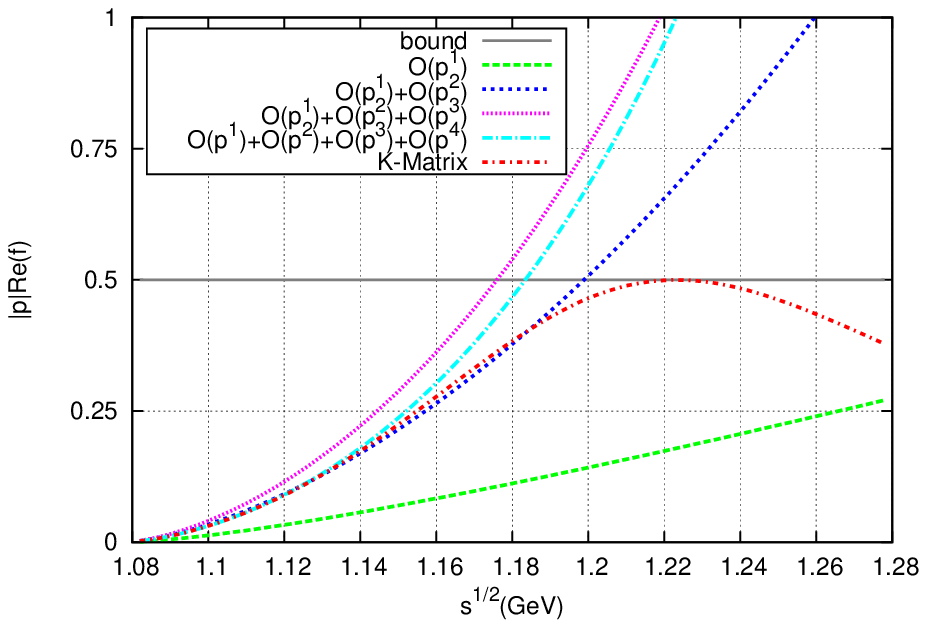}
\includegraphics[width=0.497\textwidth]{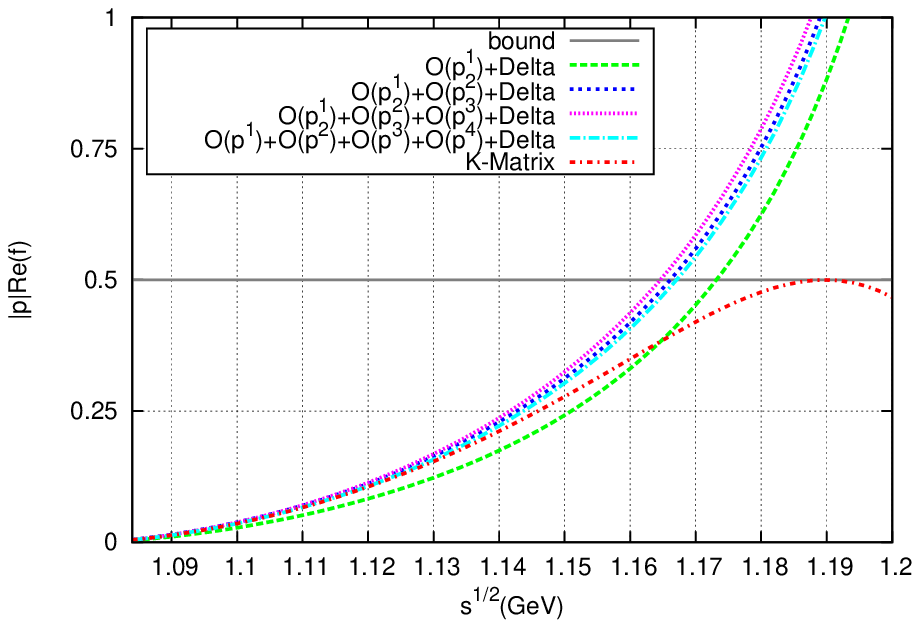}
\caption{Unitarity bound for the real part of the $P_{33}$ partial wave amplitude. Left: without the $\Delta$(1232) contribution; Right: with the $\Delta$(1232) contribution. The results from Table \ref{tabp4s2} are adopted for plotting.}\label{p33UB}
\end{figure}

 Nevertheless, the advantage of the unitarized amplitude 
 can be shown by plotting the real part of the unitarized amplitude, the one of the perturbative amplitude and the unitary bound together, e.g. see Fig.~\ref{p33UB} for the $P_{33}$ partial wave.

\section{The effect of the $\Delta(1232)$ width}
\label{secdelwidth}
\begin{figure}[!h]
\centering
\includegraphics[width=0.6\textwidth]{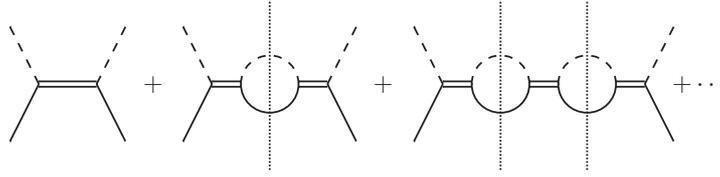}
\caption{Resummation to generate the $\Delta(1232)$ width. The solid, dashed and double solid lines represent the nucleon, pion and $\Delta(1232)$, respectively. The vertical dotted lines show that the nucleon and pion in the loops are on mass shell.}\label{reply11}
\end{figure}

In section~\ref{secDel}, when the $\Delta(1232)$ is included explicitly, the phase shift is obtained through Method 1 discussed in~\ref{secunitarity},
\bea
\delta^P=\arctan\left\{|\vec{p}|{\rm Re}f^{P}(s)\right\}=\arctan\left\{|\vec{p}|\left[f^{\Delta,Born}(s)+f^{(1)}(s)+f^{(2)}(s)+{\rm Re}f^{(3)}(s)+{\rm Re}f^{(4)}(s)\right]\right\}\ ,
\eea
where $f^{\Delta,Born}(s)$ represents the LO Born term contribution (see Eq.~\ref{LOdeltaBT}) without an explicit $\Delta$ width in the propagator.
According to the discussion shown in~\ref{secunitarity}, it is equivalent to calculate the phase shift via a K-Matrix unitarized amplitude
\bea
f^K(s)=\frac{f^{\Delta,Born}(s)+f^{(1)}(s)+f^{(2)}(s)+{\rm Re}f^{(3)}(s)+{\rm Re}f^{(4)}(s)}{1-i|\vec{p}|\left(f^{\Delta,Born}(s)+f^{(1)}(s)+f^{(2)}(s)+{\rm Re}f^{(3)}(s)+{\rm Re}f^{(4)}(s)\right)}\ .
\eea
If one forgets $f^{(1)}(s)+f^{(2)}(s)+{\rm Re}f^{(3)}(s)+{\rm Re}f^{(4)}(s)$ for a while , then
\bea
f^{K}(s)=\frac{f^{\Delta,Born}(s)}{1-i|\vec{p}|\left(f^{\Delta,Born}(s)\right)}\ .
\eea
The above equation contains an infinite resummation of Feynman diagrams shown in Fig.~\ref{reply11}. This resummation will generate the $\Delta(1232)$ width properly.

On the other hand, one can add to the $\Delta$ propagator in the $f^{\Delta,Born}(s)$ an explicit width given by Ref.\cite{pascalutsa},
\bea
\Gamma(s)=2\left(\frac{h_A}{2F_\pi}\right)^2\frac{s+m_N^2-M_\pi^2}{24\pi m_\Delta^2}|\vec{p}|^3\ .
\eea
This provides us a LO Born term $\Delta$-exchange contribution with the explicit $\Delta$ width, which we denote it by $f^{\Delta\Gamma,Born}(s)$.
Using Eq.~\ref{phaseshiftuni}, we have checked that the effect of $f^{ K}(s)$ is almost the same as that of $f^{\Delta\Gamma,Born}(s)$.

Now taking $f^{(1)}(s)+f^{(2)}(s)+{\rm Re}f^{(3)}(s)+{\rm Re}f^{(4)}(s)$ into consideration, one can observe that the resummation shown in Fig.~\ref{reply11} still exists though many new terms of different types appear. To conclude, in our work, the dominant effect of the $\Delta$ width is included through Method 1, namely through the phase shift formula Eq.~\ref{phaseshiftper} for the $\pi$-N perturbative amplitude, when performing fits to phase shift up to 1.2 GeV.

\end{document}